\documentclass[fleqn,usenatbib]{mnras}

\usepackage{newtxtext,newtxmath}

\usepackage[T1]{fontenc}

\DeclareRobustCommand{\VAN}[3]{#2}
\let\VANthebibliography\thebibliography
\def\thebibliography{\DeclareRobustCommand{\VAN}[3]{##3}\VANthebibliography}

\usepackage{graphicx}	
\usepackage{amsmath}	
\usepackage{pdflscape}
\usepackage{xspace}
\usepackage{stfloats}

\newcommand{\Sec}[1]{{\protect\hyperref[sec:#1]{Section~\ref*{sec:#1}}}}
\newcommand{\Secs}[2]{{\protect\hyperref[sec:#1]{Sections~\ref*{sec:#1}}~and~\ref{sec:#2}}}
\newcommand{\Fig}[1]{{\protect\hyperref[fig:#1]{Figure~\ref*{fig:#1}}}}

\newcommand{\subFig}[2]{{\protect\hyperref[fig:#1]{Figure~\ref*{fig:#1}~#2}}}
\newcommand{\Equ}[1]{{\protect\hyperref[equ:#1]{Equation~\ref*{equ:#1}}}}

\newcommand{\Tab}[1]{{\protect\hyperref[tab:#1]{Table~\ref*{tab:#1}}}}

\newcommand{\App}[1]{{\protect\hyperref[app:#1]{Appendix~\ref*{app:#1}}}}

\newcommand{\HII}{H{\protect\footnotesize{II}}} 

\newcommand{\magss}{{mag{\,}arcsec\ensuremath{^{-2}}}}
\newcommand{\nodata}{\protect\dots}
\newcommand{\wunits}[2]{\ensuremath{#1\,\text{#2}}}
\defcitealias{Jedrzejewski1987}{J87}

\date{Accepted 2021 September 15. Received 2021 September 15; in original form 2021 June 24}
\pubyear{2021}

\title[AutoProf]{AutoProf -- I. An automated non-parametric light profile pipeline for modern galaxy surveys}

\author[C. Stone et al.]{
Connor J. Stone,$^{1}$\thanks{connor.stone@queensu.ca}
Nikhil Arora,$^{1}$
St{\'e}phane Courteau,$^{1}$ and
Jean-Charles Cuillandre,$^{2}$
\\
$^{1}$Department of Physics, Engineering Physics \& Astronomy, Queen{'}s University, Kingston, ON K7L 3N6, Canada\\
$^{2}$AIM, CEA, CNRS, Université Paris-Saclay, Université de Paris, F-91191 Gif-sur-Yvette, France}

\begin{document}
\label{firstpage}
\pagerange{\pageref{firstpage}--\pageref{lastpage}}
\maketitle

\begin{abstract}
    We present an automated non-parametric light profile extraction pipeline called {\scriptsize AUTOPROF}.
    All steps for extracting surface brightness (SB) profiles are included in {\scriptsize AUTOPROF}, allowing streamlined analyses of galaxy images.
    {\scriptsize AUTOPROF} improves upon previous non-parametric ellipse fitting implementations with fit-stabilization procedures adapted from machine learning techniques. 
    Additional advanced analysis methods are included in the flexible pipeline for the extraction of alternative brightness profiles (along radial or axial slices), smooth axisymmetric models, and the implementation of decision trees for arbitrarily complex pipelines. 
    Detailed comparisons with widely used photometry algorithms ({\scriptsize PHOTUTILS}, {\scriptsize XVISTA}, and {\scriptsize GALFIT}) are also presented.  
    These comparisons rely on a large collection of late-type galaxy images from the PROBES catalogue.
    The direct comparison of SB profiles shows that {\scriptsize AUTOPROF} can reliably extract fainter isophotes than other methods on the same images, typically by $>2$ \magss.
    Contrasting non-parametric elliptical isophote fitting with simple parametric models also shows that two-component fits (e.g., S\'ersic plus exponential) are insufficient to describe late-type galaxies with high fidelity.
    It is established that elliptical isophote fitting, and in particular {\scriptsize AUTOPROF}, is ideally suited for a broad range of automated isophotal analysis tasks.
    {\scriptsize AUTOPROF} is freely available to the community at: \break\url{https://github.com/ConnorStoneAstro/AutoProf}.
\end{abstract}


\section{Introduction}\label{sec:intro}

Images of galaxies contain a wealth of valuable information about their structure, stellar populations, dust content, dynamics, and evolutionary history.
A common approach to extract this information from galaxy images involves modelling structural components, such as a bulge, disc, bar, and spiral arms, with pre-determined fitting functions~\citep{Hubble1930,deVauc1948,Einasto1965,sersic1968,Freeman1970}. 
While these model decompositions can be arbitrarily complex~\citep{Kent1985,Peng2002,Peng2010,Erwin2015}, their execution may be time consuming and require individual attention.  
Parametric models typically struggle with complex galaxy morphologies and manual intervention is necessary to return realistic solutions.
In practice, large galaxy surveys typically employ simplified models that can be automated \citep[e.g.,][]{Kelvin2010,Simard2011}.

A complimentary approach to analysing galaxy images involves their non-parametric\footnote{A "non-parametric" isophotal fit can use many parameters to describe the PA, ellipticity, and semimajor axis of each elliptical isophote, however these parameters have a non-trivial mapping to a structural model such as a bulge, disc, bar, etc.} representation into elliptical isophotes [contours of constant surface brightness (SB)] yielding average brightness profiles for each galaxy.
Much like the parametric methods above, the non-parametric modelling of elliptical isophotes from galaxy images has a rich history \citep[e.g.,][]{Carter1978,Kent1983,Lauer1985,Davis1985,Tody1986,Jedrzejewski1987}. 

The extraction of representative azimuthally averaged profiles of SB, position angle (PA), and ellipticity as a function of (galactocentric) radius takes advantage of the near axisymmetry of a 2D image of a galaxy, which is a projection on the sky of a complex 3D object.
Non-parametric reductions of galaxy images into 1D SB profiles can then be further decomposed into structural parameters~\citep{MacArthur2003,Ciambur2016,Bottrell2017,Gilhuly2018}.

Non-parametric models and complete pipeline automation in the era of large galaxy surveys~\citep[e.g.,][]{Amiaux2012,Ivezic2019,Mosby2020} will allow extensive representations of the data complexity available in such surveys.
However, current techniques of extracting robust non-parametric light profiles can still be time consuming and require human intervention.
For instance, \cite{Smith2021} estimated a time period of 2 yr to extract light profiles interactively for $10^5$ objects. 
A fast, flexible, automated tool to extract such galaxy light profiles will be required to take full advantage of the influx of photometric survey data in the next decade and beyond.

In this paper, we present a package called {\scriptsize AUTOPROF}, for the fast, automated, non-parametric extraction of SB profiles using elliptical isophotes.  
{\scriptsize AUTOPROF} borrows from \citet[][hereafter \citetalias{Jedrzejewski1987}]{Jedrzejewski1987}, though with the addition of regularization techniques from machine learning, and yields robust SB profiles that match light distributions, even in complex systems.
It requires no human intervention and can process a galaxy image in \wunits{{\sim}15}{s} (on a single 4GHz Intel i7-4790K processor core), though processing time varies quadratically with the galaxy size (in this case, $\sim$700 pixels).
{\scriptsize AUTOPROF} is a pipeline building code designed to take full advantage of modern galaxy surveys which allows arbitrarily complex analysis for advanced users, yet offers a robust default setup for rapid startup.
We compare {\scriptsize AUTOPROF} to a number of established photometry codes using DESI Legacy Sky Survey images~\citep{Dey2019} to demonstrate the context in which {\scriptsize AUTOPROF} is ideally suited.

The format of the paper is as follows.
\Sec{autoprofdefault} reviews the default {\scriptsize AUTOPROF} pipeline, with steps such as background, point-spread-function (PSF) estimations, centroid algorithm, isophotal fitting, light profile extraction, fit checks, and forced photometry. 
{\scriptsize AUTOPROF} offers a suite of advanced pipeline tools presented in \Sec{advancedpipelinetools}, such as star masking, wedge-sector radial sampling, lateral slicing, smooth model image reconstruction, as well as instructions to integrate and execute new functions. 
\Sec{comparisons} offers comparisons of {\scriptsize AUTOPROF} output products with those from similar isophotal fitting codes, such as  
(i) the `{\scriptsize PHOTUTILS}' implementation of the non-parametric isophotal fitting algorithm by \citetalias{Jedrzejewski1987},
(ii) the PROFILE algorithm \citep{Lauer1985,Courteau1996} within the {\scriptsize XVISTA} software package for astronomical image processing\footnote{We shall refer to the PROFILE isophotal fitting software as `{\scriptsize XVISTA}'. 
The {\scriptsize XVISTA} software package is maintained by Jon Holtzman at NMSU (New Mexico State University); http://ganymede.nmsu.edu/holtz/xvista/}, and   
(iii) the parametric galaxy image modeller {\scriptsize GALFIT} \citep{Peng2010} which decomposes galaxy images using pre-defined parametric models.
These comparisons highlight the pros and cons of model-independent and parametric light profile extractions, and our ability to scale these operations to very large multiwavelength photometric investigations. 
Main results and typical applications for {\scriptsize AUTOPROF} are summarised in \Sec{conclusions}. 
{\scriptsize AUTOPROF} is provided to the community in a flexible and modular software package. Step-by-step instructions to run {\scriptsize AUTOPROF} can be found in the documentation of the code repository: \url{https://github.com/ConnorStoneAstro/AutoProf}.

\section{The Default {\scriptsize AUTOPROF} Pipeline}
\label{sec:autoprofdefault}

{\scriptsize AUTOPROF} is an automated, fully featured, light profile extraction pipeline with modular steps.
The core {\scriptsize AUTOPROF} algorithm fits elliptical isophotes automatically to galaxy images and extracts accurate flux measurements along those ellipses.
The user constructs a configuration file which specifies the images for {\scriptsize AUTOPROF} to analyse and the steps involved in that analysis.  
Provided a galaxy image, pixel scale, and flux zero point, {\scriptsize AUTOPROF} can automatically determine the sky level, galaxy centre, fit elliptical isophotes, and extract an SB profile.
After extraction, {\scriptsize AUTOPROF} reports on the quality of the extraction and flags potentially problematic fits (see \Sec{checkfit}); this is important in the fully automated domain it is intended for.
{\scriptsize AUTOPROF} is designed to condense these and other tasks into an automated pipeline. 
This section presents {\scriptsize AUTOPROF}'s default pipeline to analyse galaxy images.

\subsection{Background Estimation}
\label{sec:backgroundestimation}

\begin{figure}
    \centering
    \includegraphics[width=0.8\columnwidth]{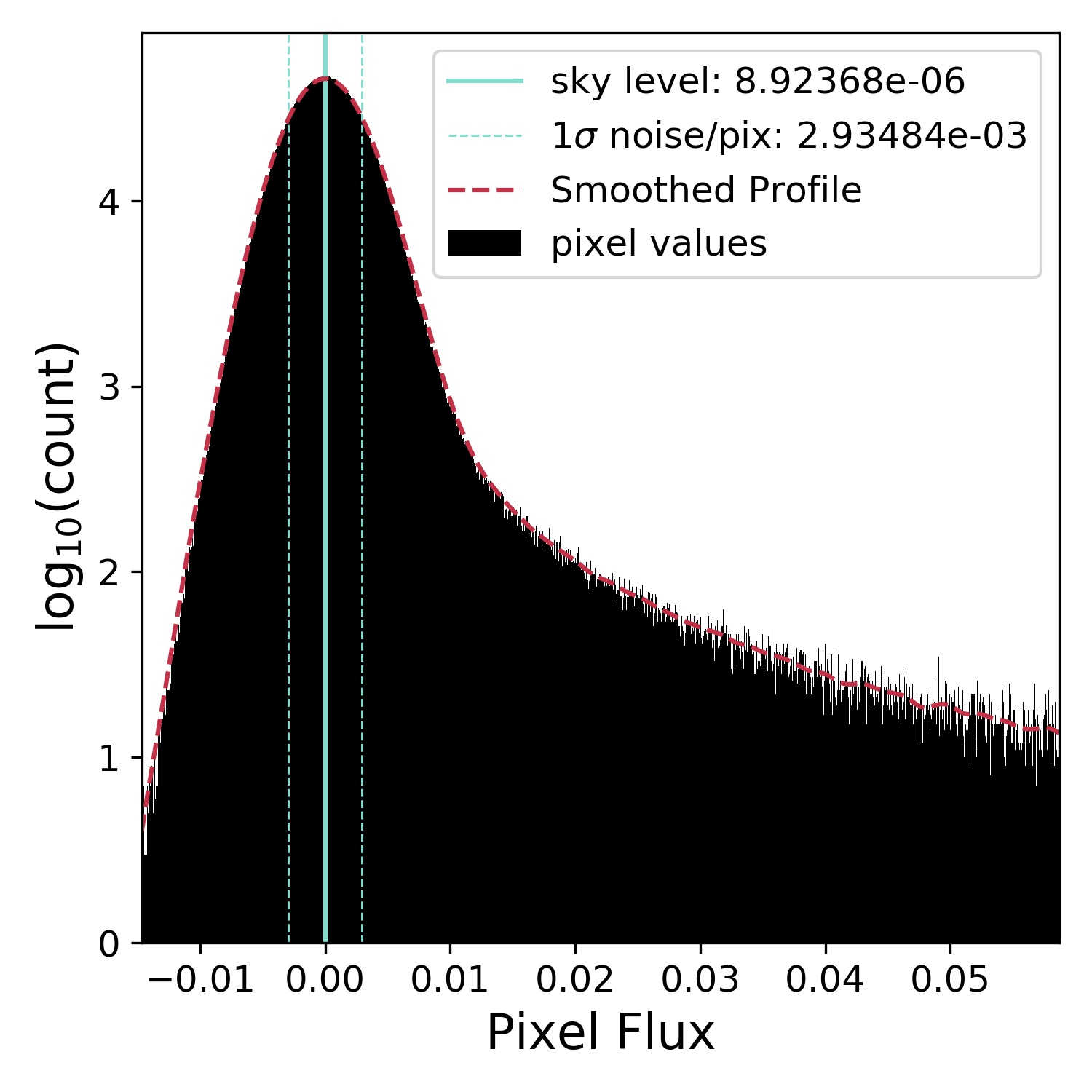}
    \includegraphics[width=0.8\columnwidth]{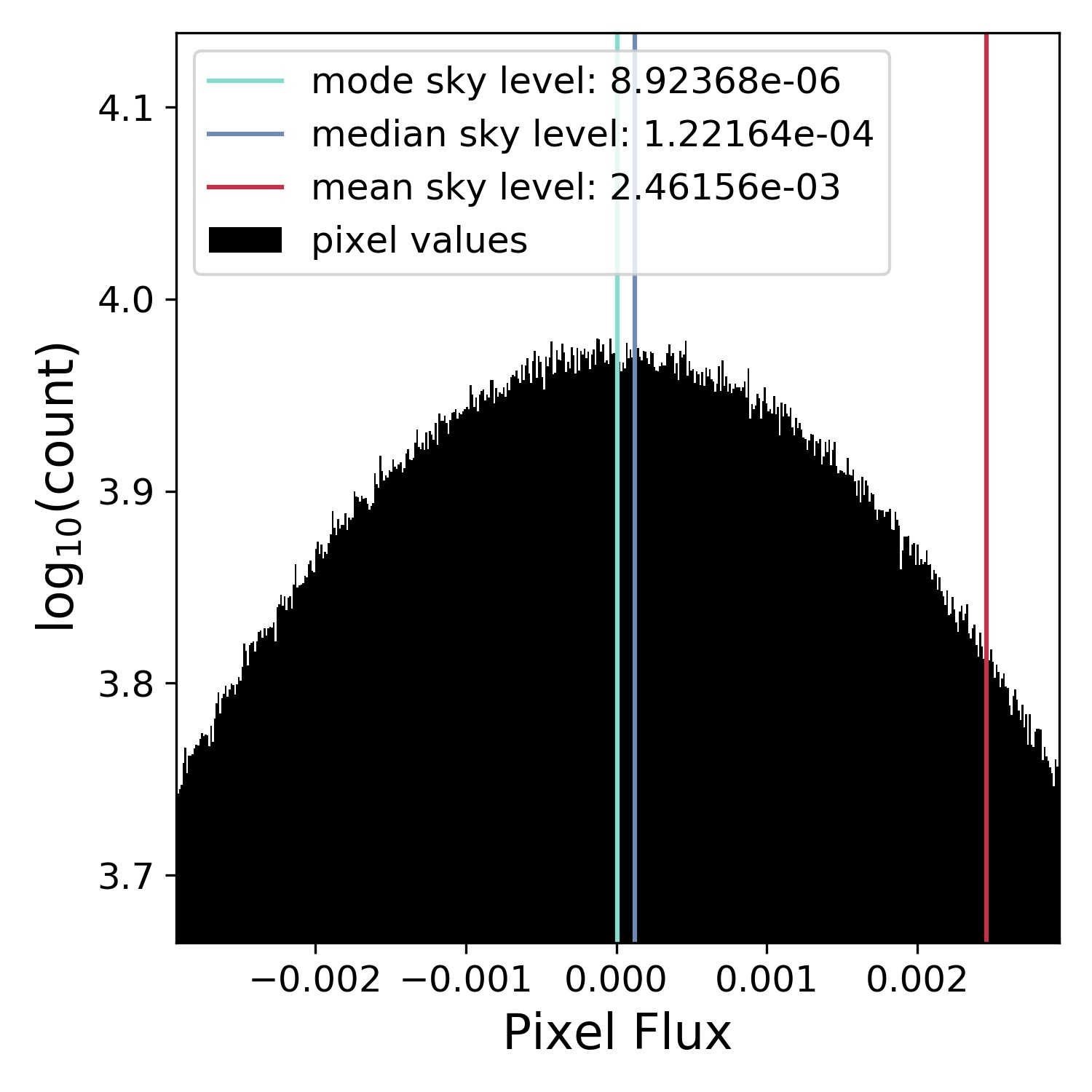}
    \caption{Background estimation procedure from the default {\scriptsize AUTOPROF} pipeline for \emph{ESO479-G1}. 
    Top figure: pixel flux values from the outer 1/5th border of the image (see \Fig{psf}).
    In black, a histogram of flux values, the red dashed line is a Gaussian smoothed profile, the blue vertical lines show the mode flux value and the $1\sigma$ scatter (as measured based on flux values below the mode flux).
    Bottom figure: zoom-in on the peak of the flux distribution.
    Also shown are mean and median values, both of which are biased high due to unmasked objects (stars, background galaxies, etc.).
    }
    \label{fig:backgroundestimation}
\end{figure}

All galaxy images contain a background signal level resulting from a combination of sky brightness, read noise, zodiacal light, diffraction, and a host of other effects.
To ensure that a light profile contains only signal from the galaxy, this background signal must be removed.
This is especially important for {\scriptsize AUTOPROF} which is designed to push faint SB limits for a given observing campaign (often more than \wunits{4}{\magss} deeper than the background noise level).
The default background calculation uses the mode of the pixel flux values, which is less affected by outliers relative to mean or median estimates. 
This method uses all pixels that are within 1/5th of the image width from the edge. 
Large images with sufficient sky background areas will be beneficial for such calculations.  
The peak of a Gaussian smoothed density profile in flux space is then identified and used as the background sky level (see \Fig{backgroundestimation}). 
The smoothing function is given by \Equ{backgroundestimation}:
\begin{equation}\label{equ:backgroundestimation}
    B = {\rm argmax}_x \sum_i e^{-\frac{(f_i - x)^2}{\sigma^2}}
\end{equation}
\noindent where $B$ is the mode background level, $f_i$ is the flux of the $i^{\rm th}$ pixel, and $\sigma$ is the smoothing length.
The smoothing length is $\sigma = \Delta_{25}^{75}/\log_{10}(\sqrt{N})$, where $\Delta_{25}^{75}$ is the interquartile range, and $N$ is the number of pixels (the precise choice of smoothing length is not very important for most tasks).
The use of mode averaging and interquartile ranges means that bright unwanted sources (stars, cosmic rays, and hot pixels) have minimal effect on the estimated background.
Large faint sources such as Galactic cirri, nearby galaxies, and intra-cluster light can bias the mode estimator. 
If these sources affect an appreciable fraction of the total number of pixels (approximately one quarter of the field of view) then the background estimator will likely be biased high.
Such areas should be avoided , masked, or a more sophisticated background estimation algorithm could be applied.
The sky noise is taken as the 68.3 percentile of flux values below the sky level ($\Delta_{31.7}^{100}$).
For properly processed astronomical images, free of systematic effects (poor detrending, faulty large scale background removal), flux values below the sky level should mostly be contributed by Poisson noise while flux values above the sky level can be contaminated by faint sources.

The default mode background estimation is accurate enough for most purposes, and is reliable in an automated framework.
However, more precise background levels can be determined with specialized clipping procedures~\citep{Ratnatunga1984, Akhlaghi2019}.
The user may take advantage of such external techniques and provide {\scriptsize AUTOPROF} with a specified background level.
{\scriptsize AUTOPROF} also includes a wrapper of the {\scriptsize PHOTUTILS} dilated source mask method, which identifies bright sources and masks sky annuli around them to remove faint tails from the image.
While effective, this method is slower than the {\scriptsize AUTOPROF} default. 

\subsection{PSF Estimation}
\label{sec:psf}

\begin{figure}
    \centering
    \includegraphics[width=0.8\columnwidth]{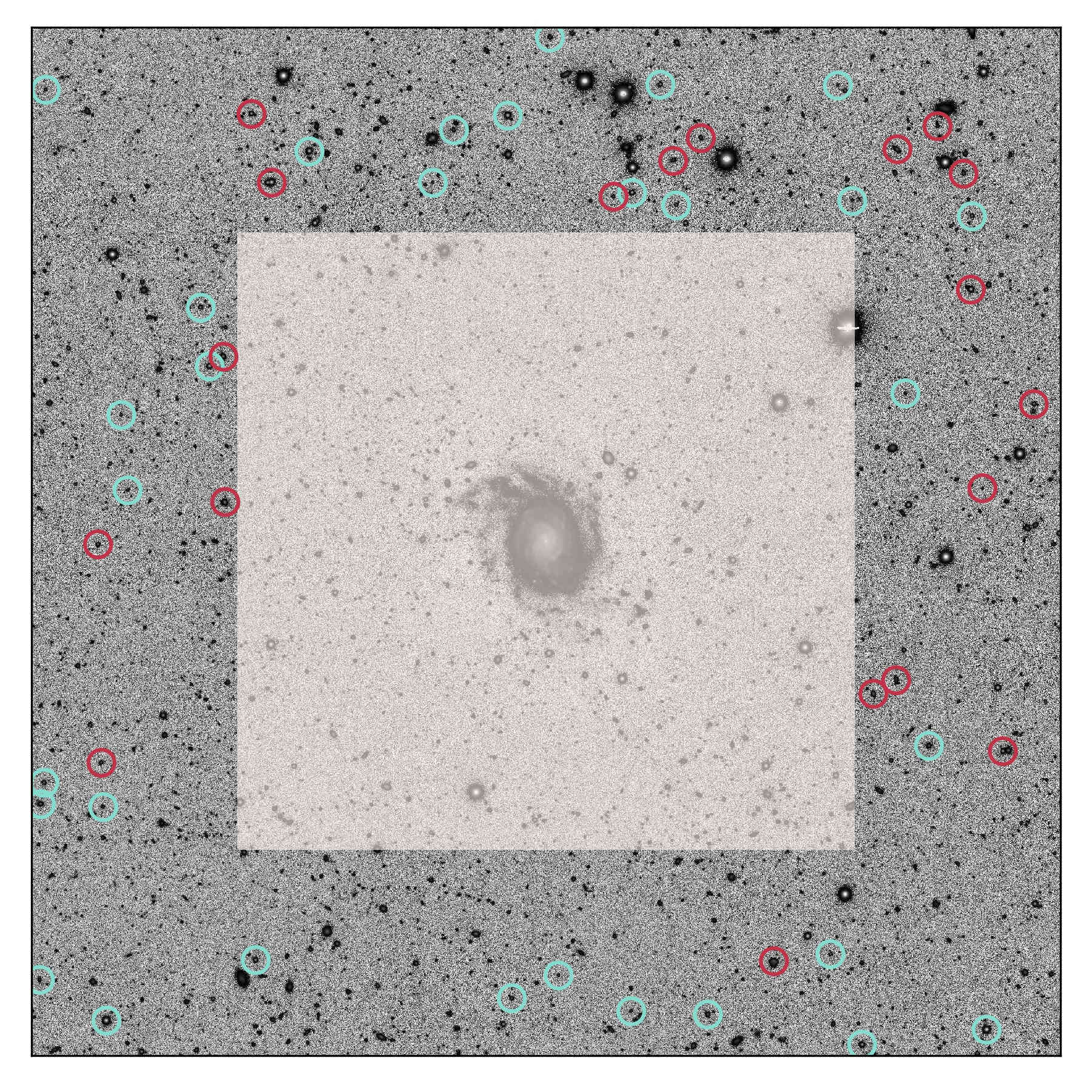}
    \caption{PSF estimation procedure from the default {\scriptsize AUTOPROF} pipeline for \emph{ESO479-G1}.
    Highlighted with circles is a selection of randomly chosen star candidates from the image.
    In red, star candidates that have been rejected based on the roundness criteria of power in low order FFT coefficients.
    In blue, star candidates that have been accepted. The PSF is taken as the FWHM, determined by a Lanczos interpolation between pixels determining the diameter at which the flux reaches half the central value.
    The shaded area is masked from the PSF analysis; this is the same region as the background estimation (see \Sec{backgroundestimation}).
    }
    \label{fig:psf}
\end{figure}

The default PSF estimation method searches for star candidates using a peak finding convolution filter.
Once selected, stellar fluxes are measured within circular apertures using a Lanczos interpolation between pixels until the flux value reaches half of the central peak value.
The full width at half-maximum (FWHM) is taken as the PSF.
While measuring the flux, {\scriptsize AUTOPROF} also tracks the low order fast-Fourier transform (FFT) coefficients of flux values around each star candidate.
Specifically, the `deformity' metric for perturbations from a circle are measured as:
\begin{equation}
    d = \sum_{i=1}^{5}\left|\frac{\mathcal{F}_i}{\tilde{\mu} + \sigma_b}\right|
\end{equation}
\noindent where $d$ is the deformity, $\mathcal{F}_i$ is the $i^{th}$ Fourier coefficient, $\tilde{\mu}$ is the median flux, and $\sigma_b$ is the background noise per pixel (\Sec{backgroundestimation}).
Note that $\tilde{\mu} +\sigma_b$ is nearly equivalent to $\mathcal{F}_0$, though it is more stable at low S/N in the presence of outliers.
Once 50 star candidates have been measured around the galaxy, a median of the "roundest" half of the star candidates is used as the image flux value; where roundness is determined by the power in low order FFT coefficients previously measured.
\Fig{psf} represents the PSF estimation procedure graphically. 

Since {\scriptsize AUTOPROF} samples flux values along isophotes, a PSF estimate is not required for its calculations.
{\scriptsize AUTOPROF} operates on galaxies of all apparent sizes from only a few dozen to many thousand of pixels across.
Primarily, the PSF estimate is used to set a meaningful scale on the image, determining the lower bound for geometrically growing radii when fitting/sampling from the image.
Thus {\scriptsize AUTOPROF} is robust to imprecise PSF estimates. 
However, the approximate PSF of inner isophotes at or below the PSF scale will be biased round due to resolution effects.

\begin{figure*}
    \centering
    \includegraphics[width=0.3\textwidth]{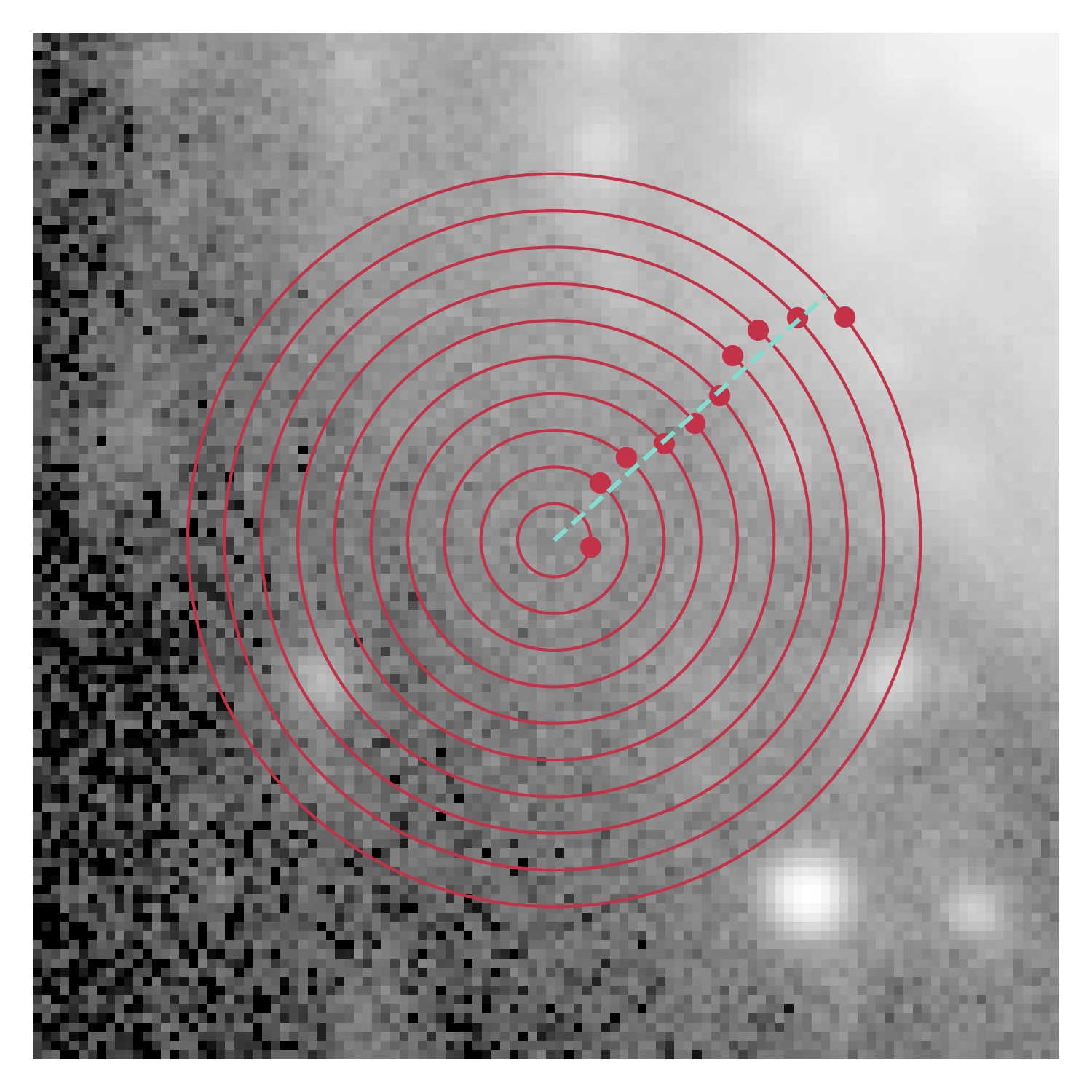}
    \includegraphics[width=0.3\textwidth]{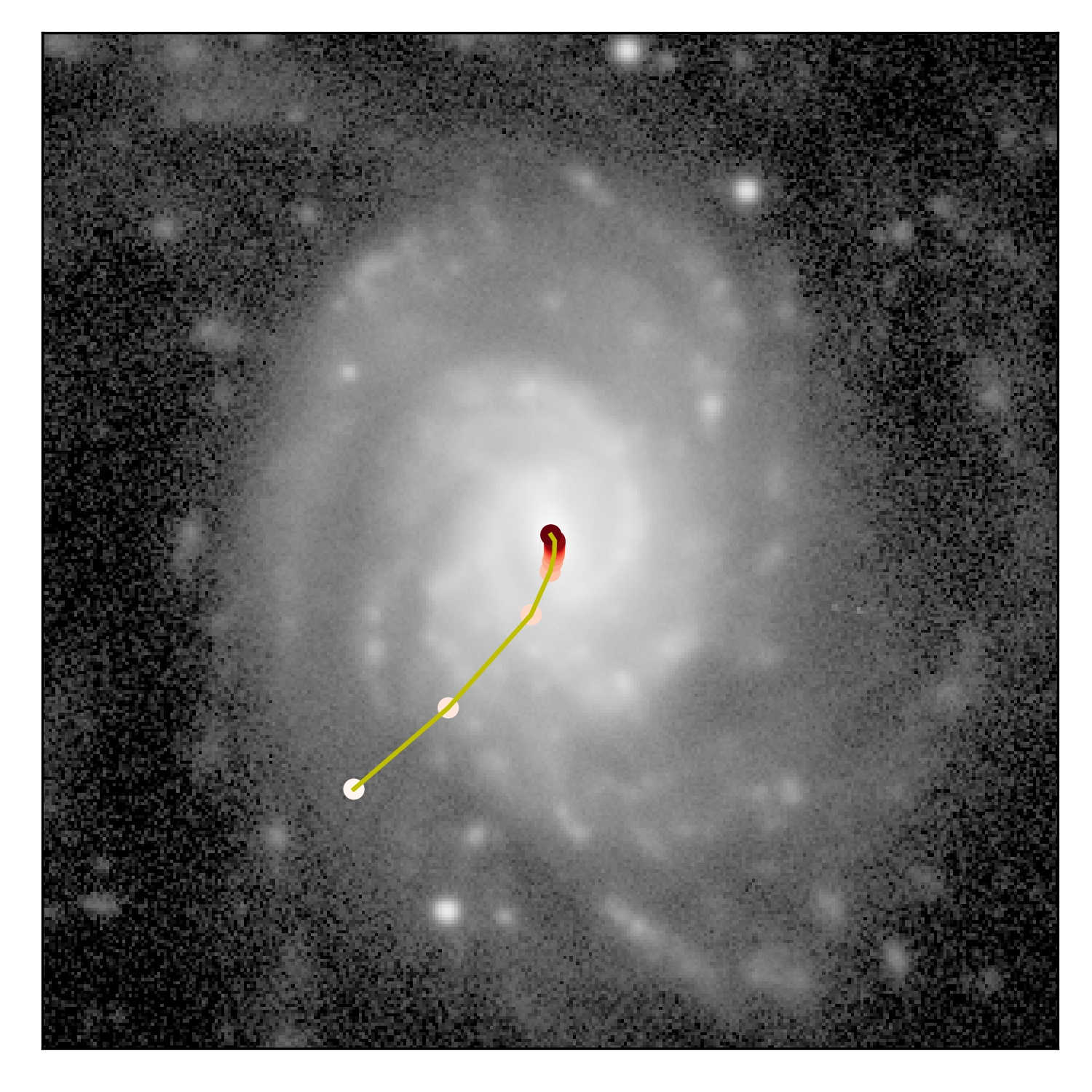}
    \includegraphics[width=0.3\textwidth]{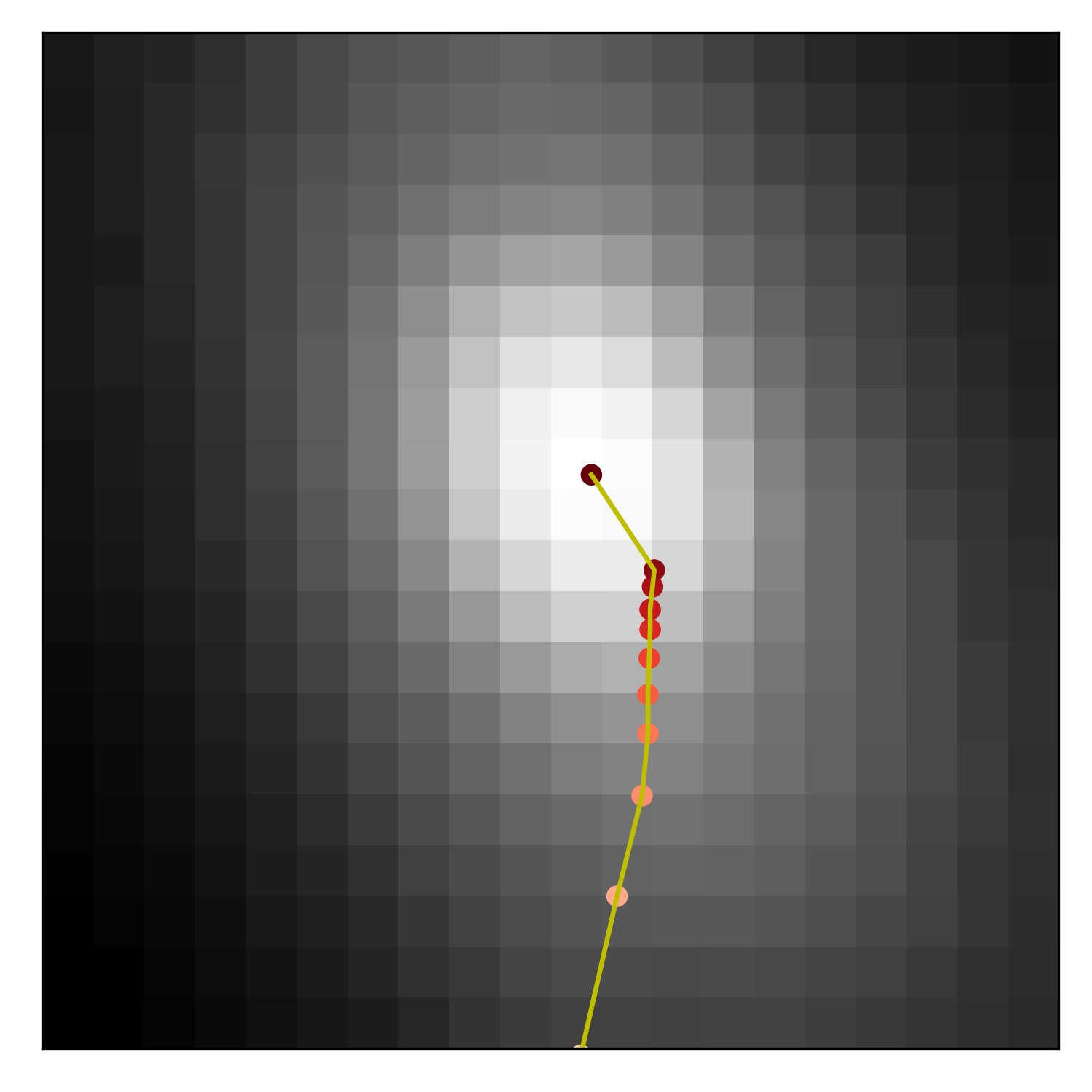}
    \caption{centre finding procedure from the default {\scriptsize AUTOPROF} pipeline for \emph{ESO479-G1}.
    Left: a single iteration of the centre finding procedure, 10 rings of flux are extracted (red rings) and the phase of $\mathcal{F}_1$ is used to provide a direction of increasing flux (red dots on the rings).
    The 10 directions are then averaged to choose a robust direction of increasing flux (blue dashed line direction) and a parabola is fit along that direction to determine the location of the maximum (blue dashed line length).
    Middle: the path taken by the iterative centre finding method. The first few steps are as large as is allowed (to the edge of the 10th ring), then smaller steps refine the centre.
    Right: a zoom-in at the centre of the galaxy showing the final few steps.
    The large last step shows the result of the centre refining with only three flux rings instead of the 10 used for the global fit; this smaller window can only be used near the global maximum as it can easily be caught in local maxima.
    }
    \label{fig:centre}
\end{figure*}

\subsection{centre Finding}
\label{sec:centre}

The {\scriptsize AUTOPROF} default centre finding algorithm iteratively updates a candidate centre until it reaches a global flux peak.
The candidate centre may be the image centre, or one supplied by the user; this allows {\scriptsize AUTOPROF} to examine any galaxy in the image.
Generally, the algorithm starts searching from the centre of the image or the user may provide an initial centre to {\scriptsize AUTOPROF} in fractional pixel coordinates.
From the initialization point, 10 circular apertures are created out to 10 times the PSF size and samples flux values around each aperture. 
An FFT is taken for the flux values around each circular aperture and the phase of $\mathcal{F}_1$ is used to determine the direction of increasing brightness. 
Taking the average direction, flux values are sampled along a line from the centre out to 10 times the PSF. 
A parabola is fit to the flux values and the centre is then updated to the maximum of the parabola. 
This is repeated until the updated steps move by half a PSF.
To refine the centre estimate at a level below 1 pixel, a Nelder-Mead simplex optimizer is used to minimize $|\mathcal{F}_1/(\tilde{\mu} + \sigma_b)|$ evaluated on three circular apertures (out to three PSF).
\Fig{centre} illustrates the iterative centre finding procedure.
In the figure, we intentionally initialize the algorithm away from the galaxy centre; the path demonstrates how the centre finding method can ignore local features to fit the global centre of the galaxy.
This method of centre finding is accurate when initialized in the main part of a galaxy, and works well for objects that are roughly centered on the image plane.
Failed centre identification is discussed in \Sec{checkfit}.

\subsection{Initialization}
\label{sec:isophoteinitialize}

\begin{figure}
    \centering
    \includegraphics[width=0.8\columnwidth]{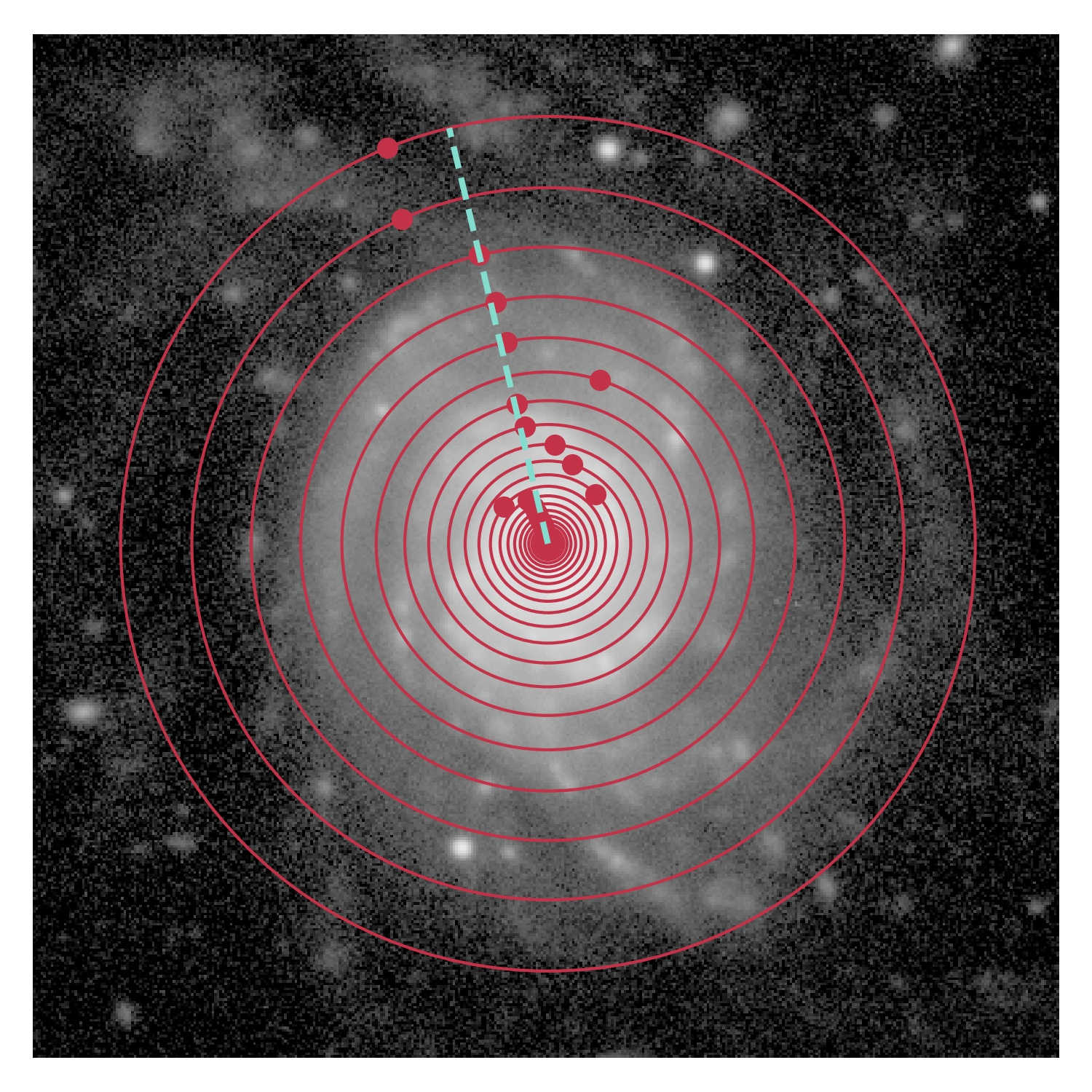}
    \includegraphics[width=0.8\columnwidth]{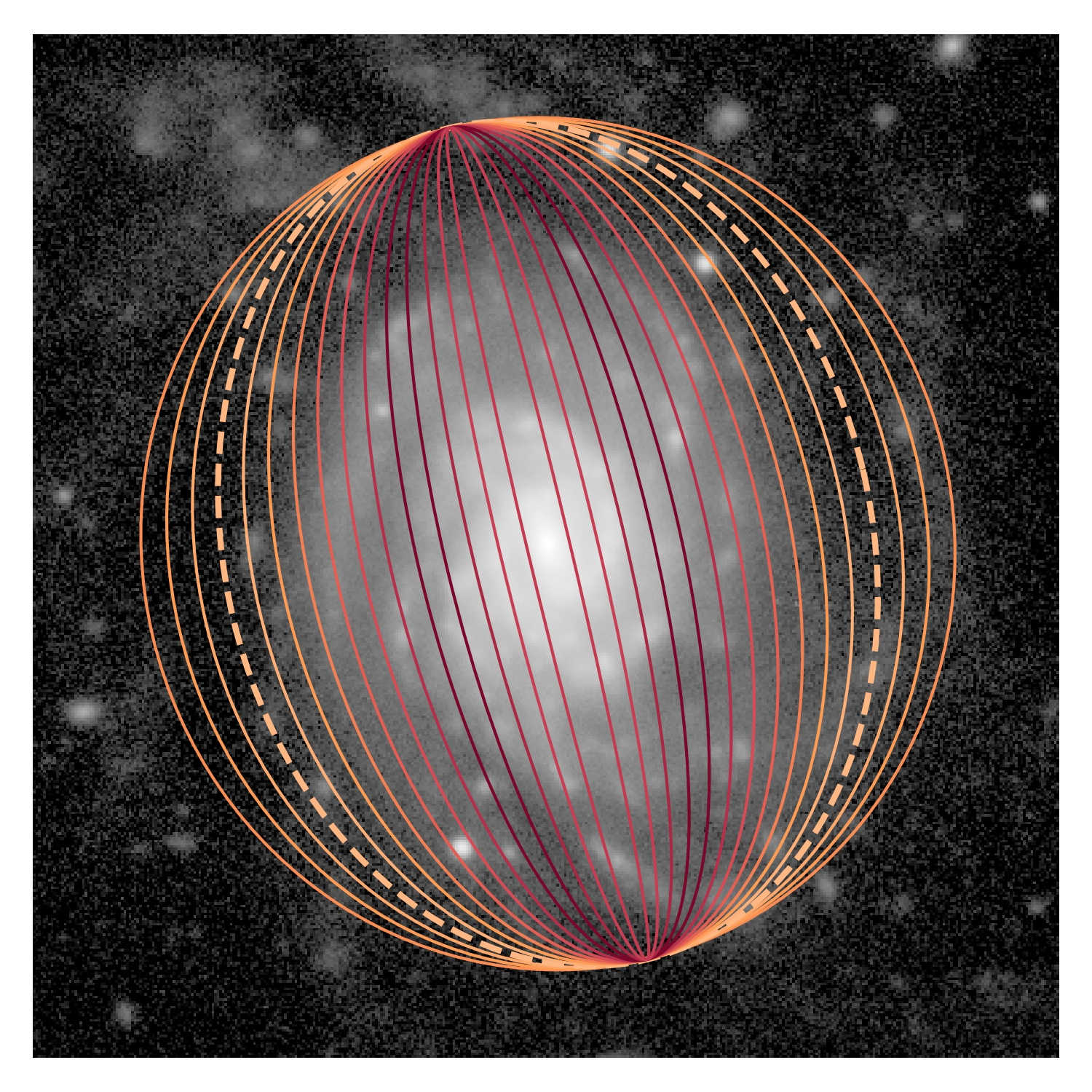}
    \caption{Isophote initialization procedure from the default {\scriptsize AUTOPROF} pipeline for \emph{ESO479-G1}.
    Top: a global PA is determined by taking rings of flux (shown as red rings) and computing the phase of the $\mathcal{F}_1$ coefficient (red dots on the rings).
    The global PA is determined as the average direction for the outer five rings.
    Bottom: the global ellipticity is determined by testing a range of ellipticity values and computing $|\mathcal{F}_2/\mathcal{F}_0|$, where brighter colours represent lower values.
    The best candidate ellipticity (shown as a dashed line) is then further optimized using a Nelder-Mead minimization routine on $|\mathcal{F}_2/\mathcal{F}_0|$.
    }
    \label{fig:isophoteinitialize}
\end{figure}

The Fourier analysis that determines the initial isophotal parameters fits the global PA and ellipticity of each ellipse in a two-step process.
First, a series of circular apertures are sampled geometrically in radius starting at the centre until they approach the background level of the image. 
An FFT is taken for the flux values around each circular aperture and the phase of $\mathcal{F}_2$ is used to determine a direction. 
The average direction for the outer isophotes is taken as the PA of the galaxy. 
Secondly, with fixed PA, the ellipticity is optimized to minimize $|\mathcal{F}_2/(\tilde{\mu} + \sigma_{b})|$.
\Fig{isophoteinitialize} illustrates the two-step process of fitting the global PA and ellipticity.

The errors on PA and ellipticity are assessed by fitting multiple isophotes with semimajor axes that are within 1 PSF length of each other in the outer part of the galaxy.
These isophotes should have near identical fits, but noise in the data and non-axisymmetric features can cause deviations.
It is those deviations that contribute to the uncertainty in PA and ellipticity as reported for the global fit.

\subsection{Isophotal Fitting}
\label{sec:isophotefitting}

With background, PSF, centroid, and an ellipse (PA and ellipticity) initialization in place, {\scriptsize AUTOPROF} can fit elliptical isophotes to the galaxy image. 
For this isophotal fitting step, the default {\scriptsize AUTOPROF} function uses a method similar to \citetalias{Jedrzejewski1987} except with a modification taken from the field of machine learning. 
The latter improves upon speed and accuracy of the fitted results (see below).
Machine learning is also used in other galaxy image analysis software, such as {\scriptsize DEEPLEGATO} \citep{Tuccillo2018} and {\scriptsize PIX2PROF} \citep{Smith2021}. 
These methods incorporate more comprehensive machine learning-based pipelines, making them fast and accurate within their trained domain.
{\scriptsize AUTOPROF}, on the other hand, provides a rich toolkit with additional user customizability and alternative output products such as those presented in \Sec{advancedpipelinetools}.
Thus {\scriptsize AUTOPROF} is ideally suited for a broad range of galaxy photometry tasks.

A series of elliptical isophotes with the global PA and ellipticity determined in \Sec{isophoteinitialize} are constructed by growing geometrically in semimajor axis length from the centre (calculated in \Sec{centre}) until they approach the background level. 
The algorithm iteratively updates the PA and ellipticity of each isophote individually (in a random order) for many rounds. 
Each round cycles between three options: optimizing PA, ellipticity, and PA/ellipticity simultaneously. 
To optimize the parameters, a Monte Carlo method is used.
New values for the current parameter (PA, ellipticity, or both) are randomly sampled with a small scatter around their current value. 
The "loss" is now computed for the new parameters (PA, ellipticity, or both) and compared to the loss before scattering.
The loss is a combination of the relative amplitude of $\mathcal{F}_2$\footnote{For comparison with the \citetalias{Jedrzejewski1987} method discussed in \Sec{photutils}, we note that ${\rm Im}(\mathcal{F}_2) = A_2$ and ${\rm Re}(\mathcal{F}_2) = B_2$.} and a regularization term. 
The regularization term is borrowed from machine learning and penalizes adjacent isophotes for having a significantly different PA or ellipticity (using the $l_1$ norm). 
For the $i^{\rm th}$ isophote the loss is:
\begin{equation}
    \begin{aligned}
        l = &\frac{|\mathcal{F}_2|}{\tilde{\mu} + \sigma_b}\left(1 + \frac{|e_i - e_{i-1}|}{1-e_{i-1}} + \frac{|e_i - e_{i+1}|}{1-e_{i+1}}\right. \\
        & \left.+\frac{|p_i - p_{i-1}|}{0.2} + \frac{|p_i - p_{i+1}|}{0.2}\right)
    \end{aligned}
\end{equation}
\noindent where $l$ is the loss to be minimized, $N$ is the number of flux values sampled along the isophote, $e_i$ and $p_i$ are the ellipticity and PA of the $i$th isophote respectively.
The first factor in the loss function $\left(\frac{|\mathcal{F}_2|}{\tilde{\mu}+\sigma_b}\right)$ is nearly equivalent to the \citetalias{Jedrzejewski1987} optimization value, except that the denominator is adapted for robustness of the fit.
The coupling between isophotes in the second factor takes the form of the $l_1$ norm~\citep{Shalev2014}; the denominators ($1-e_{i\pm 1}$ for $e_i$ and $0.2$ for $p_i$) are chosen to normalize the scale for the impact of parameter deviations.
Thus, all the isophotes are coupled and tend to fit smoothly varying profiles.

For an intuitive understanding of the regularization term, consider an example where all the isophotes have been initialized to the same PA and ellipticity.
If the $i$th isophote was to change PA by $0.2$ radians ($11.5^{\circ}$), then the regularization term would equal $3$; meaning that the $\frac{|\mathcal{F}_2|}{\tilde{\mu}+\sigma_b}$ term would have to drop by more than a factor of $3$ for the update to be accepted.
While it is rare for two adjacent isophotes to differ by so much, this example demonstrates how the regularization term prevents the optimization routine from making extreme updates for minimal improvements in the fit.
The exact choice of $0.2$ as the regularization scale for PAs is not critical and near identical solutions will result from different choices of that scale. 
The user can also adjust the impact of the regularization term with a simple scale parameter; increasing the parameter for smoother fits, and decreasing it to allow for wider variability.

\begin{figure}
    \centering
    \includegraphics[width = 0.9\columnwidth]{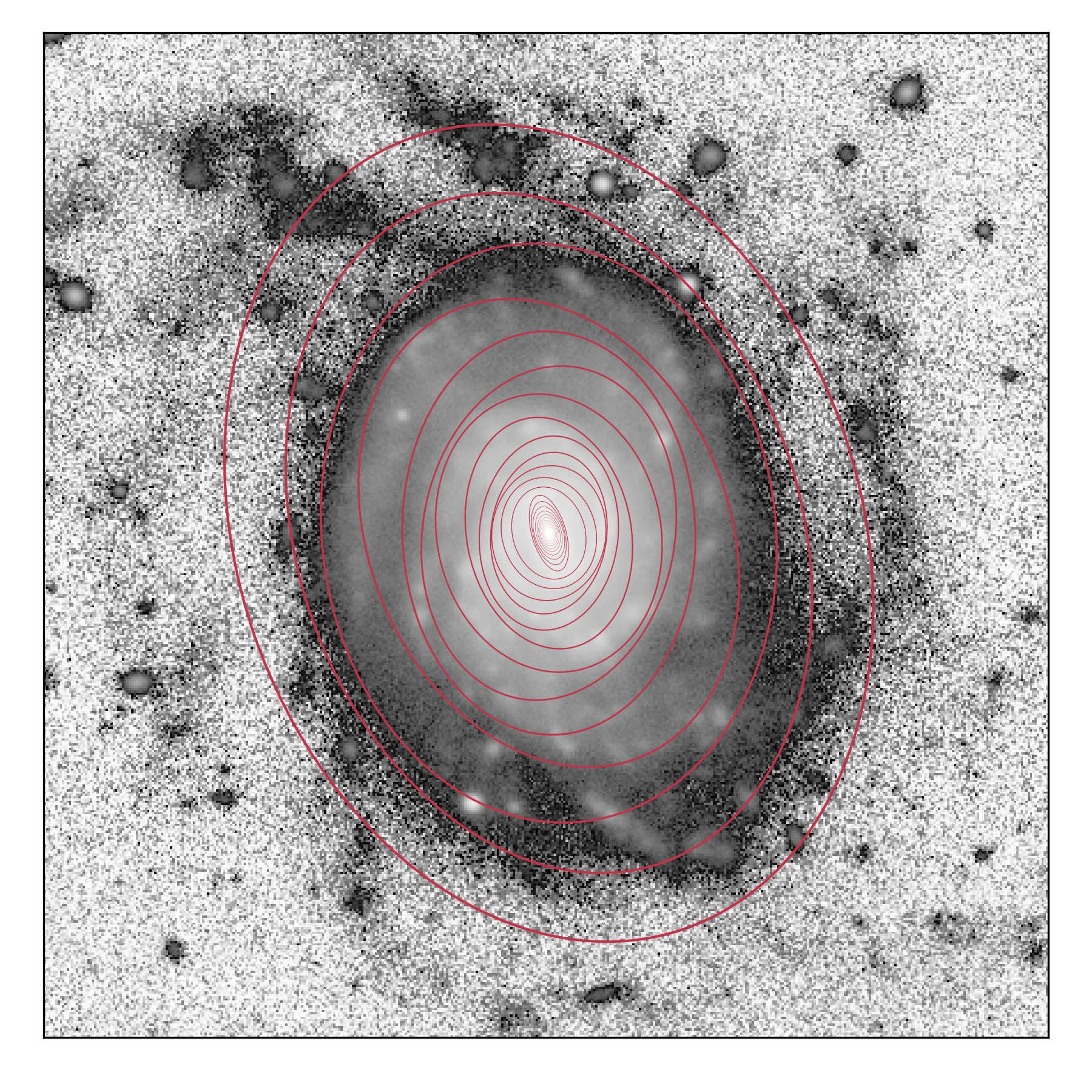}
    \caption{Example isophotal fitting solution for \emph{ESO479-G1}. {\scriptsize AUTOPROF} tracks complex axisymmetric features even in the presence of many non-axisymmetric components such as spiral arms, {\HII} clouds, and foreground stars.}
    \label{fig:examplefit}
\end{figure}

If the optimization completes three rounds without any isophotes updating, the profile is assumed to have converged. 
\Fig{examplefit} shows an example isophotal solution (red solid lines) on a complex galaxy image.
An uncertainty for each PA and ellipticity is determined by re-fitting 10 more isophotes at $\pm 5\%$ of the semimajor axis length and taking the scatter in the fitted values. 

Some codes enable ``deformed'' elliptical isophote fitting through higher order Fourier modes.
This is especially useful for early-type galaxies which display discy or boxy structures, though its implementation can be taxing in an automated context. 
{\scriptsize AUTOPROF} does not currently include this capability, though the power in Fourier modes around the elliptical isophotes can be computed to arbitrary order.

Overlapping galaxies (merging or projected) will break the assumptions in this fitting scheme and may cause undefined behaviour in the resulting outputs.
As {\scriptsize AUTOPROF} is not designed for the treatment of overlapping systems, such configurations are better suited for other methods (e.g. \Sec{galfit}).
{\scriptsize AUTOPROF} includes diagnostic checks (see \Sec{checkfit}) that aim at detecting such systems. 

\subsection{SB Profile Extraction}
\label{sec:sbextraction}

Extracting reliable flux measurements from an image and maximizing S/N can be a challenging process.
{\scriptsize AUTOPROF} approaches these goals in three steps.
For galactocentric radii less than five times the PSF, the flux values change rapidly from pixel to pixel as central regions of galaxies are bright and centrally peaked.
For this reason, {\scriptsize AUTOPROF} uses a Lanczos interpolation (default Lanczos-5 but this is user adjustable) between pixels to sample flux values exactly along each specified elliptical isophote~\citep{Shannon1949, Burger2010}.
This method samples pixel fluxes accurately, albeit slowly, along an isophote.
For intermediate radii, where pixels have S/N $>$ 2, the flux values are not assumed to change rapidly and a half pixel offset should not significantly affect the SB measurement.
In this regime, fluxes are sampled from the nearest neighbouring pixel at regular angular separations (eccentric anomaly).
For larger radii, where the S/N drops considerably, {\scriptsize AUTOPROF} samples a band of pixels around an isophote whose width is by default \wunits{5}{per cent} of the isophote radius.
This method enables {\scriptsize AUTOPROF} to maximize S/N and reach Low Surface Brightness (LSB) regimes. 
In the default configuration, isophote semimajor radii grow by \wunits{10}{per cent} so it is guaranteed that no pixel is counted twice (which would thwart the rigorous statistical interpretation of flux measurements and their uncertainty).
The user may adjust the band width and radii growth scales arbitrarily and allow for overlapping isophotes; this produces artificially smoother profiles as the flux measurements will be correlated. 
This may or may not be an issue depending on the application.
Also of note, because brightness profiles typically follow power-laws with semimajor axis, a large band relative to the power-law scale can bias flux measurements due to the inclusion of pixels which are not precisely at the desired radius.
Here is another reason why band averaging is only used in the outskirts by default.
Modelling the local power-law in brightness can allow users to mitigate the band averaging bias~\citep{Lauer1986}, though this feature is not included in {\scriptsize AUTOPROF}.

\begin{figure}
    \centering
    \includegraphics[width = 0.9\columnwidth]{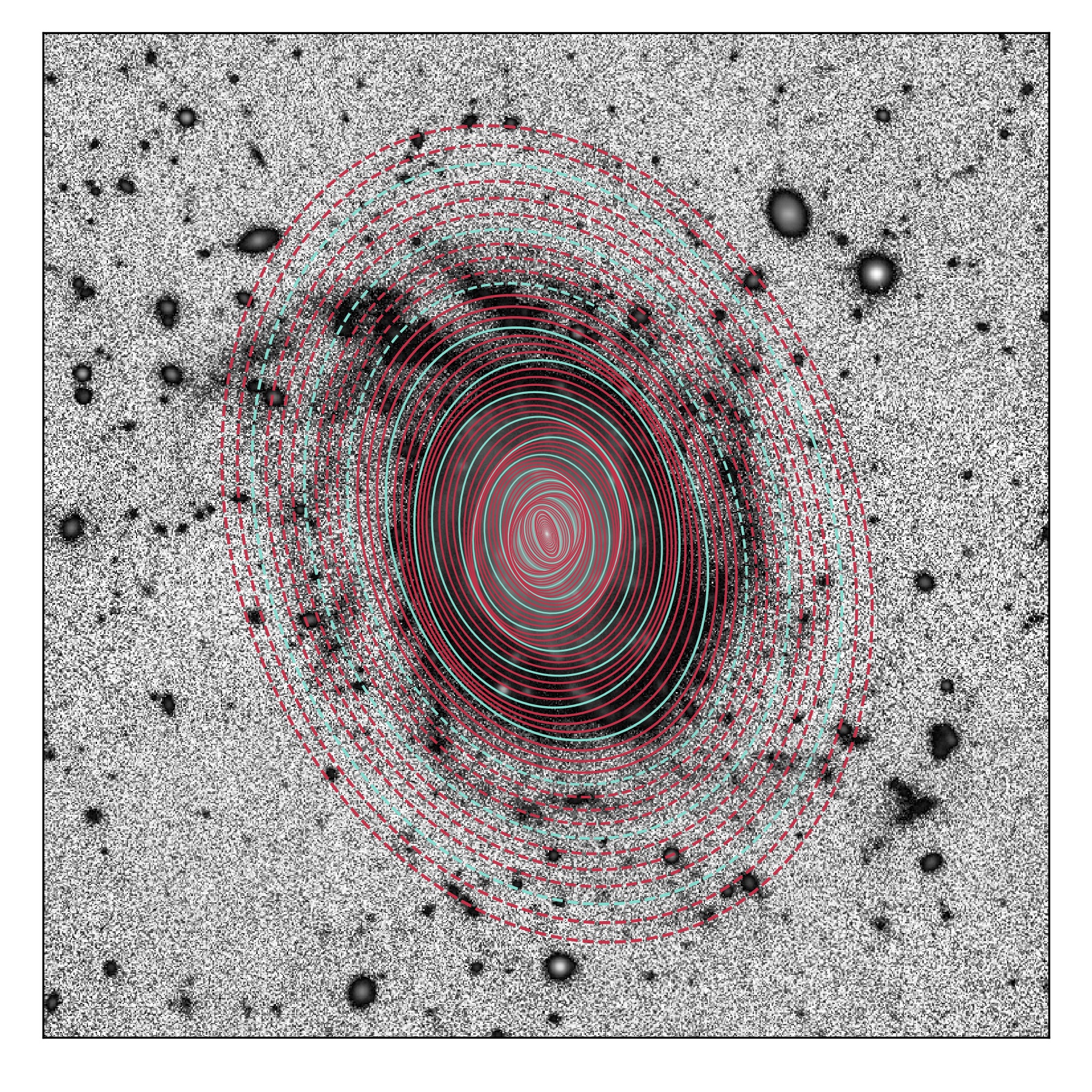}
    \includegraphics[width = 0.9\columnwidth]{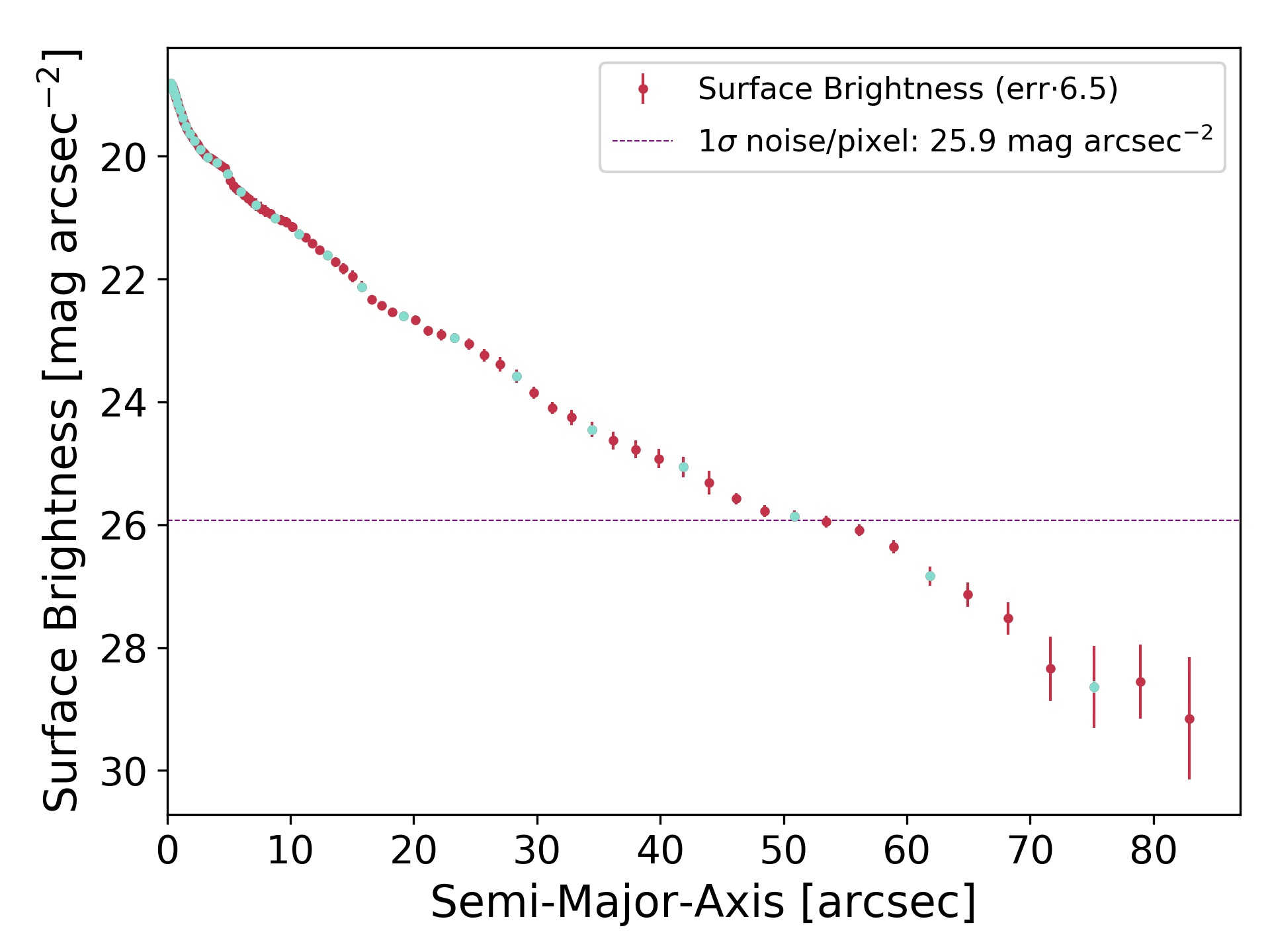}
    \caption{Diagnostic plots for the final SB profile of \emph{ESO479-G1}. Top: sampling ellipses used for extraction of isophotal flux values. The dashed ellipses indicate extrapolation of the ellipticity and PA profile from the fitted values (extrapolated values are kept constant). 
    Bottom: extracted median SB profile; cyan points (every fourth isophote) correspond to the cyan ellipses in the top figure facilitating visual comparison. The error bars have been scaled up to make them more visible; the scale factor is shown in the figure legend. The purple horizontal dotted line shows the estimated 1$\sigma$ noise per pixel from \Sec{backgroundestimation}.}
    \label{fig:photometry}
\end{figure}

Once a band of pixels is selected, {\scriptsize AUTOPROF} computes the average flux using the median by default (though mean and mode are available).
Additionally, {\scriptsize AUTOPROF} can perform sigma-clipping to remove bright sources along an isophote; users may take advantage of this feature in especially crowded fields (foreground stars, cluster environments, background galaxies on deep images, etc.).
A segmentation map such as from {\scriptsize SEXTRACTOR}~\citep{Bertin1996} or other masks can also be provided to remove unwanted sources from the flux extraction procedure.
A carefully masked image should yield more reliable galaxy flux estimates (see \Sec{mask}), though sigma-clipping may produce comparable results.

Tests of forced photometry on generic regions of sky data (including cirri, stars, background galaxies, etc.) show that {\scriptsize AUTOPROF} can regularly achieve robust SB levels of \wunits{{\sim}28}{\magss} in the DESI $g$- and $r$-bands, and can often go even deeper in a clean environment (no large neighbour, no large contaminant such as a bright star).
To achieve this level requires the sigma clipping feature; masks are also more effective.
The limiting brightness is certainly a function of the sky environment and care must be taken when interpreting the results. 
 
In order to track the total light within an isophote, {\scriptsize AUTOPROF} automatically includes two estimates of the total light; (i) an integral of the SB profile and, (ii) a direct pixel summation.
The direct pixel summation includes all non-masked pixels (masked pixels are flagged and never replaced by a numerical value).
The SB profile integration is performed as:
\begin{equation}
    f = 2\pi\int_0^R rI(r)q(r)dr 
\end{equation}
\noindent where $f$ is the total flux, $R$ is the final semimajor axis of the integral, $r$ is the semimajor axis over which the integral is performed, $I(r)$ is the flux as a function of semimajor axis, and $q(r)$ is the axis ratio as a function of semimajor axis.
The integral is performed with the trapezoid method and has no treatment for overlapping isophotes.
In principle, direct pixel summation should produce the most accurate total flux measurements.
However, in practice, foreground objects and overlapping galaxies may alter this outcome.
If these external sources are not carefully masked, the SB integral method is favoured for more reliable total flux measurements.

With the final isophotal solution in place, {\scriptsize AUTOPROF} in its default configuration generates a number of output products.
The primary product, an SB profile, is a text file with a .prof extension.
Such an SB profile can also be represented graphically, as shown in \Fig{photometry}. 
\Tab{profcolumns} gives a full description of the columns in the .prof output file; this information is also found in the GitHub code repository.
In addition to the SB profile, {\scriptsize AUTOPROF} generates a number of useful parameters for data quality checks and specific use cases.
These extra checks and global image parameters are written to an auxiliary .aux file of the same name as the profile.
\begin{table*}
\caption{Surface brightness profile output columns (.prof file)}
\label{tab:profcolumns}
\begin{tabular}{ccl}
\hline 
Column & Units & Description \\ 
(1) & (2) & (3)\\
\hline 
R & arcsec & Isophote semimajor axis length \\
SB & \magss & Median SB along isophote \\
SB\_e & \magss & Uncertainty on SB estimate \\
totmag & mag & Total magnitude enclosed in isophote, computed by integrating SB profile \\
totmag\_e & mag & Uncertainty in totmag estimate propagated through integral \\
ellip & \nodata & Ellipticity of isophote ($1-b/a$) \\
ellip\_e & \nodata & Uncertainty in ellipticity estimate determined by local variability \\
pa & deg & PA of isophote relative to positive $y$-axis (increasing counter-clockwise) on the image \\
pa\_e & deg & Uncertainty in PA estimate determined by local variability \\
pixels & count & Number of unmasked pixels sampled along isophote or within band \\
maskedpixels & count & Number of masked pixels rejected along isophote or within band \\
totmag\_direct & mag & Total magnitude enclosed in current isophote by direct pixel flux summation \\
\hline
\end{tabular}
\end{table*}

{\scriptsize AUTOPROF} can output a number of diagnostic plots for each operation.
These plots allow the user to identify pathological cases that may disrupt isophotal fitting.
Many of the figures in this paper are indeed examples of such diagnostic plots.
A detailed description of all diagnostic plots can be found in the {\scriptsize GITHUB} code repository.
In general, the plots favour diagnostic purposes over aesthetics and may magnify spurious features of the data.

\subsection{Checking The Fit}
\label{sec:checkfit}

Various checks are applied to ensure that the isophotal fit has reached an acceptable solution.
This is necessary in an automated analysis domain as large samples cannot be examined individually by a human, yet challenging or pathological cases may require extra attention.
The latter may take multiple forms.  
For small to mid-sized samples, a human operator may triage the identified problematic cases.
For larger samples, one can take advantage of {\scriptsize AUTOPROF}'s decision tree pipeline construction to modify settings and iterate until a suitable fit is achieved or it becomes clear that no reliable fit is possible.
The checks described here are designed to flag cases in which the final fitted isophotes are not well described by a constant SB plus noise.
This can happen for a variety of reasons including a fit which failed to converge, a misaligned centre, a bright foreground object, or a galaxy whose light distribution is poorly modeled by elliptical isophotes~\citep{Ciambur2015}.
Here we describe checks which {\scriptsize AUTOPROF} performs in its default configuration. 
Users may also implement their own checks for a particular application. 

First, {\scriptsize AUTOPROF} compares the interquartile range of flux values along an isophote versus the median flux value.
If over \wunits{20}{per cent} of the isophote have such variability, then the fit is flagged as having possibly failed.
Secondly, {\scriptsize AUTOPROF} computes an FFT along the isophote and examines the power in the first-order mode.
If \wunits{80}{per cent} of the isophotes have $|\mathcal{F}_1/(\tilde{\mu} + \sigma_b)| > 0.05$ or if \wunits{30}{per cent} have $|\mathcal{F}_1/(\tilde{\mu} + \sigma_b)| > 0.1$, then the fit is flagged.
Thirdly, the same checks are performed except with the second Fourier coefficient ($\mathcal{F}_2$).
The percentages and thresholds above were chosen by expert assessment with the goal of successfully capturing all cases that would be visually flagged for re-fitting. 
The flags may also capture a number of acceptable fits and may thus not be suitable for all situations/surveys. These checks are included to assist the user, however it may be necessary to create custom checks for a given application.
These checks can serve as a starting point for more specialized versions.

\subsection{Forced Photometry}

Forced photometry is the process of taking the isophotal solution (a profile of ellipticity and PA as a function of galactocentric radius) from one image and applying it to another image (generally taken at a different photometric bandpass).
Forced photometry is critical for properly comparing colour measurements, since it forces consistent measurements at each pixel.
It can also save on computation time as a galaxy should have a similar structure in other bands, ignoring the effects of dust, stellar populations, PSF, and sensitivity.
Forced photometry can be used in cases where low S/N precludes the possibility of finding a robust fit, but where one expects a measurable signal.

In order to apply forced photometry, the user must provide a configuration file with the image file to process and a previously fit {\scriptsize AUTOPROF} profile. 
The default forced photometry pipeline is a modification of the default pipeline described above. 
The background level and PSF are re-calculated for each new image 
as in \Secs{backgroundestimation}{psf}, 
as these typically change from image to image. 
For centre finding, {\scriptsize AUTOPROF} uses the centre from the forcing profile, though a new centre can be fitted if needed.  
The global isophotal fitting in \Sec{isophoteinitialize} is skipped and the forcing profile values are used instead.
The isophote fitting step is no longer needed and skipped as well.
The isophote extraction step proceeds as in \Sec{sbextraction} except that it now reads the PA and ellipticity values from the forcing profile and uses those as the set of isophotes to extract.
The fitting checks in \Sec{checkfit} are not run as no fitting is performed.
The default forced photometry pipeline can, of course, be adapted and extended just as the regular photometry pipeline.

\section{Advanced Pipeline Tools}
\label{sec:advancedpipelinetools}

We now describe several analysis tools that complement the default {\scriptsize AUTOPROF} toolkit.
These optional pipeline steps can be included in the configuration file.
Users may also design their own pipeline by adding, removing, or reordering the steps in the default {\scriptsize AUTOPROF} pipeline.
This may benefit special cases, fine tuning the analysis or adapting to the type of galaxy being analysed. 
The default functions, coupled with a flexible pipeline, make {\scriptsize AUTOPROF} well suited for a wide range of small and large scale galaxy photometry applications.

\subsection{Star Masking}\label{sec:mask}

\begin{figure}
    \centering
    \includegraphics[width = 0.9\columnwidth]{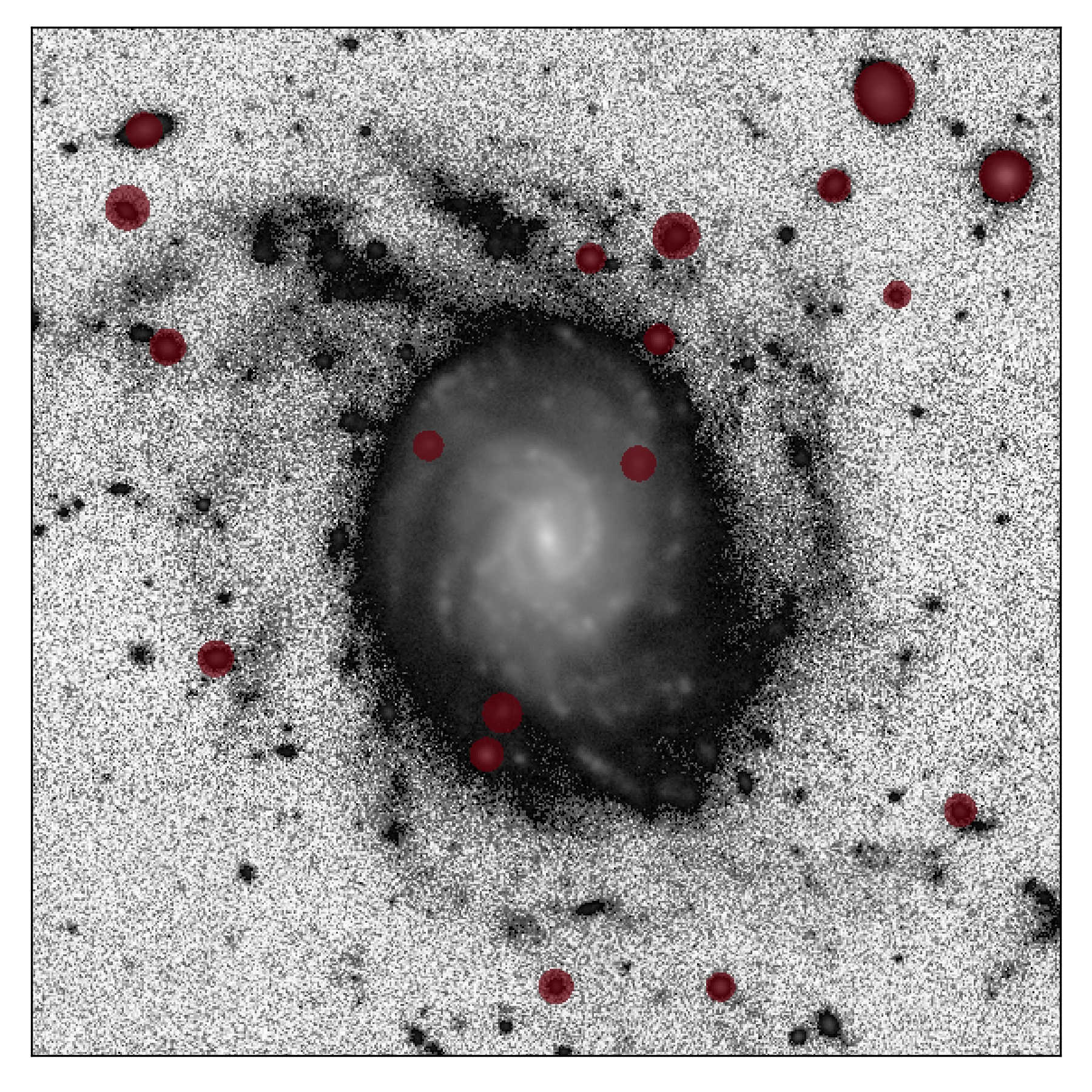}
    \caption{Example star mask method applied to \emph{ESO479-G1}. The circular masks are adjusted in size based on the fitted FWHM for each star candidate. Typically two FWHM will be masked, though for bright objects this aperture increases logarithmicaly with the peak flux.}
    \label{fig:starmask}
\end{figure}

In general, {\scriptsize AUTOPROF} can operate effectively without a mask as all procedures rely on robust mode, median, and percentile estimates of flux.
However, the total luminosity, as reflected in the parameter `totmag\_direct' in \Tab{profcolumns}, is directly affected by a lack of masking.
Densely populated fields likely require that the galaxy image be masked appropriately before {\scriptsize AUTOPROF} processing to achieve reliable flux estimates. 
Built into {\scriptsize AUTOPROF} is a star finding algorithm (also used in \Sec{psf}) which identifies peaks using an edge detecting convolutional filter.
{\scriptsize AUTOPROF} can automatically mask any identified star using the star finder (called `starmask').
An example of such a mask is given in \Fig{starmask}, here all identified star candidates are masked regardless of deformity score (see \Sec{psf}) and the size of the mask is adjusted according to the individually fit FWHM.
Note that the {\scriptsize AUTOPROF} star mask focuses on objects brighter than 10 times the noise of the background level.  For more detailed masks on very faint stellar or extragalactic sources which dominate the background in modern imaging surveys, the user may take advantage of more specialized masking tools~\citep[e.g.,]{Bertin1996, Akhlaghi2019}. 

A mask can be provided by specifying the path to a fits file containing an image with the same dimensions as the galaxy image being studied.
Any non-zero value in the provided file is considered a masked pixel. 
In the event that the user provides a segmentation map, such as the output from {\scriptsize SEXTRACTOR}~\citep{Bertin1996}, {\scriptsize AUTOPROF} will first identify the segmentation ID of the main galaxy using the centre from \Sec{centre} and ensure that this ID is set to zero (i.e., not masked).
The various mask options can be added sequentially via a logical-or operation.

\subsection{Radial Profiles}\label{sec:radialsampling}

\begin{figure}
    \centering
    \includegraphics[width = 0.9\columnwidth]{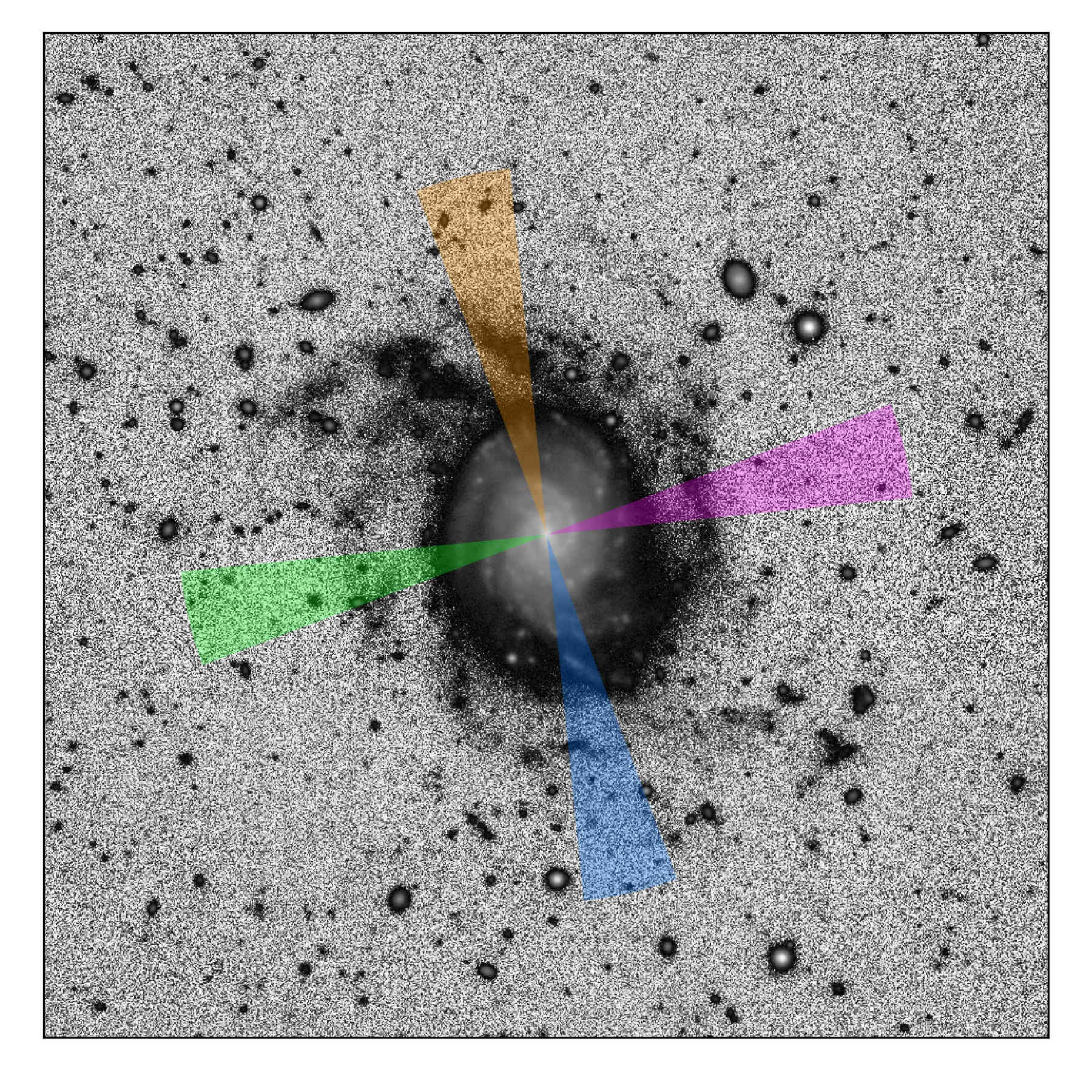}
    \includegraphics[width = 0.9\columnwidth]{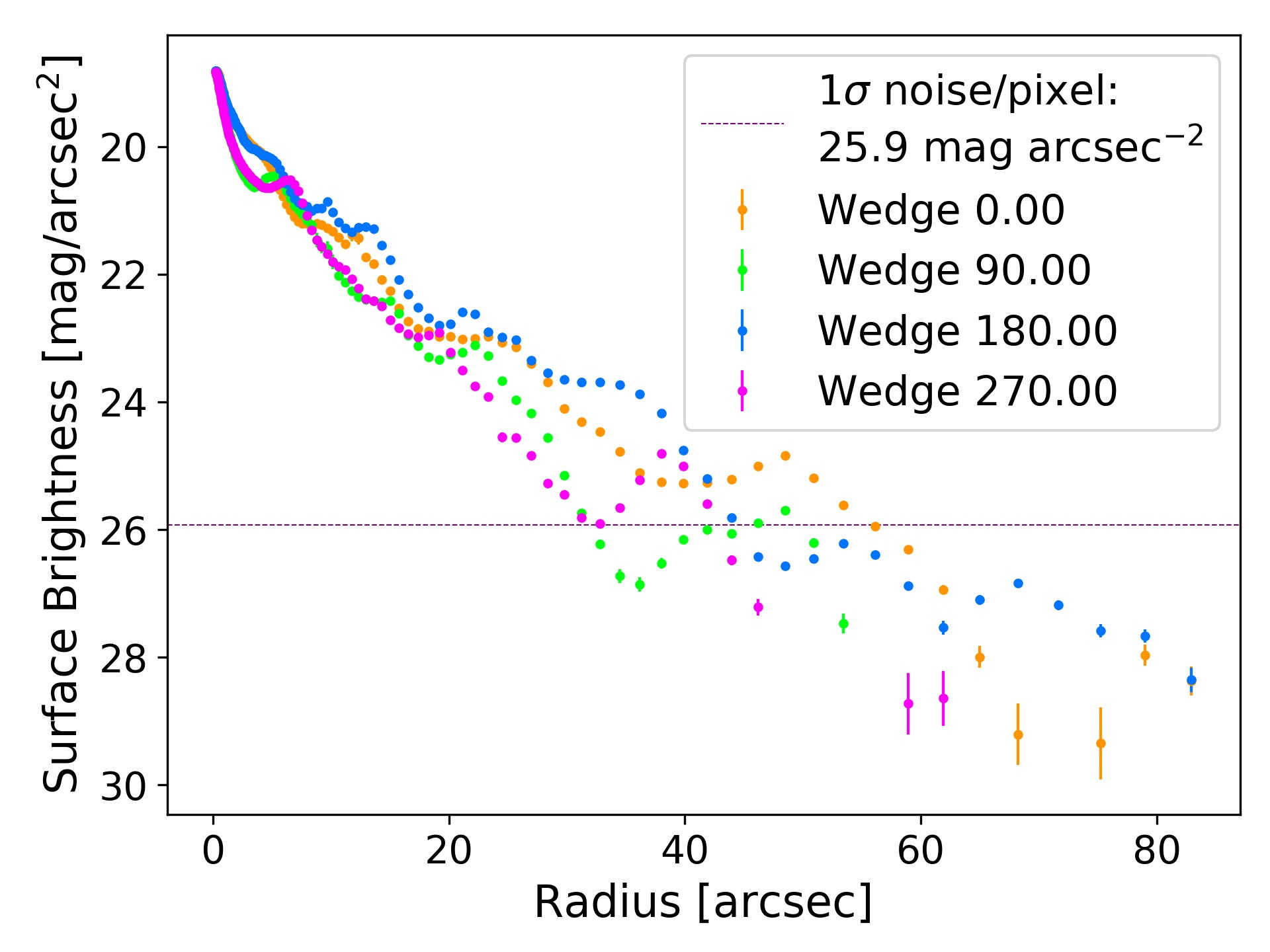}
    \caption{Example radial profile wedges for \emph{ESO479-G1} demonstrating which parts of the image are sampled for each wedge. Top: visualization of sample extraction; the blue/orange wedges trace the semimajor axis, while the pink/green wedges trace the semiminor axis. Bottom: resulting radial SB profiles showing features along each wedge.}
    \label{fig:radialsample}
\end{figure}

This complimentary tool in {\scriptsize AUTOPROF} (called `radialprofiles') involves the extraction of light profiles along any axes (PAs) radiating from the galaxy centre~\citep{Courteau2011}.
Radial profiles (or wedges) typically trace the major or minor-axis of the galaxy (e.g., \Fig{radialsample}), though any orientation is possible.
These radial profiles can be used to resolve axisymmetries that would otherwise be lost in azimuthal averaging\footnote{Standard isophotal analysis, or surface photometry, is not valid for edge-on galaxies as the entire disc is projected along the line-of-sight.}, examine dust distributions, disentangle thin and thick discs, and so on \citep[e.g.,][]{Comeron2018}.   
The computation of wedge profiles proceeds similarly to profile extraction in \Sec{sbextraction} except that flux measurements are a function of radius and angle instead of radius alone.
By default the radial wedges are sampled in four directions relative to the global PA of the galaxy, where each wedge is \wunits{15}{deg} wide as shown in \Fig{radialsample}.
The number of wedges, orientation, and the width of the wedges can be defined by the user.
Each profile will be sampled at the same radii as the semimajor axis values in the primary SB profile.
Wedges with exponentially growing widths~\citep{Courteau2011} that maintain constant S/N ratios are also available. 
Similarly, wedges with variable PAs that follow complex features near the centre (bar or spiral arms) can also be modeled. 
Radial wedges are provided as an advanced option and are not activated by default.

\subsection{Axial Profiles}
\label{sec:axialprofiles}

\begin{figure}
    \centering
    \includegraphics[width = 0.9\columnwidth]{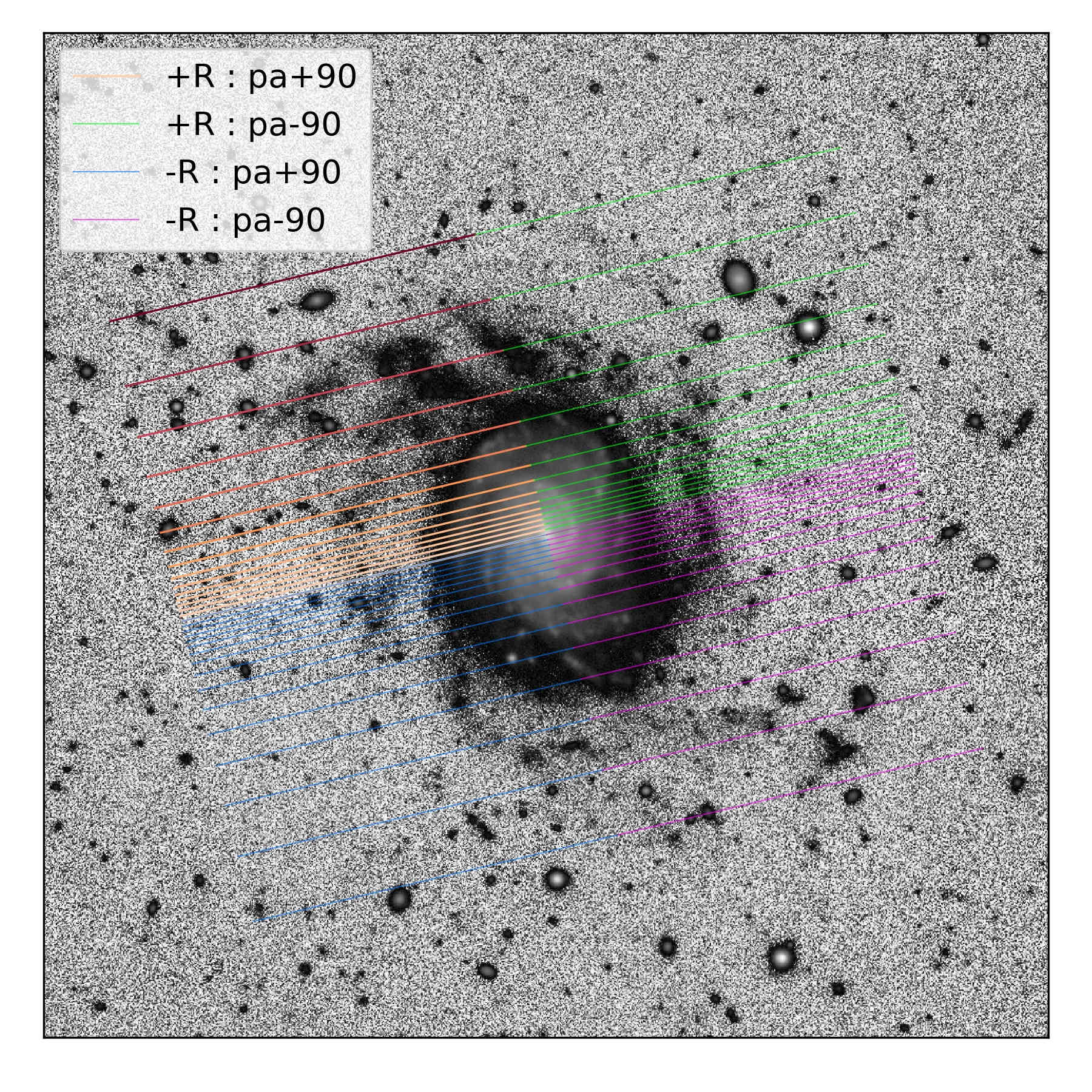}
    \includegraphics[width = 0.9\columnwidth]{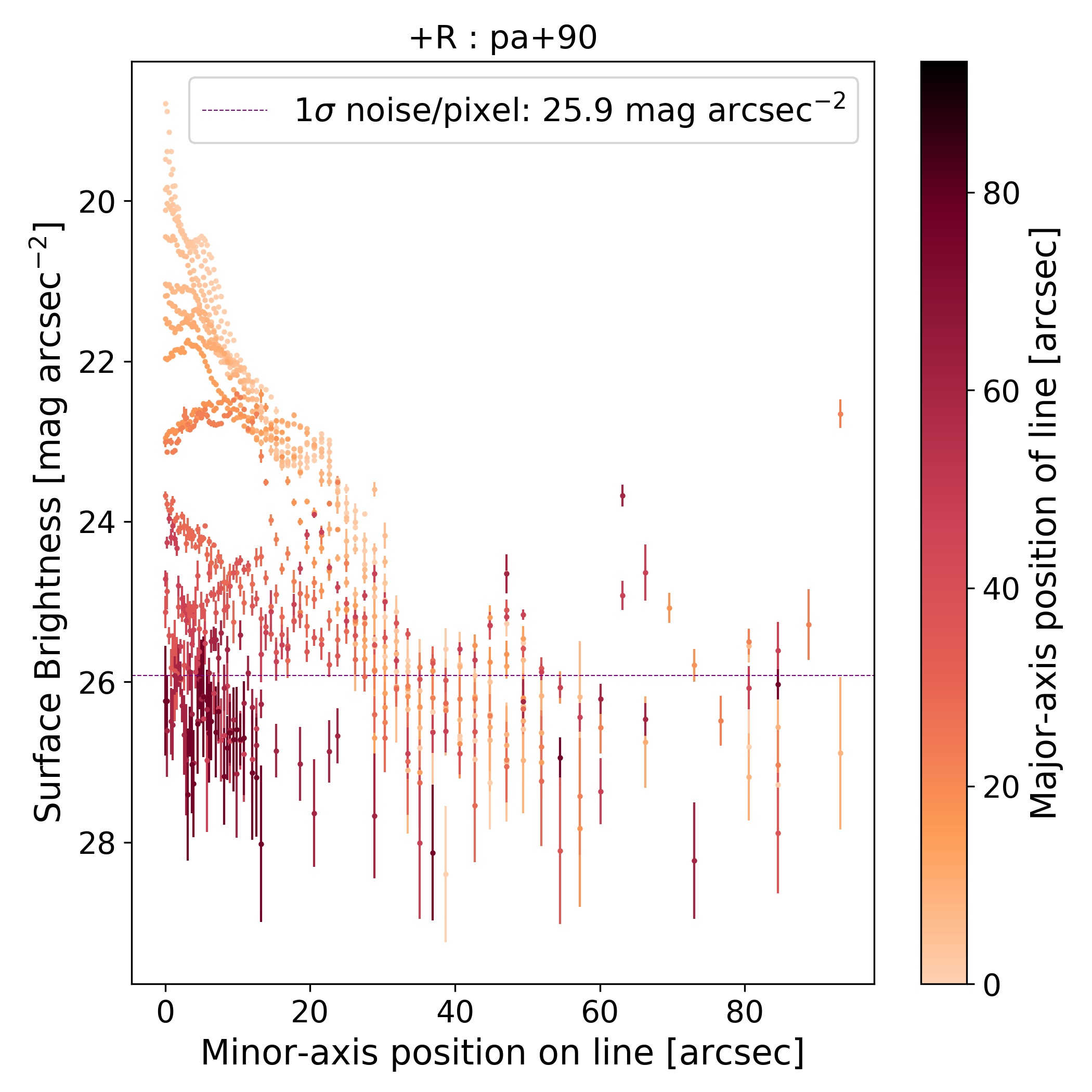}
    \caption{Example of {\scriptsize AUTOPROF} axial profiles on \emph{ESO479-G1}. Top: projection of axial profile lines parallel to the minor-axis on the galaxy image. Colours indicate what quadrant of the galaxy is being sampled. Bottom: resulting set of axial profiles for one quadrant (orange lines of top panel) of the top figure. The x-axis is the distance from (in this case) the major-axis of the galaxy, and the colour scale now indicates the starting point for each sampling line from the galaxy centre along the major-axis.}
    \label{fig:orthogonalsampling}
\end{figure}

Another tool provided by {\scriptsize AUTOPROF}, called `axialprofiles', is the extraction of axial SB profiles~\citep{Comeron2018} along lines parallel to a chosen axis (typically the major or minor axes). 
These slices can be used to study the same features as radial wedges but keeping their widths constant and allowing offsets from the galaxy centre. 

\Fig{orthogonalsampling} displays every fifth sampling line and resulting profiles for one quadrant on \emph{ESO479-G1}.
The spacing between axial profiles (and their width) grows geometrically by default, allowing {\scriptsize AUTOPROF} to collect more signal in fainter regions of the galaxy.
Along each axial profile, the bin sizes grow geometrically to increase S/N as the profile moves away from the primary axis (as can be seen by the spacing between points in \Fig{orthogonalsampling}).

\subsection{Ellipse Model}

\begin{figure*}
    \centering
    \includegraphics[width = 0.32\textwidth]{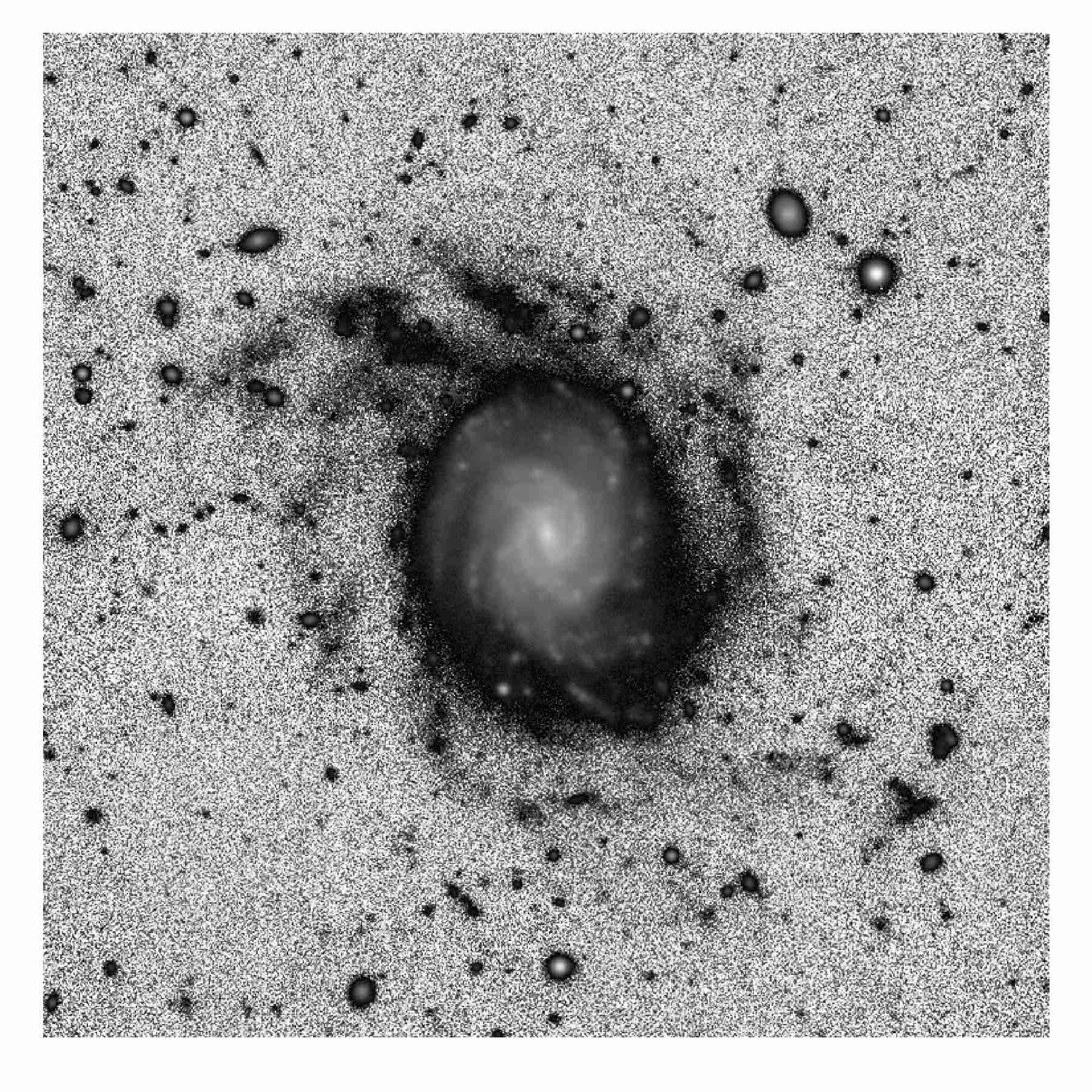}
    \includegraphics[width = 0.32\textwidth]{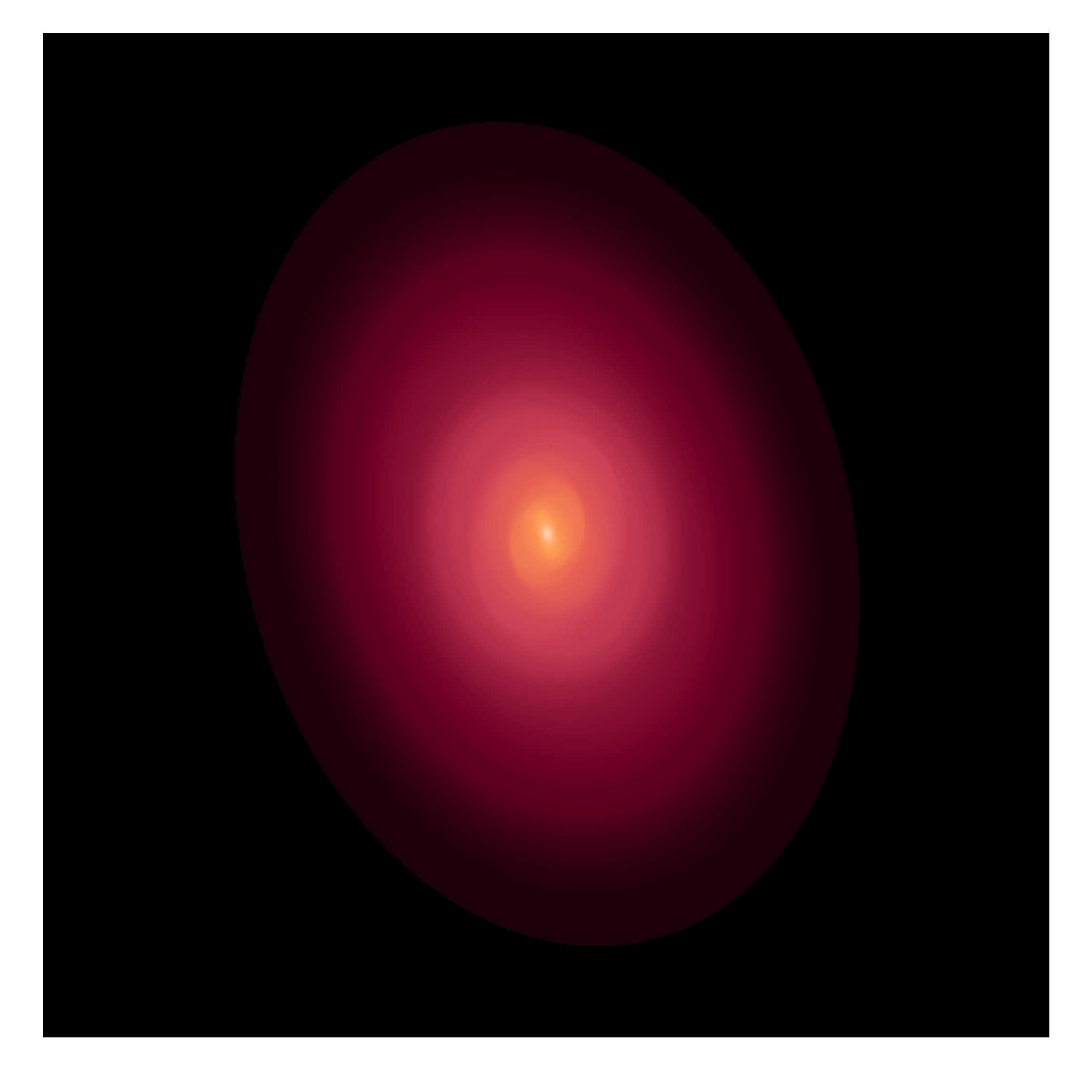}
    \includegraphics[width = 0.32\textwidth]{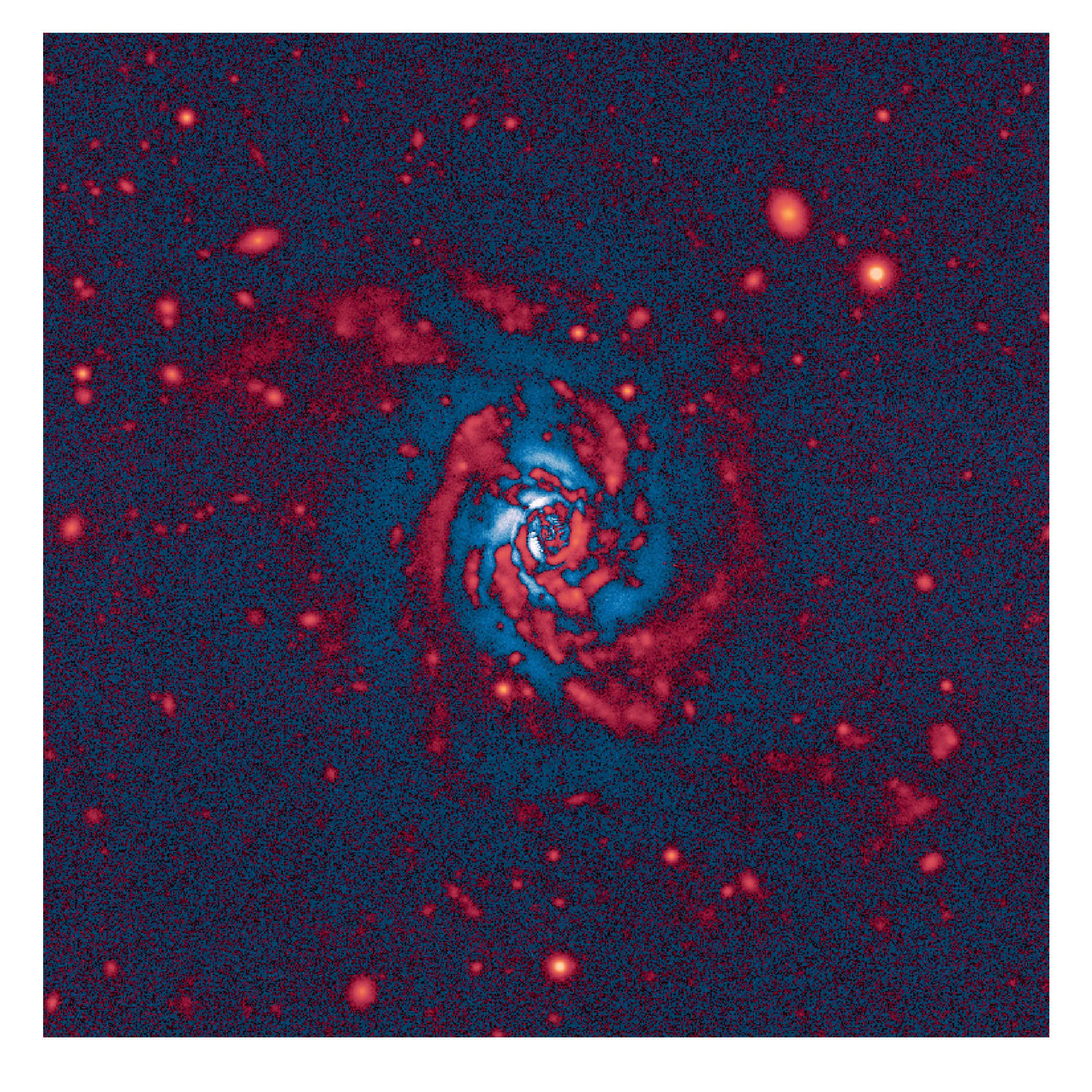}
    \caption{Left: image of \emph{ESO479-G1}. Middle: example smooth ellipse model of flux values for \emph{ESO479-G1} demonstrating the constructed model. Colours follow a log scale flux and show how the smoothed model handles variable PA and ellipticity values along the profile. Right: residual plot with ellipse model subtracted from original image. Blue colours indicate negative values and red colours indicate log-scale positive values.}
    \label{fig:ellipsemodel}
\end{figure*}

Another {\scriptsize AUTOPROF} tool (called `ellipsemodel') generates smooth 2D models of galaxies based on {\scriptsize AUTOPROF}'s 1D SB profile; this is saved to a FITS file with the same dimensions as the original image. 
There are two possible ellipse model modes.
In the simplest mode, the image is stretched along the minor-axis until the global ellipse fit (\Sec{isophoteinitialize}) is circular in the transformed space.
The `SB\_fix' column of the SB profile is interpolated at the radial location of each pixel and assigned the corresponding interpolated flux value.
The more general ellipse model shown in \Fig{ellipsemodel} is constructed like the fixed ellipse model.
The latter is repeated for each isophote in the SB profile using its individual PA and ellipticity values.
Each pixel uses the corresponding values from the closest ellipse once all iterations have been completed.

For both the fixed and the general ellipse models, no extrapolation of the SB profile is performed.
Instead all flux values beyond the outer isophote are set to zero.
This can be seen in the middle image of \Fig{ellipsemodel} in the sharp transition to the black background.

\subsection{Fourier Mode Analysis}
\label{sec:fouriermodeanalysis}

{\scriptsize AUTOPROF} includes two tools for examining higher order information on fitted isophotes.
The first and simplest available method measures the Fourier mode power spectrum along an ellipse.
During the isophote brightness extraction step (see \Sec{sbextraction}) {\scriptsize AUTOPROF} can perform an FFT decomposition along each ellipse and measure the power in each mode. 
These are reported relative to the zeroth coefficient, which roughly corresponds to the mean flux along an isophote; the real and imaginary parts of the coefficient are given as $\frac{B_n}{F_0}$ and $\frac{A_n}{F_0}$ respectively. 
The fourth mode, $F_4$, is commonly used to measure the boxyness/disciness of an isophote, whilst the low odd modes, $F_1$ and $F_3$, are commonly used to measure asymmetry in a galaxy~\citep{Peng2010}.

In some cases, it is not sufficient to measure the Fourier modes along elliptical isophotes;
instead a fit is needed to include these perturbations in the isophote.
This can occur for boxy/discy galaxies, or very lopsided galaxies where the contours of constant SB depart from an elliptical shape.
{\scriptsize AUTOPROF} provides built in functionality to fit isophotes as ellipses with user selected cosine perturbations.
The perturbation for each mode is described by two parameters, $A_m$ and $\phi_m$, which control the amplitude and phase respectively for each mode $m$.
The exact form of these perturbations is given in \Equ{fouriermodeperturbations}.

\begin{equation}\label{equ:fouriermodeperturbations}
    \begin{aligned} 
        R(\theta) &= R_0(\theta)e^{\sum_m A_m\cos(m(\theta + \phi_m))},    
    \end{aligned}
\end{equation}

\noindent where $R_0$ is the radius for a traditional ellipse and $R$ is the radius of the perturbed isophote.
Note that the sum is only performed over user selected modes, thus it is possible to construct an isophote using only e.g. fourth order perturbations.
In other implementations the form of the perturbation in \Equ{fouriermodeperturbations} would be $R_0(\theta)(1 + \sum_m A_m\cos(m(\theta + \phi_m)))$ assuming $A_m$ is small~\cite[e.g.][]{Peng2010}.
However this has the undesirable feature of becoming negative if an $A_m$, or some combination of them, is greater than 1.
Whilst this is unlikely to occur in real cases, an optimization algorithm may fail if part of the parameter space is invalid.
Thus we use $e^{\sum_m A_m\cos(m(\theta_m + \phi_m))}$ which is always positive and for small amplitudes, it is equivalent to other implementations ($e^x = 1 + x + \mathcal{O}(x^2)$).

As more parameters are added to the fitting procedure, {\scriptsize AUTOPROF} can model more complex systems.
However the increased parameter space naturally also increases the likelihood of fitting spurious image features, causing worse performance overall.
To combat this, we extend the loss function described in \Sec{isophotefitting} to now include regularization terms for the cosine perturbations:

\begin{equation}
    \begin{aligned}
        l = &\frac{\sum_{2,m}[|\mathcal{F}_m|]}{\tilde{\mu} + \sigma_b}\left(1 + \frac{|e_i - e_{i-1}|}{1-e_{i-1}} + \frac{|e_i - e_{i+1}|}{1-e_{i+1}}\right. \\
        & \left.+\frac{|p_i - p_{i-1}|}{0.2} + \frac{|p_i - p_{i+1}|}{0.2} + \sum_m\left[\frac{|A_{m,i} - A_{m,i-1}|}{0.2}\right.\right. \\
        & \left.\left.  + \frac{|A_{m,i} - A_{m,i+1}|}{0.2} + \frac{|\phi_{m,i} - \phi_{m,i-1}|}{0.1m} + \frac{|\phi_{m,i} - \phi_{m,i+1}|}{0.1m}\right] \right).
    \end{aligned}
\end{equation}

\noindent Note that the first sum now includes the user selected modes, $m$, and the second Fourier mode; the sum in the regularization term only goes over the user selected modes.
For the $\phi_{m,i}$ terms in the regularization scale, the factor $m$ in the denominator reflects the reduced angular range of the higher mode perturbations (a full cycle completes every $\frac{2\pi}{m}$ radians).
All other standard {\scriptsize AUTOPROF} methods will automatically adapt to the new higher order isophotes.
The isophote flux sampling procedures described in \Sec{sbextraction} will now accept the more complex isophote shapes, which may result in better error performance as the brightness is more uniform along each isophote.
However, the interpretation of the measurements (and even the semimajor axis) can be more challenging.

\begin{figure}
    \centering
    \includegraphics[width = 0.9\columnwidth]{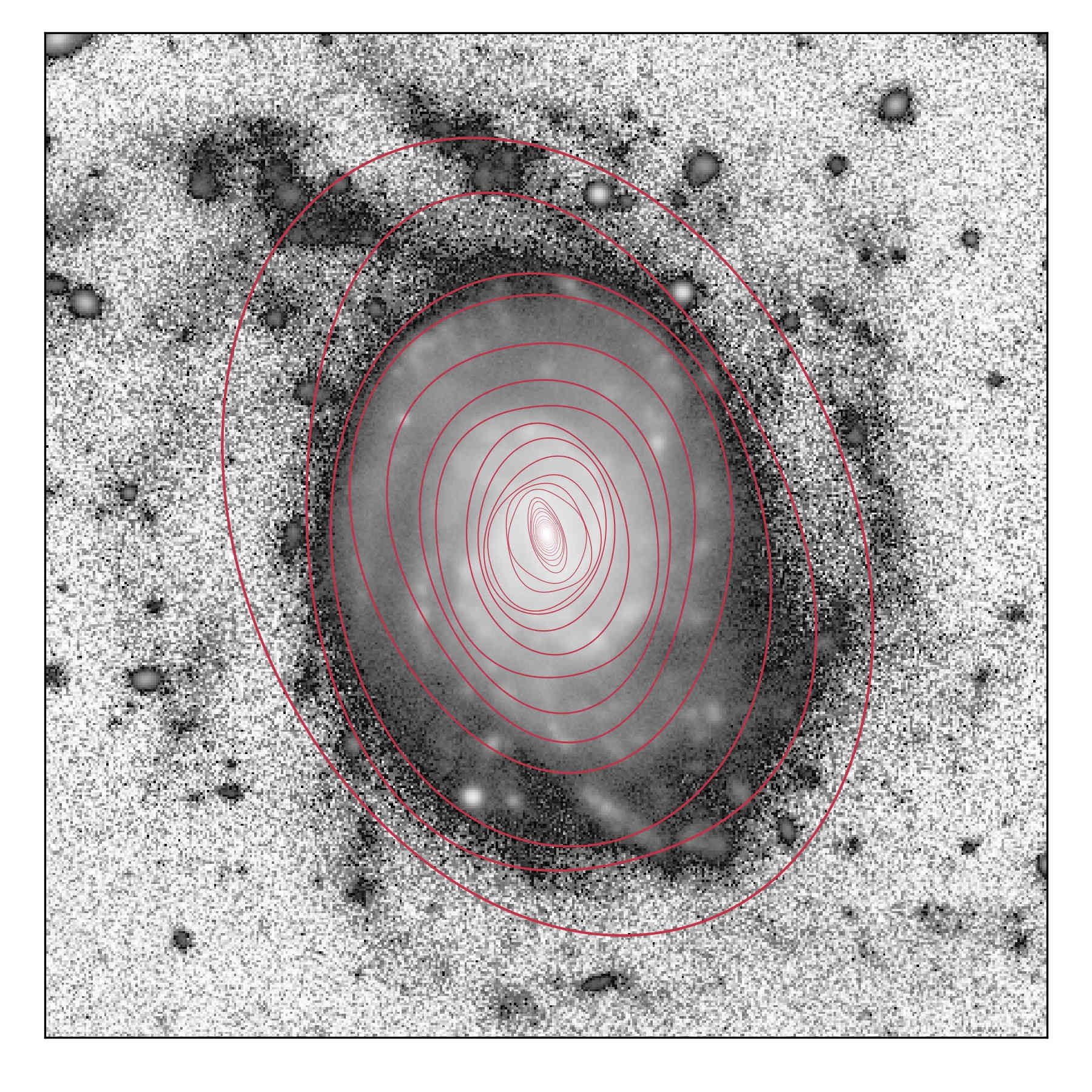}
    \caption{ Example of fitted isophotes with higher mode perturbations, formatted as in \Fig{examplefit}. Isophotes follow contours of constant SB more closely, at the cost of increased complexity.
    Here, {\scriptsize AUTOPROF} has fitted first, third, and fourth order perturbations to a standard ellipse, with the outer isophotes primarily using the first and third modes to model the galaxy asymmetry, while the inner isophotes primarily use the fourth mode to track spiral features.}
    \label{fig:fourierperturbations}
\end{figure}

The resulting fitted isophotes are stable, even with high frequency modes.
However, the runtime for fits with these perturbations is considerably longer, often taking twice as long to converge.
The resulting isophotes follow contours of constant SB more closely than ellipses alone, as can be seen in \Fig{fourierperturbations} which was fitted using modes 1, 3, and 4.
There is a clear lopsidedness to the galaxy at larger radii; 
at small radii, the perturbations allow {\scriptsize AUTOPROF} to follow the spiral arms contours.
This added complexity may or may not provide added insight.
In general, users should use the cosine mode fitting with intention, only fitting when a signal is expected.
For example, with the fourth mode to early type galaxies, or with the first and third modes for searches of lopsided features (i.e. tidal streams).
The second mode is also partly degenerate with the ellipticity and should thus be avoided in typical scenarios.

The two methods available in {\scriptsize AUTOPROF} for evaluating higher order information should in principle yield identical results; in practice, this is not the case.
The fitted perturbations sample new parts of an image compared to simply measuring the Fourier amplitudes along an ellipse, thus the two techniques are evaluated on different information.
Further, the fitted perturbations only model the selected modes, and thus they may "wash out" some information which could be expressed in other modes.
Each method will also be impacted by noise and foreground objects differently.
Finally, attempting to use both methods simultaneously will invalidate the Fourier amplitude measurement method as it will no longer apply to an ellipse, but rather to a more complex isophote.
Thus it is best to select the method which best suits a given application and solely apply it to the images.

\subsection{Decision Tree Pipelines}

\begin{figure*}
    \centering
    \includegraphics[width = 0.9\textwidth]{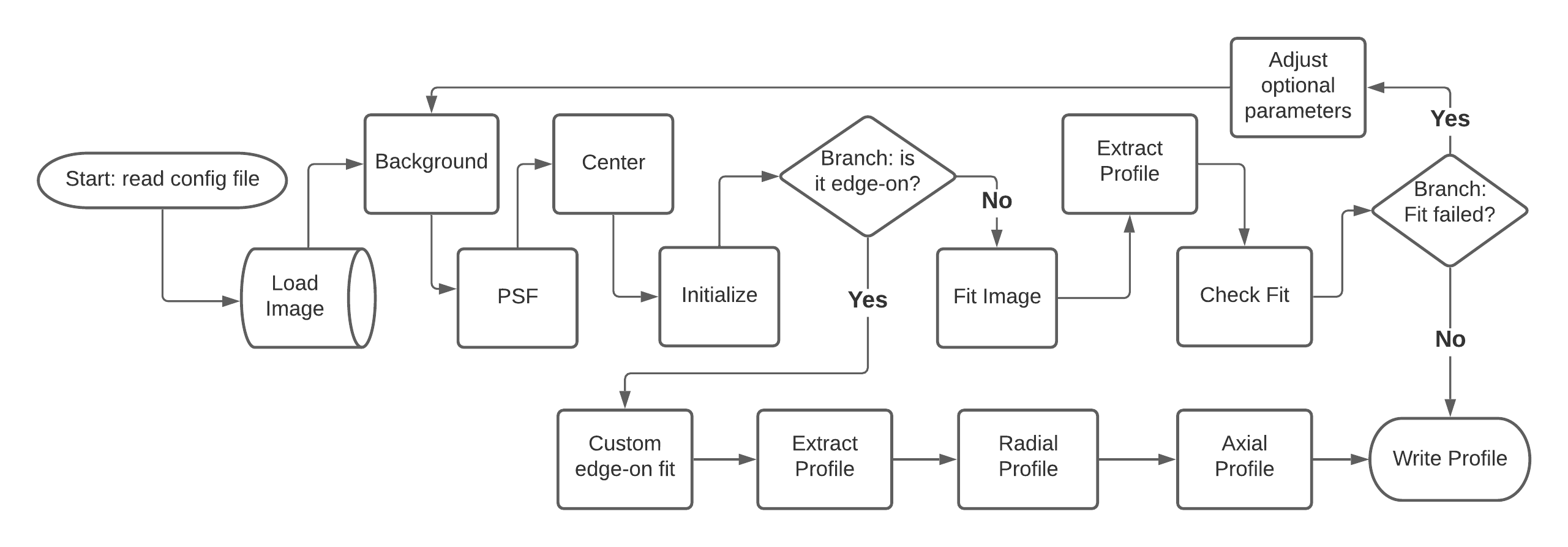}
    \caption{Example of an {\scriptsize AUTOPROF} decision tree flowchart. 
    There are two instances where the "branches" make a decision (indicated by rhombus blocks). 
    If an edge-on galaxy is encountered, a specialized analysis pipeline can be applied. 
    If the fitting procedure in the standard pipeline fails, the branch can update parameters and perform the fit again. 
    The decision tree is fully adaptable and expandable.}
    \label{fig:decisiontree}
\end{figure*}

In addition to {\scriptsize AUTOPROF}'s toolkit, users can implement decision trees in order to run specific applications on separate galaxy images. 
For instance, one may wish to extract different data from edge-on vs face-on galaxies, early- versus late-type galaxies, crowded fields vs isolated galaxies, large vs small extent on the sky, and so on.
Another application is to re-analyse a galaxy which fails the fit checks (\Sec{checkfit}) using alternate settings in an attempt to recover an accurate ellipse  model.
An example flow chart, shown in \Fig{decisiontree}, demonstrates two types of decision tree choices. 
The decision tree is provided through the configuration file and an example is included with the code to support initial development.
{\scriptsize AUTOPROF} decision trees can be arbitrarily complex and integrate user-defined functions.

\section{Comparisons With Other Methods}
\label{sec:comparisons}

To demonstrate the accuracy of {\scriptsize AUTOPROF} data products, we compare our results to similar surface photometry image analysis software.
For these comparisons, we compare with the non-parametric packages {\scriptsize PHOTUTILS}~\citep{photutils} and {\scriptsize XVISTA}~\citep{Lauer1985}, as well as the parametric decomposition software {\scriptsize GALFIT}~\citep{Peng2010}. This is not an exhaustive comparison of isophotal fitting codes, others include {\scriptsize ELLIPSE}~\citep{Tody1986}, {\scriptsize GIM2D}~\citep{Simard2002}, {\scriptsize PYMORPH}~\citep{Vikram2010}, {\scriptsize GALAPAGOS}~\citep{Barden2012}, {\scriptsize IMFIT}~\citep{Erwin2015}, and {\scriptsize ISOFIT}~\citep{Ciambur2015} to name a few.
Clearly, modelling galaxy light profiles is an active field in extra-galactic astronomy.

To establish the relative performance of {\scriptsize AUTOPROF} and other methods, we take advantage of 1387 late-type galaxies from the `Photometry and Rotation Curve Observations from Extragalactic Surveys' (PROBES) catalogue~\citep{Stone2019,Stone2021}. 
The PROBES galaxy images were extracted from the DESI-LIS Data Release 9 \citep{Dey2019}, which provided deep photometry in $g,r,$ and $z$ photometric bands. 
All results are computed in the $r$-band unless otherwise indicated.
The comparisons below should serve to adequately present {\scriptsize AUTOPROF} in the context of available codes.

Our comparisons have also included measures of the relative run times for each method. 
Their interpretation is however thwarted by operational considerations unique to each method (e.g. we initialize {\scriptsize PHOTUTILS} with {\scriptsize AUTOPROF} pipeline functions).
The runtime also varied considerably with user tunable parameters and the conditions of each image.
Still, the general trend shows that the automated approaches ({\scriptsize PHOTUTILS}, {\scriptsize AUTOPROF}, and {\scriptsize GALFIT}) have comparable runtimes, while interactive methods ({\scriptsize XVISTA}) are slower by orders of magnitude~\citep{Smith2021}

\subsection{{\scriptsize PHOTUTILS}}
\label{sec:photutils}

The {\scriptsize PHOTUTILS} package includes a variety of methods for analysis of astronomical images~\citep{photutils}.
Included in the package is an implementation of the iterative isophotal fitting method by \citetalias{Jedrzejewski1987}, which performs a weighted least squares fit of the intensities along an isophote according to:

\begin{equation}
    I(\phi) = I_0 + A_1\sin(\phi) + B_1\cos(\phi) + A_2\sin(2\phi) + B_2\cos(2\phi)
\end{equation}

\noindent where $I(\phi)$ is the intensity at angle $\phi$ around the isophote, $I_0$ is the average intensity around the isophote, and the $A_n,B_n$ are Fourier coefficients with n indicating the order of the coefficient.
The properties of the isophote are iteratively updated until the Fourier coefficient amplitudes are below \wunits{4}{per cent} of the RMS scatter of the intensities (or the maximum number of iterations is reached).

\begin{figure}
    \centering
    \includegraphics[width = \columnwidth]{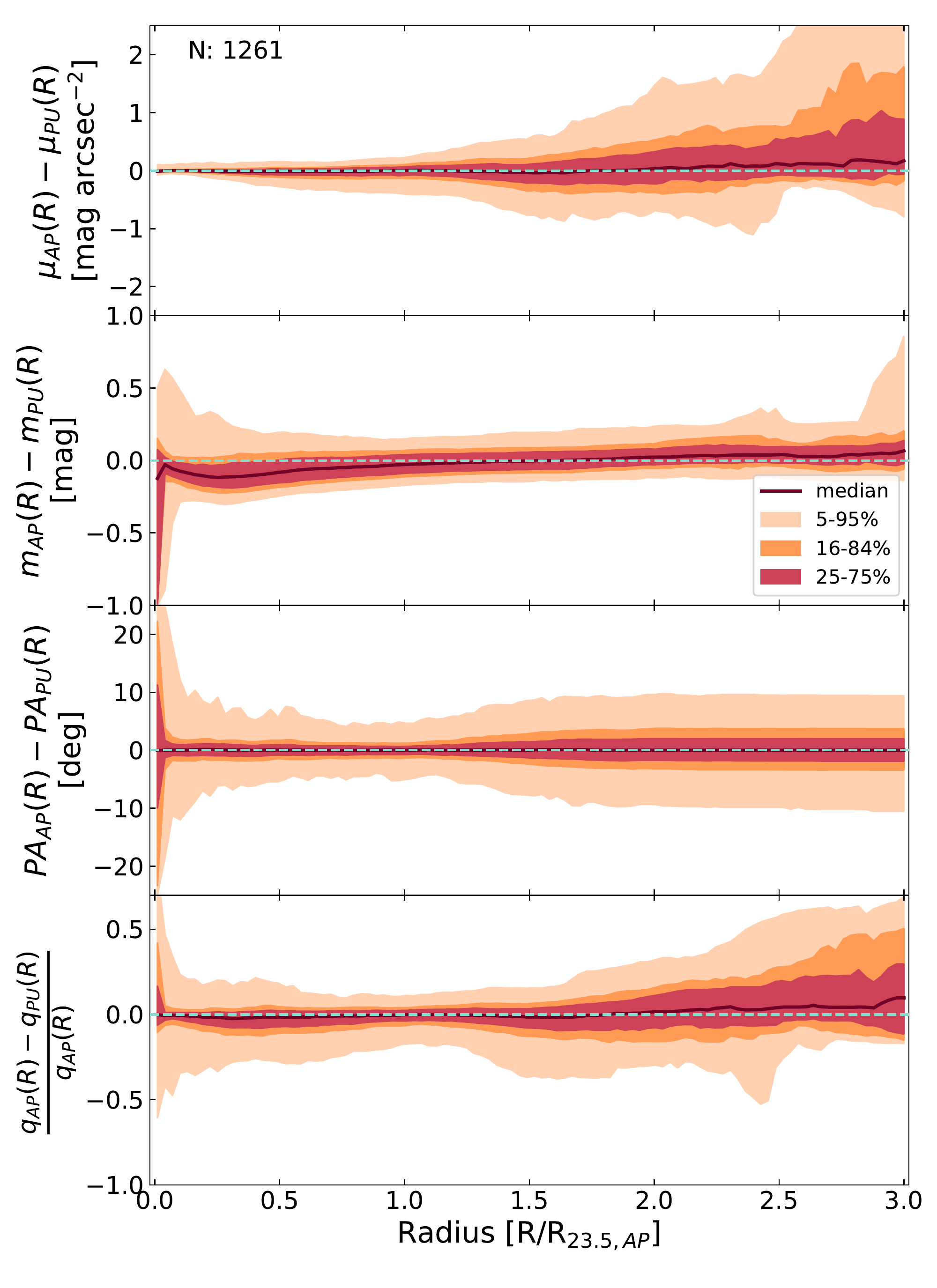}
    \caption{Comparison of photometric profiles from {\scriptsize AUTOPROF} and {\scriptsize PHOTUTILS}. Subplots show differences in the SB, curves of growth, PA, and axis ratio measurements, respectively. 
    The solid dark line represents the median deviation as a function of radius, and the cyan dashed line gives zero as a reference point. 
    The contours enclose 50, 68.3, and 90 per cent of the parameter comparisons at each radius, as indicated in the legend. 
    The number of compared galaxies, $N$, is shown in the upper left corner of the upper panel. 
    }
    \label{fig:compareprofilesphotutils}
\end{figure}

This technique reliably produces isophotes for smooth systems (e.g., typical elliptical galaxies).
However, the possible presence of strong non-axisymmetric features violates the method's assumptions and can produce chaotic isophotal solutions.
Interactive corrections of such chaotic isophotal solutions are enabled in {\scriptsize XVISTA}, as discussed in \Sec{xvista}. 

For the {\scriptsize PHOTUTILS} implementation of the isophotal solving technique in \citetalias{Jedrzejewski1987}, an initial ellipse with a reasonable first guess for the PA and ellipticity are required.
To perform a robust comparison, we make use of the {\scriptsize AUTOPROF} pipeline, replacing the isophotal fitting scheme in \Sec{isophotefitting} and the isophote extraction scheme in \Sec{sbextraction} with their {\scriptsize PHOTUTILS} counterparts.
The flexibility of the {\scriptsize AUTOPROF} pipeline design allows us to easily interchange individual segments for the sake of comparisons.
This makes for a clean comparison between the two methods as all pre-processing steps are identical.

\Fig{compareprofilesphotutils} shows a comparison of the SB profiles from {\scriptsize PHOTUTILS} and {\scriptsize AUTOPROF} applied to all PROBES galaxies; including only cases where both methods successfully converged on a solution. 
The four difference panels show that the median fitted value for SB, PA, and axis ratios are nearly identical for {\scriptsize AUTOPROF} and {\scriptsize PHOTUTILS} out to $2R_{23.5}$. 
This is expected as both methods are based on the same \citetalias{Jedrzejewski1987} method.
The curve of growth comparison shows {\scriptsize PHOTUTILS} as dimmer near the centre and systematically brighter past ${\sim}1.5R_{23.5}$. 
The extra light is due to the {\scriptsize PHOTUTILS} curve of growth being based on pixel flux summation which will be biased high from unmasked sources.
Beyond $2R_{23.5}$, {\scriptsize PHOTUTILS} reports systematically narrower and brighter isophotes.
Our {\scriptsize AUTOPROF} analysis included sigma-clipping of the SB measurements, removing most of the brightness bias due to foreground sources, yielding dimmer SB measurements.
Some deviations in the curve of growth and axis ratio also exist at very small radii. 
Indeed, the inner few isophotes can be significantly impacted by small pixel fluctuations causing swings in PA and ellipticity values.
{\scriptsize AUTOPROF} combats these small radii effects by using its regularization scale and only fitting down to one PSF radius, however the issues occur for both {\scriptsize AUTOPROF} and {\scriptsize PHOTUTILS} fits.
The choice of interpolation scheme also plays a role in profile differences for the central regions; 
{\scriptsize AUTOPROF} uses Lanczos interpolation while {\scriptsize PHOTUTILS} uses a less accurate Bilinear method.
Comparison showed that the two methods can return differing flux estimates by over 5 percent, although it varies considerably from case to case.

At large radii, beyond $R_{23.5}$, the scatter for the axis ratio and SB comparison increases markedly due largely to differences in low S/N behavior.
The \citetalias{Jedrzejewski1987} method can acquire deviations from non-axisymmetric features (e.g., unmasked stars may play a role at these low SB levels), whereas {\scriptsize AUTOPROF} is stabilized by the regularization term.  
\App{fitcomparisons} shows specific cases of disagreement between the methods.

\begin{figure}
    \centering
    \includegraphics[width = \columnwidth]{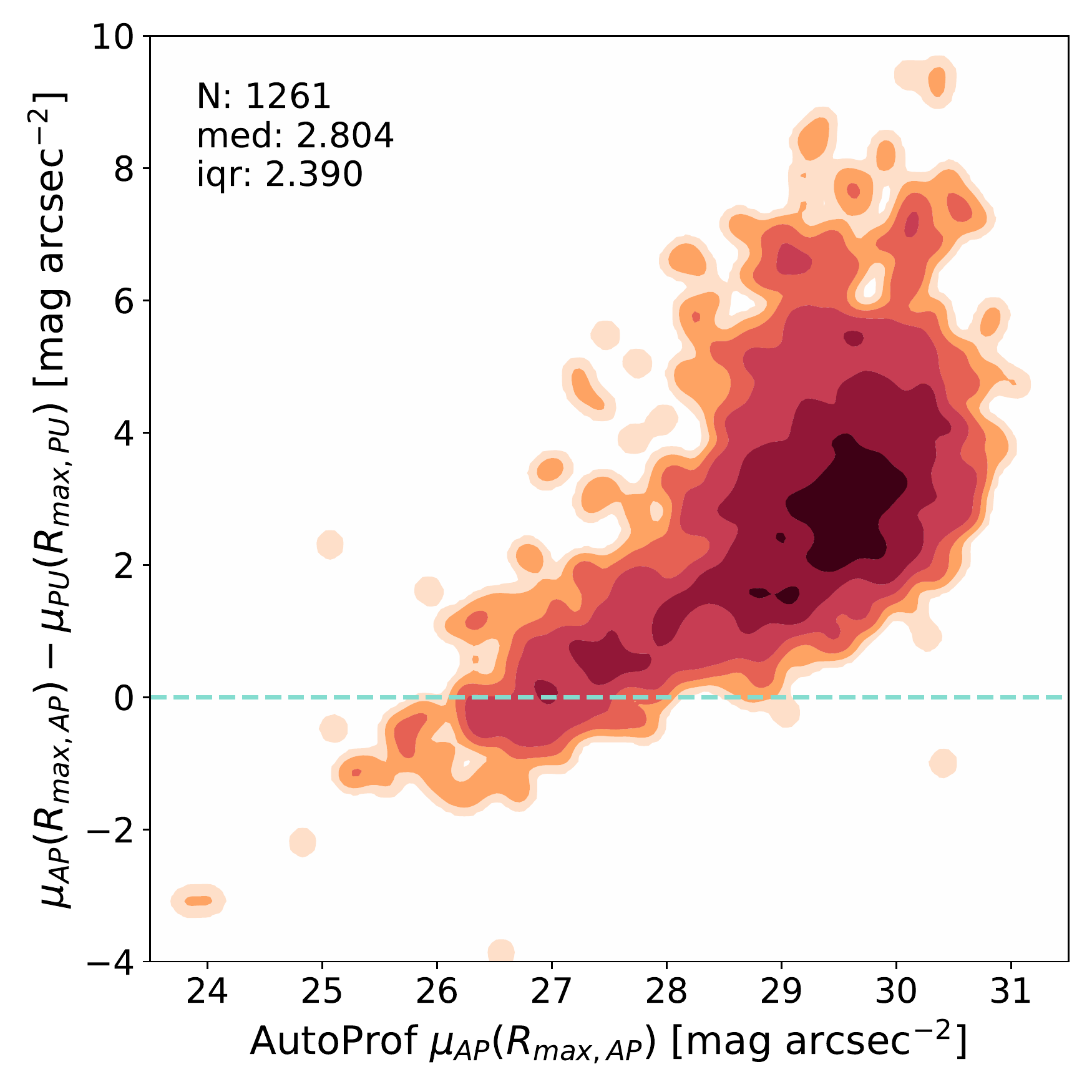}
    \caption{Comparison of photometric depth achieved by {\scriptsize AUTOPROF} and {\scriptsize PHOTUTILS}. The $x$-axis shows the deepest isophotes in the {\scriptsize AUTOPROF} photometry before reaching an error limit of \wunits{0.2}{\magss}. 
    The $y$-axis shows the SB difference between the {\scriptsize AUTOPROF} and {\scriptsize PHOTUTILS} deepest isophotes (before reaching an error of \wunits{0.2}{\magss}). 
    The contours represent the density (log scale) of galaxies. The cyan dashed line indicates zero difference in photometric depth.}
    \label{fig:comparedepthphotutils}
\end{figure}

{\scriptsize AUTOPROF} applies band averaging to determine the average flux for outer isophotes, the degree to which this technique improves the $S/N$ is dependent on the size of the band and of the galaxy on the sky.
The band averaging method can produce significant gains over direct sampling along an isophote.
To quantify the improvement, we can compare the photometric depth achieved by {\scriptsize AUTOPROF} and {\scriptsize PHOTUTILS}.
in \Fig{comparedepthphotutils}, we show deepest SB levels for {\scriptsize AUTOPROF} and {\scriptsize PHOTUTILS} before reaching a nominal error limit of \wunits{0.2}{\magss}.
Thanks largely to the band averaging method, {\scriptsize AUTOPROF} can reliably achieve a \wunits{2}{\magss} improvement over {\scriptsize PHOTUTILS} and in many cases reach even deeper.  
Factors other than the band averaging also contribute to achieving deep SB levels. 
{\scriptsize AUTOPROF} determines its errors with $\frac{\Delta_{16}^{84}}{2\sqrt{N}}$, whereas {\scriptsize PHOTUTILS} uses $\frac{\sigma}{\sqrt{N}}$.
For Gaussian noise these measures are identical, however in the presence of outliers (foreground stars and background galaxies) the {\scriptsize PHOTUTILS} errors are biased high relative to {\scriptsize AUTOPROF}. 
Some of the more extreme SB differences (e.g., $\Delta{\rm\magss}\approx 8$) stem from error-based stopping criteria in {\scriptsize PHOTUTILS} fits, which can terminate a profile in the central regions if a foreground star or an irregular galaxy feature (e.g., bright {\HII} region, spiral arms, or dust lanes) is encountered. 

\subsection{{\scriptsize XVISTA}}
\label{sec:xvista}

The {\scriptsize XVISTA} photometry package\footnote{ http://astronomy.nmsu.edu/holtz/xvista/index.html} offers a wide range of image processing tools.
We focus on the galaxy surface photometry analysis, which uses low-order Fourier coefficients (much like \citetalias{Jedrzejewski1987}) to iteratively fit elliptical contours of constant SB~\citep{Lauer1985,Courteau1996,McDonald2011, Hall2012,Gilhuly2018}.
Included in {\scriptsize XVISTA}'s surface photometry package are extra steps for manual adjustment of the isophotal solution after basic profile fitting.
Here, the user examines the radial profiles of SB, PA, and ellipticity to identify problematic regions, and interpolates between robust regions to fill problematic areas.
The resulting profiles represent the large-scale structure of the galaxy with high fidelity, without the influence from small non-axisymmetric features (and guards against the possibility of crossing isophotes).
Provided close attention, {\scriptsize XVISTA} allows for an accurate non-parametric elliptical representation of the light distribution of a galaxy.
Profile interpolations may also remove some fine structure in the fit, but the global light distribution is always preserved~\citep{Lauer2005}.
The primary drawback of this method is the lack of scalability if human intervention is involved.

\begin{figure}
    \centering
    \includegraphics[width = \columnwidth]{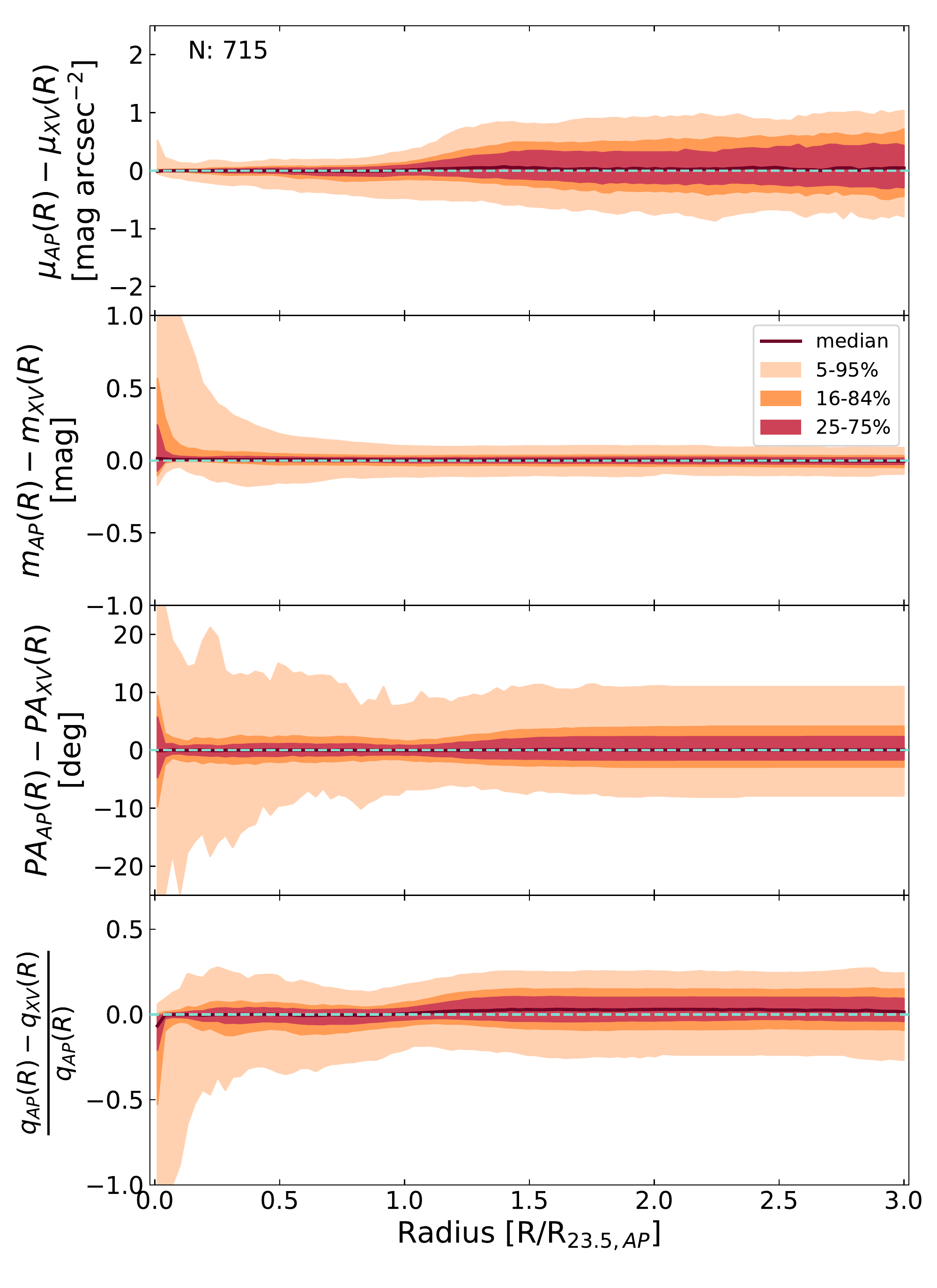}
    \caption{Same as \Fig{compareprofilesphotutils} for comparing radial profiles from {\scriptsize AUTOPROF} and {\scriptsize XVISTA}.}
    \label{fig:compareprofilesxvista}
\end{figure}

We have applied the {\scriptsize XVISTA} surface photometry analysis to a sample of $722$ PROBES galaxies for comparison with {\scriptsize AUTOPROF} profiles. These were all extracted uniformly by one of us (NA). 
The {\scriptsize XVISTA} surface photometry package includes its own methods for background subtraction, star masking (also done manually), SB extraction, etc~\citep{Courteau1996}.
Thus our comparison with {\scriptsize XVISTA} will be influenced by differences in all of these steps, as well as the actual isophotal fitting routine.
{\scriptsize XVISTA} and {\scriptsize AUTOPROF} are indeed fully independent of each other.
Despite these caveats, \Fig{compareprofilesxvista} shows remarkable agreement between {\scriptsize AUTOPROF} and {\scriptsize XVISTA}.
Our SB estimates are reliably equivalent to within less than \wunits{0.3}{\magss} out to the outskirts ($\sim 3R_{23.5}$) of each galaxy.
There is a slight bias beyond $R_{23.5}$ for {\scriptsize AUTOPROF} to have dimmer isophotes (by \wunits{0.03}{\magss}) which may be explained by both {\scriptsize AUTOPROF} and {\scriptsize XVISTA} applying the technique of extrapolating the outer fitted isophote, each using a different point to select the last isophote.
The slight brightness difference could also arise from a different masking procedures, where {\scriptsize XVISTA} uses manually created object masks and {\scriptsize AUTOPROF} uses sigma clipping.
The curve of growth and PA profiles are remarkably tight over the entire range tested.
Note that {\scriptsize XVISTA} nominally defines PAs relative to the positive $y$-axis and increasing clockwise; we inverted this for the sake of comparison.
The axis ratio comparison shows a slight systematic shift beyond $R_{23.5}$ indicating that {\scriptsize AUTOPROF} finds rounder isophotes by approximately \wunits{2}{per cent} (median deviation past $R_{23.5}$).

\begin{figure}
    \centering
    \includegraphics[width = \columnwidth]{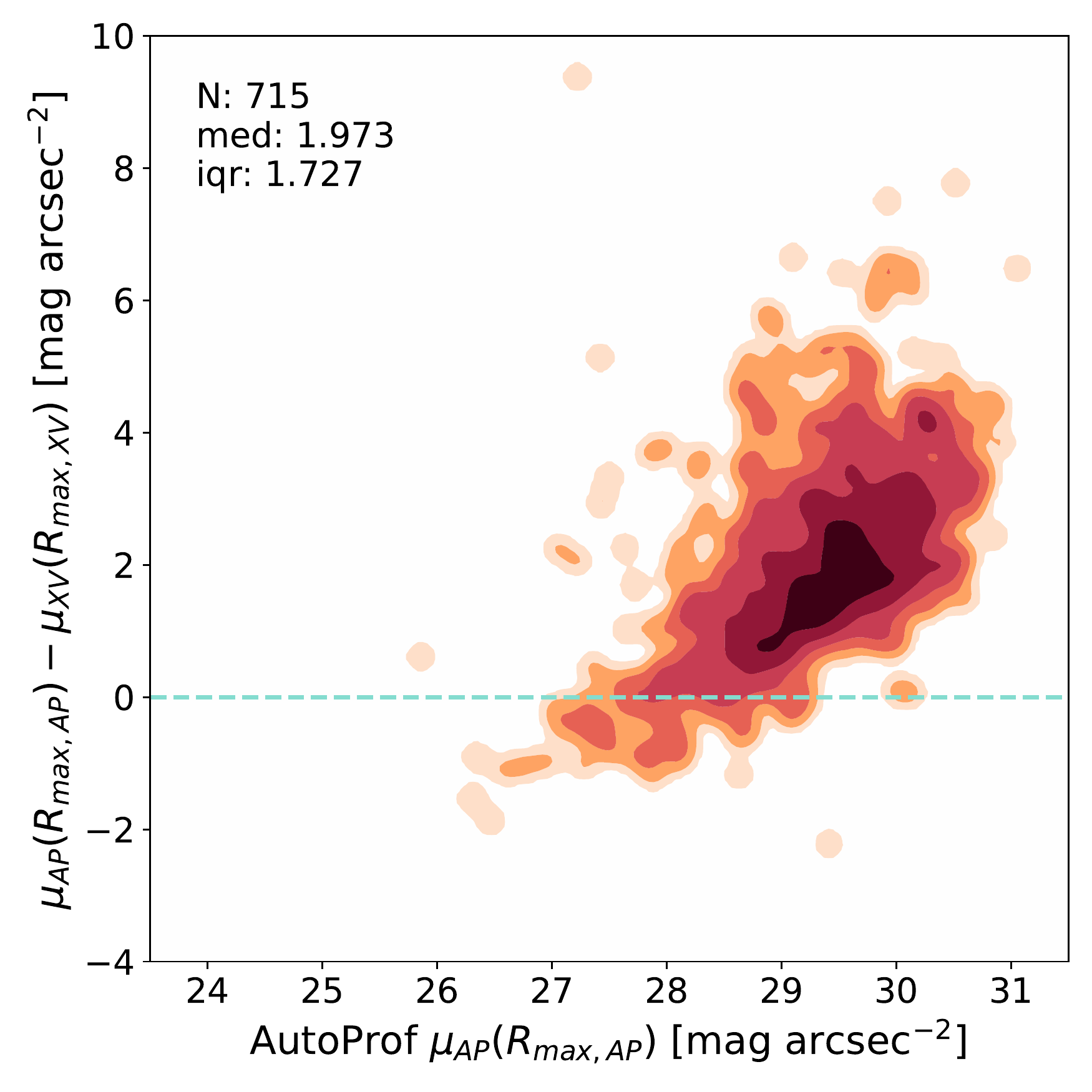}
    \caption{Comparison of photometric depths achieved by {\scriptsize AUTOPROF} and {\scriptsize XVISTA}; formatted as in \Fig{comparedepthphotutils}.}
    \label{fig:comparedepthxvista}
\end{figure}

A comparison of LSB features is also most instructive. 
\Fig{comparedepthxvista} shows a comparison of photometric depths between {\scriptsize XVISTA} and {\scriptsize AUTOPROF} before encountering a SB error of \wunits{0.2}{\magss} limit.
For the same data, {\scriptsize AUTOPROF}'s SB profiles reach deeper SB levels since {\scriptsize AUTOPROF} uses band averaging in the outskirts (\Sec{sbextraction}). 
{\scriptsize AUTOPROF} typically achieves deeper photometry by \wunits{\sim 1}{\magss} and can often reach deeper levels by more than \wunits{2}{\magss}.
Note that {\scriptsize XVISTA} computes errors similarly to {\scriptsize PHOTUTILS} using $\sigma/\sqrt{N}$ and is biased high by unmasked sources.

\subsection{{\scriptsize GALFIT}}
\label{sec:galfit}

{\scriptsize GALFIT} performs structural decompositions of galaxy images with parametric 2D light distributions~\citep{Peng2010}.
Comparisons of structural parameters derived with {\scriptsize AUTOPROF} and {\scriptsize GALFIT} have already been performed in \citet{Arora2021} using {\scriptsize PYMORPH}, here we will compare profiles directly.
{\scriptsize GALFIT}'s parametric approach to modelling galaxy images is fundamentally different from {\scriptsize AUTOPROF}, {\scriptsize PHOTUTILS}, and {\scriptsize XVISTA}.
Through least squares optimization, {\scriptsize GALFIT} can extract an arbitrary number of parametric models to fit a galaxy image.
These models include the exponential disc~\citep{deVaucouleurs1958, Freeman1970,Kormendy1978,Kent1991,vanderkruit2011},
S{\'e}rsic~\citep{sersic1968},
Moffat~\citep{Moffat1969}, 
Nuker~\citep{Lauer1995}, and others.
A PA and ellipticity are fit for each component globally.
{\scriptsize GALFIT} models are typically azimuthally represented as an ellipse, though they can be transformed to fit boxy, pointed, twisted, and more shapes.
{\scriptsize GALFIT} can also create especially detailed models \footnote{see \url{https://users.obs.carnegiescience.edu/peng/work/galfit/README.pdf}}, more so than elliptical non-parametric fits.
With manual fine-tuning and well chosen initial conditions, {\scriptsize GALFIT} models can describe a galaxy's light distribution with high fidelity, relative to the other methods considered here ({\scriptsize AUTOPROF}, {\scriptsize PHOTUTILS}, and {\scriptsize XVISTA}).
The user can extract detailed information about individual features such a bulge, bar, spiral arm, etc, by first modelling the whole galaxy then either focusing on a single component or subtracting various components by choice. 
As such, one may investigate galaxy structure in a fundamentally different way than in elliptical non-parametric fits, making the methods highly complimentary.
However, the process of constructing such a detailed model is time consuming and its efficient and reliable implementation for large surveys can be challenging.
For large surveys, {\scriptsize GALFIT} has been used in an automated fashion with one or two component models to fit galaxy images~\citep[e.g.,][]{Kelvin2010,Simard2011}.
Such models are well suited for low resolution images, and for overlapping systems where all sources in the image can be simultaneously modelled.

Since {\scriptsize AUTOPROF} is intended for scalability without user input on a per galaxy level, our comparison of the two methods uses simple two-component {\scriptsize GALFIT} models consisting of combinations of S\'ersic and exponential disc profiles.
We have applied those {\scriptsize GALFIT} parametric models to the our PROBES sample of galaxies.
The {\scriptsize GALFIT} fitted models are sensitive to parameter initialization, owing partly to the overly simplistic choice of the two component models.
To start the models with the closest approximation to their final values, we used the results from the {\scriptsize AUTOPROF} pipeline to construct {\scriptsize GALFIT} configuration files.
The non-parametric {\scriptsize AUTOPROF} SB profiles were decomposed into S\'ersic and exponential components using a least squares optimization to the 1D profiles.
The resulting parameters were then given as initialization for {\scriptsize GALFIT}, along with the global PA and ellipticity fitted to the image.
{\scriptsize GALFIT} is also sensitive to un-modeled image components.  
We therefore applied {\scriptsize SEXTRACTOR}~\citep{Bertin1996} to build a segmentation map and mask all non-galaxy components of the image (mostly stars and background galaxies).
The initialization for the component centres were taken from the {\scriptsize AUTOPROF} pipeline.
We constructed a PSF for each image using a median of many flux-normalized stars (selected using the {\scriptsize AUTOPROF} star finder from \Sec{psf}).
Each galaxy was modeled with (i) an exponential disc, (ii) a S\'ersic function, (iii) an exponential disc and a S\'ersic, and (iv) a doubleS\'ersic.  
The adopted model was the one with lowest $\chi^2/{\rm dof}$ as reported by the {\scriptsize GALFIT} fitting routine~\citep[e.g.,][]{Simard2011,Gilhuly2018}. 

\begin{figure}
    \centering
    \includegraphics[width = \columnwidth]{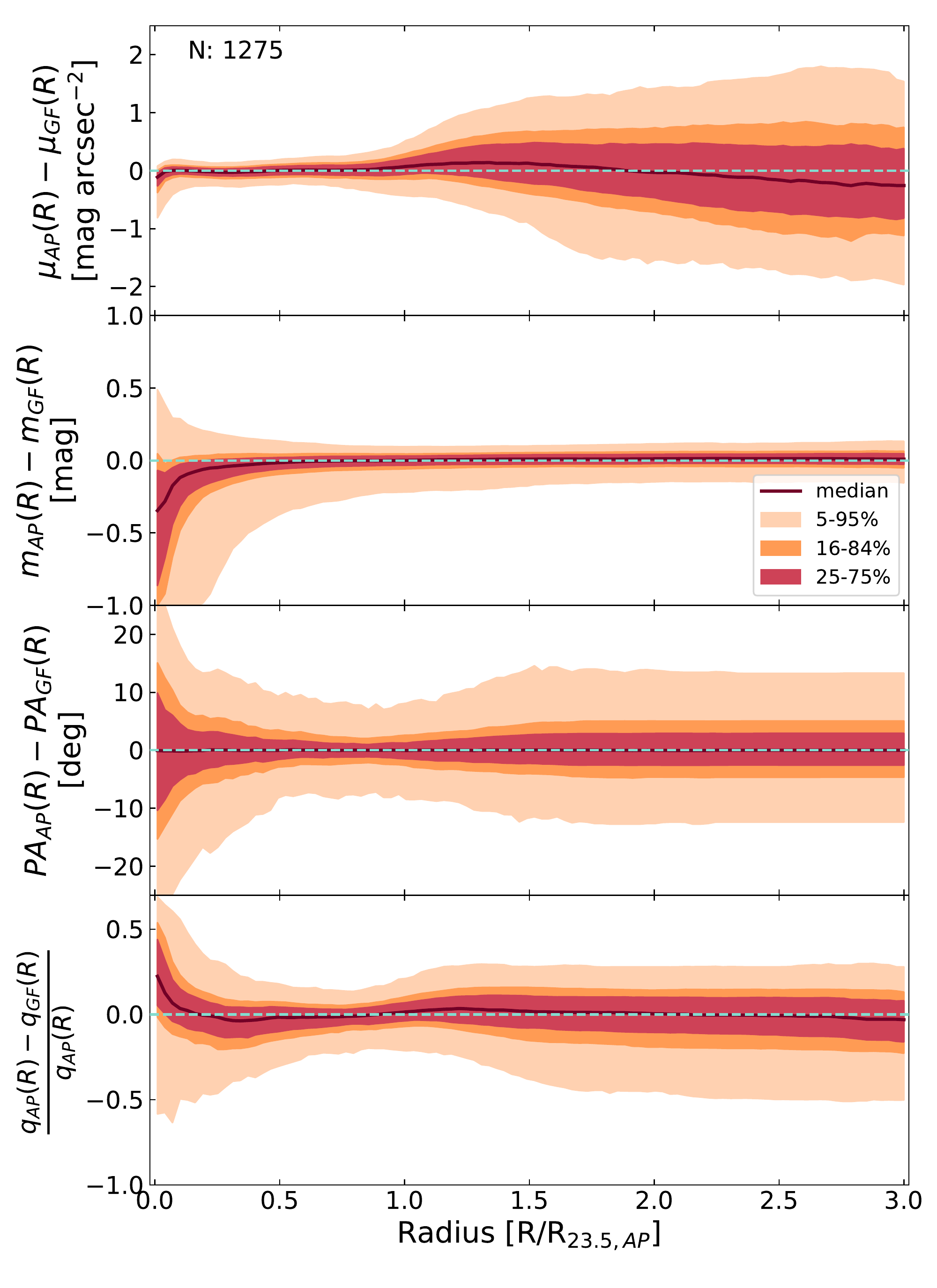}
    \caption{Same as \Fig{compareprofilesphotutils} for the comparison of radial profiles from {\scriptsize AUTOPROF} and {\scriptsize GALFIT}.}
    \label{fig:compareprofilesgalfit}
\end{figure}

{\scriptsize GALFIT} models must be converted into SB profiles for proper comparison with {\scriptsize AUTOPROF}'s output.
This is accomplished by applying {\scriptsize AUTOPROF} to the final, idealized, {\scriptsize GALFIT} model.
\Fig{compareprofilesgalfit} shows a comparison of the resulting photometry profiles. 
Within $R_{23.5}$ the match between {\scriptsize AUTOPROF} and {\scriptsize GALFIT} is good.
Larger systematic deviations are seen beyond $R_{23.5}$. 
At most radii, the scatter around the zero line is larger than in the comparisons with {\scriptsize PHOTUTILS} and {\scriptsize XVISTA};
this is to be expected given our choice of a two component model.
Simple models fail to capture the full complexity of the late-type galaxy population in PROBES; a multicomponent model could yield better results at the expense of scalability. 
While the curve of growth comparison is poor in the centre, it converges quickly and remains good out to $3R_{23.5}$.
A mismatch at the centre is expected and occurs on the PSF scale (roughly $0.2R_{23.5}$ on average) where isophotes are rounded. 
PA comparisons show no systematic differences throughout the profiles, however the random scatter is considerably larger than with {\scriptsize PHOTUTILS} and {\scriptsize XVISTA} (see \Secs{photutils}{xvista}).
The PA scatter is most pronounced near the centre where non-axisymmetric structures such as bars can have PAs offset from the global PA; {\scriptsize GALFIT} could only recover the global PA using simple two component models.
The axis ratio profiles show {\scriptsize AUTOPROF} finding systematically rounder values near the centre as expected due to the PSF scale ($\sim 0.2R_{23.5}$) which affects {\scriptsize AUTOPROF} results, but not {\scriptsize GALFIT} since it models the PSF. 
At large radii where axis ratios are more stable, {\scriptsize AUTOPROF} and {\scriptsize GALFIT} agree with no bias, albeit with a larger scatter than {\scriptsize PHOTUTILS} and {\scriptsize XVISTA}.
Overall, both methods adequately match the global light distribution of each galaxy.  
However, {\scriptsize GALFIT} is restricted to two "rigid" model components and thus cannot account for variations in PA and ellipticity.
Being unable to account for those, {\scriptsize GALFIT} must `split the difference' which ultimately leads to a larger scatter in all profile parameters.
More detailed multicomponent {\scriptsize GALFIT} models can match galaxy images with greater fidelity and likely better agreement with {\scriptsize AUTOPROF} radial profiles of SB, PA, and ellipticity. 
An advantage of the non-parametric ellipse fitting approach in {\scriptsize AUTOPROF} (and similar codes) is that it is nearly equivalent to fitting many parametric models simultaneously allowing for more complete representation of the light distribution.

\section{Conclusion}
\label{sec:conclusions}

We have presented the robust automated isophotal solving pipeline, {\scriptsize AUTOPROF} for the extraction of valuable structural information from galaxy images. 
The pipeline is easily implemented and offers a suite of robust default and optional tools for SB profile extractions and related methods.
The {\scriptsize AUTOPROF} pipeline is highly extensible and can be adapted for a variety of applications. 
The {\scriptsize AUTOPROF} code is available freely to the community at: \break  \url{https://github.com/ConnorStoneAstro/AutoProf}.

We have compared {\scriptsize AUTOPROF} with similar widely used surface photometry codes ({\scriptsize PHOTUTILS}, {\scriptsize XVISTA}, and {\scriptsize GALFIT}) and found generally superb agreement, especially with other non-parametric solutions.
However, the ease of implementation, automation, and ability to tease out faint signals in galaxy's outskirts sets {\scriptsize AUTOPROF} apart from other softwares. 
Indeed, it is shown that {\scriptsize AUTOPROF}, with its band averaging technique and other numerical treatments, yields SB profiles that typically reach 2-3 {\magss} deeper than other similar software for a given SB error. 
{\scriptsize AUTOPROF} is also the only method considered here which fully automates all analysis steps, making it the fastest to implement for new users.

We find that overly simplistic parametric models with only one or two components cannot adequately describe the light distribution of a galaxy, yielding larger scatter in fitted parameters than can be achieved with non-parametric approaches.
Overall, {\scriptsize AUTOPROF} balances the need for complex (spatially resolved) information available in each galaxy image with the desire for fast, automated, and efficient representations of a galaxy.
For large-scale applications, {\scriptsize AUTOPROF} produces reliable and extended brightness profiles.
{\scriptsize AUTOPROF} is thus ideally suited for modern large-scale investigations of the detailed statistical nature of galaxy formation and evolution.

\section*{Data Availability}
The data underlying this article will be shared on reasonable request to the corresponding author(s).

\section*{Acknowledgements}

We are grateful to the Natural Sciences and Engineering Research Council of Canada, the Ontario Government, and Queen's University for support through various generous scholarships and grants.
Ivo Busko, Tod Lauer, and Chien Peng have provided especially valuable comments regarding our comparisons with {\scriptsize PHOTUTILS}, {\scriptsize XVISTA}, and {\scriptsize GALFIT}; their contributions are duly acknowledged. 
Sean Begy, Sim{\'o}n D{\'i}az Garc{\'i}a, Matt Frosst, and Mike Smith are thanked for comments that improved {\scriptsize AUTOPROF}'s implementation.

\bibliographystyle{mnras}
\bibliography{AutoProf}

\begin{thebibliography}{}
\makeatletter
\relax
\def\mn@urlcharsother{\let\do\@makeother \do\$\do\&\do\#\do\^\do\_\do\%\do\~}
\def\mn@doi{\begingroup\mn@urlcharsother \@ifnextchar [ {\mn@doi@}
  {\mn@doi@[]}}
\def\mn@doi@[#1]#2{\def\@tempa{#1}\ifx\@tempa\@empty \href
  {http://dx.doi.org/#2} {doi:#2}\else \href {http://dx.doi.org/#2} {#1}\fi
  \endgroup}
\def\mn@eprint#1#2{\mn@eprint@#1:#2::\@nil}
\def\mn@eprint@arXiv#1{\href {http://arxiv.org/abs/#1} {{\tt arXiv:#1}}}
\def\mn@eprint@dblp#1{\href {http://dblp.uni-trier.de/rec/bibtex/#1.xml}
  {dblp:#1}}
\def\mn@eprint@#1:#2:#3:#4\@nil{\def\@tempa {#1}\def\@tempb {#2}\def\@tempc
  {#3}\ifx \@tempc \@empty \let \@tempc \@tempb \let \@tempb \@tempa \fi \ifx
  \@tempb \@empty \def\@tempb {arXiv}\fi \@ifundefined
  {mn@eprint@\@tempb}{\@tempb:\@tempc}{\expandafter \expandafter \csname
  mn@eprint@\@tempb\endcsname \expandafter{\@tempc}}}

\bibitem[\protect\citeauthoryear{{Akhlaghi}}{{Akhlaghi}}{2019}]{Akhlaghi2019}
{Akhlaghi} M.,  2019, arXiv e-prints, \href
  {https://ui.adsabs.harvard.edu/abs/2019arXiv190911230A} {p. arXiv:1909.11230}

\bibitem[\protect\citeauthoryear{{Amiaux} et~al.,}{{Amiaux}
  et~al.}{2012}]{Amiaux2012}
{Amiaux} J.,  et~al., 2012, in {Clampin} M.~C.,  {Fazio} G.~G.,  {MacEwen}
  H.~A.,   {Oschmann} Jacobus~M. J.,  eds,  Society of Photo-Optical
  Instrumentation Engineers (SPIE) Conference Series Vol. 8442, Space
  Telescopes and Instrumentation 2012: Optical, Infrared, and Millimeter Wave.
  p. 84420Z (\mn@eprint {arXiv} {1209.2228}), \mn@doi{10.1117/12.926513}

\bibitem[\protect\citeauthoryear{{Arora}, {Stone}, {Courteau}  \&
  {Jarrett}}{{Arora} et~al.}{2021}]{Arora2021}
{Arora} N.,  {Stone} C.,  {Courteau} S.,   {Jarrett} T.~H.,  2021, \mn@doi
  [\mnras] {10.1093/mnras/stab1430}, \href
  {https://ui.adsabs.harvard.edu/abs/2021MNRAS.505.3135A} {505, 3135}

\bibitem[\protect\citeauthoryear{{Barden}, {H{\"a}u{\ss}ler}, {Peng},
  {McIntosh}  \& {Guo}}{{Barden} et~al.}{2012}]{Barden2012}
{Barden} M.,  {H{\"a}u{\ss}ler} B.,  {Peng} C.~Y.,  {McIntosh} D.~H.,   {Guo}
  Y.,  2012, \mn@doi [\mnras] {10.1111/j.1365-2966.2012.20619.x}, \href
  {https://ui.adsabs.harvard.edu/abs/2012MNRAS.422..449B} {422, 449}

\bibitem[\protect\citeauthoryear{{Bertin} \& {Arnouts}}{{Bertin} \&
  {Arnouts}}{1996}]{Bertin1996}
{Bertin} E.,  {Arnouts} S.,  1996, \mn@doi [\aaps] {10.1051/aas:1996164}, \href
  {https://ui.adsabs.harvard.edu/abs/1996A&AS..117..393B} {117, 393}

\bibitem[\protect\citeauthoryear{{Bottrell}, {Torrey}, {Simard}  \&
  {Ellison}}{{Bottrell} et~al.}{2017}]{Bottrell2017}
{Bottrell} C.,  {Torrey} P.,  {Simard} L.,   {Ellison} S.~L.,  2017, \mn@doi
  [\mnras] {10.1093/mnras/stx276}, \href
  {https://ui.adsabs.harvard.edu/abs/2017MNRAS.467.2879B} {467, 2879}

\bibitem[\protect\citeauthoryear{{Bradley} et~al.,}{{Bradley}
  et~al.}{2020}]{photutils}
{Bradley} L.,  et~al., 2020, {astropy/photutils: 1.0.0},
  \mn@doi{10.5281/zenodo.4044744}

\bibitem[\protect\citeauthoryear{Burger \& Burge}{Burger \&
  Burge}{2010}]{Burger2010}
Burger W.,  Burge M.,  2010, Principles of Digital Image Processing: Core
  Algorithms.
Undergraduate Topics in Computer Science, Springer London, \url
  {https://books.google.ca/books?id=s5CBZLBakawC}

\bibitem[\protect\citeauthoryear{{Carter}}{{Carter}}{1978}]{Carter1978}
{Carter} D.,  1978, \mn@doi [\mnras] {10.1093/mnras/182.4.797}, \href
  {https://ui.adsabs.harvard.edu/abs/1978MNRAS.182..797C} {182, 797}

\bibitem[\protect\citeauthoryear{{Ciambur}}{{Ciambur}}{2015}]{Ciambur2015}
{Ciambur} B.~C.,  2015, \mn@doi [\apj] {10.1088/0004-637X/810/2/120}, \href
  {https://ui.adsabs.harvard.edu/abs/2015ApJ...810..120C} {810, 120}

\bibitem[\protect\citeauthoryear{{Ciambur}}{{Ciambur}}{2016}]{Ciambur2016}
{Ciambur} B.~C.,  2016, \mn@doi [\pasa] {10.1017/pasa.2016.60}, \href
  {https://ui.adsabs.harvard.edu/abs/2016PASA...33...62C} {33, e062}

\bibitem[\protect\citeauthoryear{{Comer{\'o}n}, {Salo}  \&
  {Knapen}}{{Comer{\'o}n} et~al.}{2018}]{Comeron2018}
{Comer{\'o}n} S.,  {Salo} H.,   {Knapen} J.~H.,  2018, \mn@doi [\aap]
  {10.1051/0004-6361/201731415}, \href
  {https://ui.adsabs.harvard.edu/abs/2018A&A...610A...5C} {610, A5}

\bibitem[\protect\citeauthoryear{{Courteau}}{{Courteau}}{1996}]{Courteau1996}
{Courteau} S.,  1996, \mn@doi [\apjs] {10.1086/192281}, \href
  {https://ui.adsabs.harvard.edu/#abs/1996ApJS..103..363C} {103, 363}

\bibitem[\protect\citeauthoryear{{Courteau}, {Widrow}, {McDonald},
  {Guhathakurta}, {Gilbert}, {Zhu}, {Beaton}  \& {Majewski}}{{Courteau}
  et~al.}{2011}]{Courteau2011}
{Courteau} S.,  {Widrow} L.~M.,  {McDonald} M.,  {Guhathakurta} P.,  {Gilbert}
  K.~M.,  {Zhu} Y.,  {Beaton} R.~L.,   {Majewski} S.~R.,  2011, \mn@doi [\apj]
  {10.1088/0004-637X/739/1/20}, \href
  {https://ui.adsabs.harvard.edu/abs/2011ApJ...739...20C} {739, 20}

\bibitem[\protect\citeauthoryear{{Davis}, {Cawson}, {Davies}  \&
  {Illingworth}}{{Davis} et~al.}{1985}]{Davis1985}
{Davis} L.~E.,  {Cawson} M.,  {Davies} R.~L.,   {Illingworth} G.,  1985,
  \mn@doi [\aj] {10.1086/113723}, \href
  {https://ui.adsabs.harvard.edu/abs/1985AJ.....90..169D} {90, 169}

\bibitem[\protect\citeauthoryear{{Dey} et~al.,}{{Dey} et~al.}{2019}]{Dey2019}
{Dey} A.,  et~al., 2019, \mn@doi [\aj] {10.3847/1538-3881/ab089d}, \href
  {https://ui.adsabs.harvard.edu/abs/2019AJ....157..168D} {157, 168}

\bibitem[\protect\citeauthoryear{{Einasto}}{{Einasto}}{1965}]{Einasto1965}
{Einasto} J.,  1965, Trudy Astrofizicheskogo Instituta Alma-Ata, \href
  {https://ui.adsabs.harvard.edu/abs/1965TrAlm...5...87E} {5, 87}

\bibitem[\protect\citeauthoryear{{Erwin}}{{Erwin}}{2015}]{Erwin2015}
{Erwin} P.,  2015, \mn@doi [\apj] {10.1088/0004-637X/799/2/226}, \href
  {https://ui.adsabs.harvard.edu/abs/2015ApJ...799..226E} {799, 226}

\bibitem[\protect\citeauthoryear{{Freeman}}{{Freeman}}{1970}]{Freeman1970}
{Freeman} K.~C.,  1970, \mn@doi [\apj] {10.1086/150474}, \href
  {http://adsabs.harvard.edu/abs/1970ApJ...160..811F} {160, 811}

\bibitem[\protect\citeauthoryear{{Gilhuly} \& {Courteau}}{{Gilhuly} \&
  {Courteau}}{2018}]{Gilhuly2018}
{Gilhuly} C.,  {Courteau} S.,  2018, \mn@doi [\mnras] {10.1093/mnras/sty756},
  \href {https://ui.adsabs.harvard.edu/#abs/2018MNRAS.477..845G} {477, 845}

\bibitem[\protect\citeauthoryear{{Hall}, {Courteau}, {Dutton}, {McDonald}  \&
  {Zhu}}{{Hall} et~al.}{2012}]{Hall2012}
{Hall} M.,  {Courteau} S.,  {Dutton} A.~A.,  {McDonald} M.,   {Zhu} Y.,  2012,
  \mn@doi [\mnras] {10.1111/j.1365-2966.2012.21290.x}, \href
  {http://adsabs.harvard.edu/abs/2012MNRAS.425.2741H} {425, 2741}

\bibitem[\protect\citeauthoryear{{Hubble}}{{Hubble}}{1930}]{Hubble1930}
{Hubble} E.~P.,  1930, \mn@doi [\apj] {10.1086/143250}, \href
  {https://ui.adsabs.harvard.edu/abs/1930ApJ....71..231H} {71, 231}

\bibitem[\protect\citeauthoryear{{Ivezi{\'c}} et~al.,}{{Ivezi{\'c}}
  et~al.}{2019}]{Ivezic2019}
{Ivezi{\'c}} {\v Z}.,  et~al., 2019, \mn@doi [\apj] {10.3847/1538-4357/ab042c},
  \href {http://adsabs.harvard.edu/abs/2019ApJ...873..111I} {873, 111}

\bibitem[\protect\citeauthoryear{{Jedrzejewski}}{{Jedrzejewski}}{1987}]{Jedrzejewski1987}
{Jedrzejewski} R.~I.,  1987, \mn@doi [\mnras] {10.1093/mnras/226.4.747}, \href
  {https://ui.adsabs.harvard.edu/abs/1987MNRAS.226..747J} {226, 747}

\bibitem[\protect\citeauthoryear{{Kelvin}, {Driver}, {Robotham}, {Hill}  \&
  {Cameron}}{{Kelvin} et~al.}{2010}]{Kelvin2010}
{Kelvin} L.,  {Driver} S.,  {Robotham} A.,  {Hill} D.,   {Cameron} E.,  2010,
  in {Debattista} V.~P.,  {Popescu} C.~C.,  eds,  American Institute of Physics
  Conference Series Vol. 1240, Hunting for the Dark: the Hidden Side of Galaxy
  Formation. pp 247--248, \mn@doi{10.1063/1.3458501}

\bibitem[\protect\citeauthoryear{{Kent}}{{Kent}}{1983}]{Kent1983}
{Kent} S.~M.,  1983, \mn@doi [\apj] {10.1086/160803}, \href
  {https://ui.adsabs.harvard.edu/abs/1983ApJ...266..562K} {266, 562}

\bibitem[\protect\citeauthoryear{{Kent}}{{Kent}}{1985}]{Kent1985}
{Kent} S.~M.,  1985, \mn@doi [\apjs] {10.1086/191066}, \href
  {https://ui.adsabs.harvard.edu/#abs/1985ApJS...59..115K} {59, 115}

\bibitem[\protect\citeauthoryear{{Kent}, {Dame}  \& {Fazio}}{{Kent}
  et~al.}{1991}]{Kent1991}
{Kent} S.~M.,  {Dame} T.~M.,   {Fazio} G.,  1991, \mn@doi [\apj]
  {10.1086/170413}, \href
  {https://ui.adsabs.harvard.edu/abs/1991ApJ...378..131K} {378, 131}

\bibitem[\protect\citeauthoryear{{Kormendy} \& {Bruzual A.}}{{Kormendy} \&
  {Bruzual A.}}{1978}]{Kormendy1978}
{Kormendy} J.,  {Bruzual A.} G.,  1978, \mn@doi [\apjl] {10.1086/182729}, \href
  {https://ui.adsabs.harvard.edu/abs/1978ApJ...223L..63K} {223, L63}

\bibitem[\protect\citeauthoryear{{Lauer}}{{Lauer}}{1985}]{Lauer1985}
{Lauer} T.~R.,  1985, \mn@doi [\apjs] {10.1086/191011}, \href
  {https://ui.adsabs.harvard.edu/abs/1985ApJS...57..473L} {57, 473}

\bibitem[\protect\citeauthoryear{{Lauer}}{{Lauer}}{1986}]{Lauer1986}
{Lauer} T.~R.,  1986, \mn@doi [\apj] {10.1086/164752}, \href
  {https://ui.adsabs.harvard.edu/abs/1986ApJ...311...34L} {311, 34}

\bibitem[\protect\citeauthoryear{{Lauer} et~al.,}{{Lauer}
  et~al.}{1995}]{Lauer1995}
{Lauer} T.~R.,  et~al., 1995, \mn@doi [\aj] {10.1086/117719}, \href
  {https://ui.adsabs.harvard.edu/abs/1995AJ....110.2622L} {110, 2622}

\bibitem[\protect\citeauthoryear{{Lauer} et~al.,}{{Lauer}
  et~al.}{2005}]{Lauer2005}
{Lauer} T.~R.,  et~al., 2005, \mn@doi [\aj] {10.1086/429565}, \href
  {https://ui.adsabs.harvard.edu/abs/2005AJ....129.2138L} {129, 2138}

\bibitem[\protect\citeauthoryear{{MacArthur}, {Courteau}  \&
  {Holtzman}}{{MacArthur} et~al.}{2003}]{MacArthur2003}
{MacArthur} L.~A.,  {Courteau} S.,   {Holtzman} J.~A.,  2003, \mn@doi [\apj]
  {10.1086/344506}, \href
  {https://ui.adsabs.harvard.edu/abs/2003ApJ...582..689M} {582, 689}

\bibitem[\protect\citeauthoryear{McDonald, Courteau, Tully  \&
  Roediger}{McDonald et~al.}{2011}]{McDonald2011}
McDonald M.,  Courteau S.,  Tully R.~B.,   Roediger J.,  2011, \mn@doi [\mnras]
  {10.1111/j.1365-2966.2011.18519.x}, 414, 2055

\bibitem[\protect\citeauthoryear{{Moffat}}{{Moffat}}{1969}]{Moffat1969}
{Moffat} A.~F.~J.,  1969, \aap, \href
  {https://ui.adsabs.harvard.edu/abs/1969A&A.....3..455M} {3, 455}

\bibitem[\protect\citeauthoryear{Mosby et~al.,}{Mosby et~al.}{2020}]{Mosby2020}
Mosby G.,  et~al., 2020, \mn@doi [Journal of Astronomical Telescopes,
  Instruments, and Systems] {10.1117/1.JATIS.6.4.046001}, 6, 1

\bibitem[\protect\citeauthoryear{{Peng}, {Ho}, {Impey}  \& {Rix}}{{Peng}
  et~al.}{2002}]{Peng2002}
{Peng} C.~Y.,  {Ho} L.~C.,  {Impey} C.~D.,   {Rix} H.-W.,  2002, \mn@doi [\aj]
  {10.1086/340952}, \href
  {https://ui.adsabs.harvard.edu/abs/2002AJ....124..266P} {124, 266}

\bibitem[\protect\citeauthoryear{{Peng}, {Ho}, {Impey}  \& {Rix}}{{Peng}
  et~al.}{2010}]{Peng2010}
{Peng} C.~Y.,  {Ho} L.~C.,  {Impey} C.~D.,   {Rix} H.-W.,  2010, \mn@doi [\aj]
  {10.1088/0004-6256/139/6/2097}, \href
  {https://ui.adsabs.harvard.edu/abs/2010AJ....139.2097P} {139, 2097}

\bibitem[\protect\citeauthoryear{{Ratnatunga} \& {Newell}}{{Ratnatunga} \&
  {Newell}}{1984}]{Ratnatunga1984}
{Ratnatunga} K.~U.,  {Newell} E.~B.,  1984, \mn@doi [\aj] {10.1086/113498},
  \href {https://ui.adsabs.harvard.edu/abs/1984AJ.....89..176R} {89, 176}

\bibitem[\protect\citeauthoryear{{S\'ersic}}{{S\'ersic}}{1968}]{sersic1968}
{S\'ersic} J.~L.,  1968, {Atlas de Galaxias Australes}.
Observatorio Astronomico,Cordoba

\bibitem[\protect\citeauthoryear{Shai Shalev-Shwartz}{Shai
  Shalev-Shwartz}{2014}]{Shalev2014}
Shai Shalev-Shwartz S. B.-D.,  2014, Understanding Machine Learning: From
  Theory to Algorithms.
Cambrige University Press

\bibitem[\protect\citeauthoryear{Shannon}{Shannon}{1949}]{Shannon1949}
Shannon C.,  1949, \mn@doi [Proceedings of the IRE]
  {10.1109/JRPROC.1949.232969}, 37, 10

\bibitem[\protect\citeauthoryear{{Simard} et~al.,}{{Simard}
  et~al.}{2002}]{Simard2002}
{Simard} L.,  et~al., 2002, \mn@doi [\apjs] {10.1086/341399}, \href
  {https://ui.adsabs.harvard.edu/abs/2002ApJS..142....1S} {142, 1}

\bibitem[\protect\citeauthoryear{{Simard}, {Mendel}, {Patton}, {Ellison}  \&
  {McConnachie}}{{Simard} et~al.}{2011}]{Simard2011}
{Simard} L.,  {Mendel} J.~T.,  {Patton} D.~R.,  {Ellison} S.~L.,
  {McConnachie} A.~W.,  2011, \mn@doi [\apjs] {10.1088/0067-0049/196/1/11},
  \href {https://ui.adsabs.harvard.edu/abs/2011ApJS..196...11S} {196, 11}

\bibitem[\protect\citeauthoryear{{Smith}, {Arora}, {Stone}, {Courteau}  \&
  {Geach}}{{Smith} et~al.}{2021}]{Smith2021}
{Smith} M.~J.,  {Arora} N.,  {Stone} C.,  {Courteau} S.,   {Geach} J.~E.,
  2021, \mn@doi [\mnras] {10.1093/mnras/stab424}, \href
  {https://ui.adsabs.harvard.edu/abs/2021MNRAS.503...96S} {503, 96}

\bibitem[\protect\citeauthoryear{{Stone} \& {Courteau}}{{Stone} \&
  {Courteau}}{2019}]{Stone2019}
{Stone} C.,  {Courteau} S.,  2019, \mn@doi [\apj] {10.3847/1538-4357/ab3126},
  \href {https://ui.adsabs.harvard.edu/abs/2019ApJ...882....6S} {882, 6}

\bibitem[\protect\citeauthoryear{{Stone}, {Courteau}  \& {Arora}}{{Stone}
  et~al.}{2021}]{Stone2021}
{Stone} C.,  {Courteau} S.,   {Arora} N.,  2021, \mn@doi [\apj]
  {10.3847/1538-4357/abebe4}, \href
  {https://ui.adsabs.harvard.edu/abs/2021ApJ...912...41S} {912, 41}

\bibitem[\protect\citeauthoryear{{Tody}}{{Tody}}{1986}]{Tody1986}
{Tody} D.,  1986, in {Crawford} D.~L.,  ed.,  Society of Photo-Optical
  Instrumentation Engineers (SPIE) Conference Series Vol. 627, Instrumentation
  in astronomy VI. p.~733, \mn@doi{10.1117/12.968154}

\bibitem[\protect\citeauthoryear{{Tuccillo}, {Huertas-Company},
  {Decenci{\`e}re}, {Velasco-Forero}, {Dom{\'{\i}}nguez S{\'a}nchez}  \&
  {Dimauro}}{{Tuccillo} et~al.}{2018}]{Tuccillo2018}
{Tuccillo} D.,  {Huertas-Company} M.,  {Decenci{\`e}re} E.,  {Velasco-Forero}
  S.,  {Dom{\'{\i}}nguez S{\'a}nchez} H.,   {Dimauro} P.,  2018, \mn@doi
  [\mnras] {10.1093/mnras/stx3186}, \href
  {http://adsabs.harvard.edu/abs/2018MNRAS.475..894T} {475, 894}

\bibitem[\protect\citeauthoryear{{Vikram}, {Wadadekar}, {Kembhavi}  \&
  {Vijayagovindan}}{{Vikram} et~al.}{2010}]{Vikram2010}
{Vikram} V.,  {Wadadekar} Y.,  {Kembhavi} A.~K.,   {Vijayagovindan} G.~V.,
  2010, \mn@doi [\mnras] {10.1111/j.1365-2966.2010.17426.x}, \href
  {https://ui.adsabs.harvard.edu/abs/2010MNRAS.409.1379V} {409, 1379}

\bibitem[\protect\citeauthoryear{{de Vaucouleurs}}{{de
  Vaucouleurs}}{1948}]{deVauc1948}
{de Vaucouleurs} G.,  1948, Annales d'Astrophysique, \href
  {https://ui.adsabs.harvard.edu/abs/1948AnAp...11..247D} {11, 247}

\bibitem[\protect\citeauthoryear{{de Vaucouleurs}}{{de
  Vaucouleurs}}{1958}]{deVaucouleurs1958}
{de Vaucouleurs} G.,  1958, \mn@doi [\apj] {10.1086/146564}, \href
  {https://ui.adsabs.harvard.edu/abs/1958ApJ...128..465D} {128, 465}

\bibitem[\protect\citeauthoryear{{van der Kruit} \& {Freeman}}{{van der Kruit}
  \& {Freeman}}{2011}]{vanderkruit2011}
{van der Kruit} P.~C.,  {Freeman} K.~C.,  2011, \mn@doi [\araa]
  {10.1146/annurev-astro-083109-153241}, \href
  {http://adsabs.harvard.edu/abs/2011ARA%26A..49..301V} {49, 301}

\makeatother
\end{thebibliography}

\appendix

\section{Other Photometry Method Comparisons}

\begin{figure*}
    \centering
    \includegraphics[width = 0.33\textwidth]{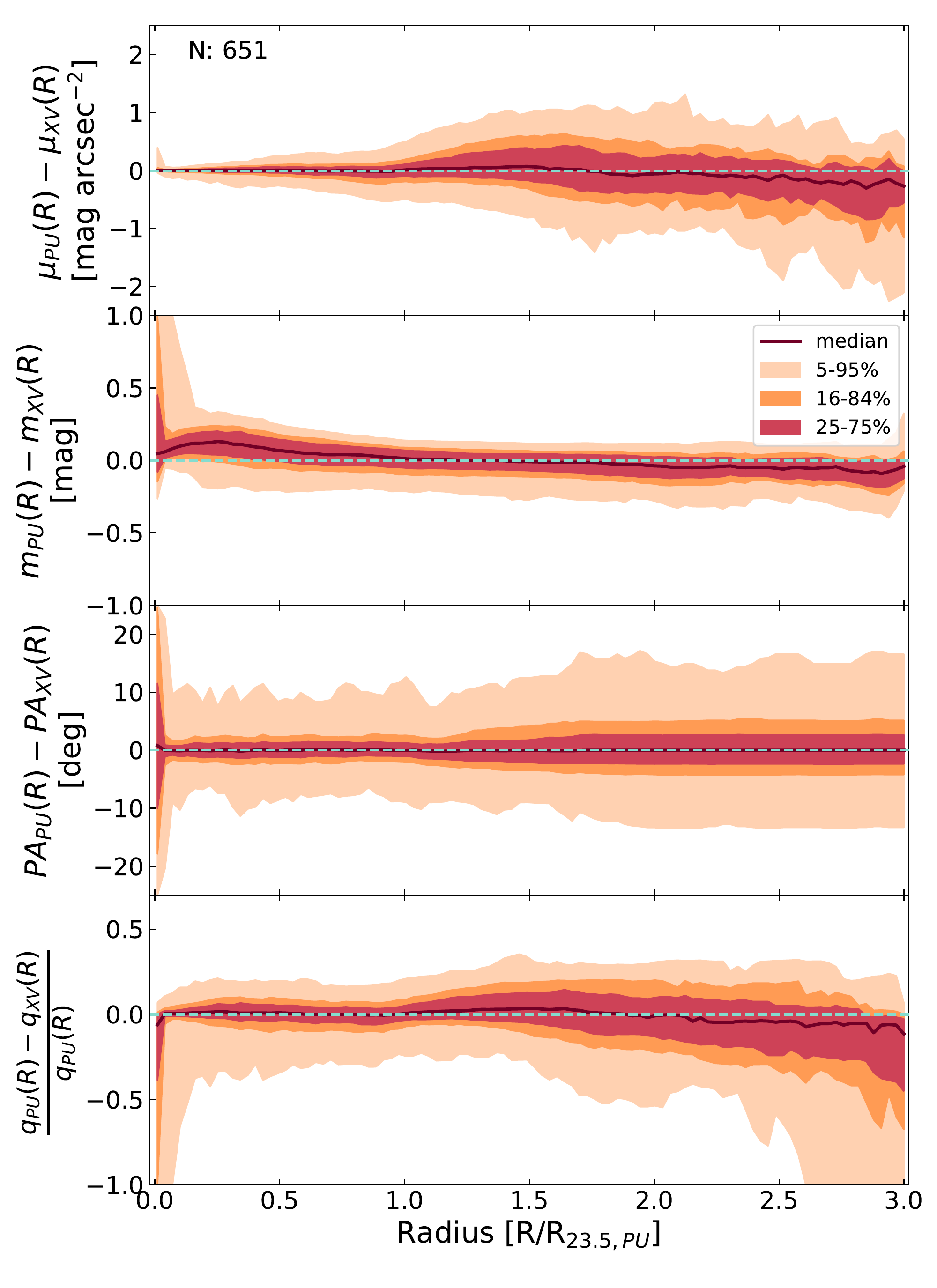}
    \includegraphics[width = 0.33\textwidth]{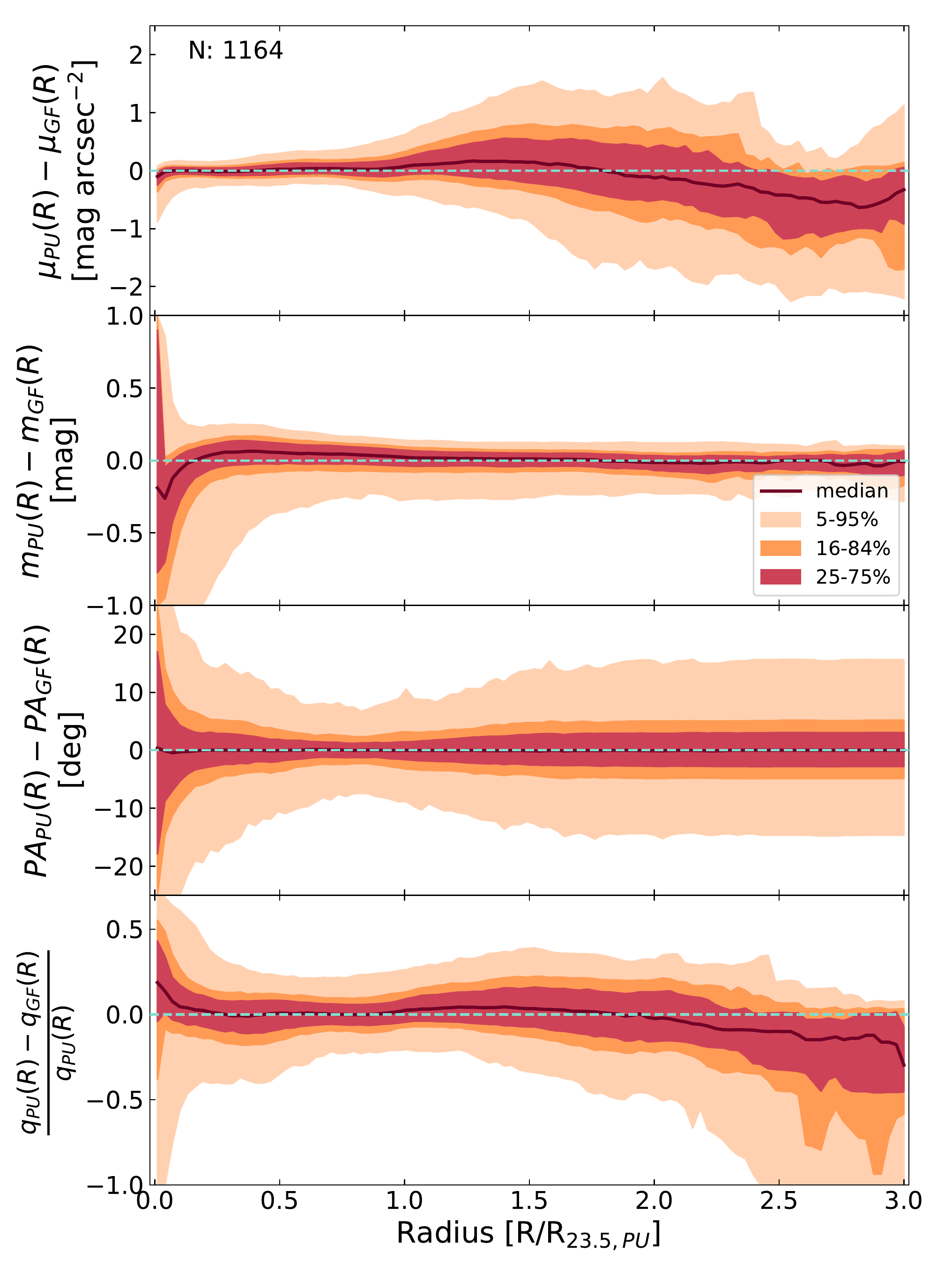}
    \includegraphics[width = 0.33\textwidth]{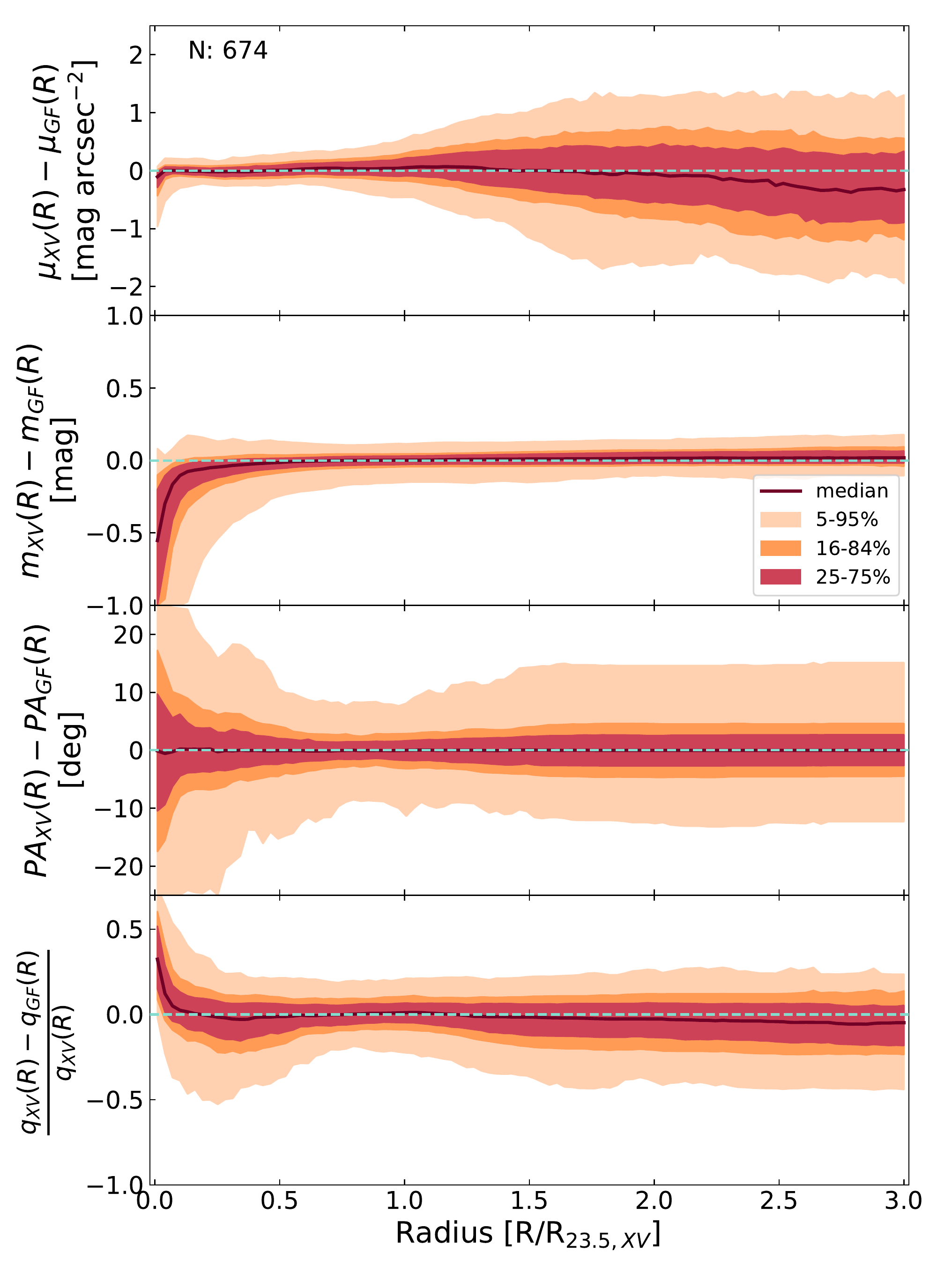}
    \caption{Comparison of radial profiles between {\scriptsize PHOTUTILS}, {\scriptsize XVISTA}, and {\scriptsize GALFIT}. Each panel formatted as in \Fig{compareprofilesphotutils}. }
    \label{fig:compareotherprofiles}
\end{figure*}

This appendix provides profile comparisons between the other photometry methods for completeness.
The comparisons in \Fig{compareotherprofiles} between {\scriptsize PHOTUTILS}, {\scriptsize XVISTA}, and {\scriptsize GALFIT} display near identical behavior as reported for {\scriptsize AUTOPROF}.
The consistency of these results demonstrates the reliability of {\scriptsize AUTOPROF} as a SB profile extracting algorithm.  
The scatter about zero can be considered as a systematic error inherent in SB profile fitting.
The SB errors reported in {\scriptsize AUTOPROF} and other codes are only statistical, and can often be quite small; users should consider systematic errors of similar scale to those presented in \Fig{compareotherprofiles} and earlier sections before establishing final SB errors.

\Fig{compareotherprofiles}(left) illustrates that SB and $q$ for {\scriptsize PHOTUTILS} and {\scriptsize XVISTA} agree within ${\sim}2R_{23.5}$ before showing a small bias (similar to findings in \Sec{photutils}).
The inner $0.1R_{23.5}$ shows again a strong discrepancy in mag, PA, and q likely due to unstable central ellipses in {\scriptsize photutils} and simplistic bi-linear interpolation between pixels.
The middle and right hand panels in \Fig{compareotherprofiles} show comparisons with {\scriptsize PHOTUTILS} and {\scriptsize XVISTA} against {\scriptsize GALFIT} demonstrating the same behaviour seen in \Sec{galfit} where there is overall agreement, though with slightly larger scatter.
These figures are consistent with the description in \Sec{galfit} that the overly simplistic parametric models given to {\scriptsize GALFIT} cannot fully describe the complexity of late-type systems.
The central axis ratio values are also larger for non-parametric analyses compared to {\scriptsize GALFIT}; this is consistent with elliptical isophotes becoming rounder towards the centre as  expected in the presence of bulges and a circularizing PSF.
For low resolution features on scales of the PSF, the forward modelled {\scriptsize GALFIT} solution is superior.

\section{Example Fit Comparisons}
\label{app:fitcomparisons}

\begin{figure*}
    \centering
    \includegraphics[width=\textwidth]{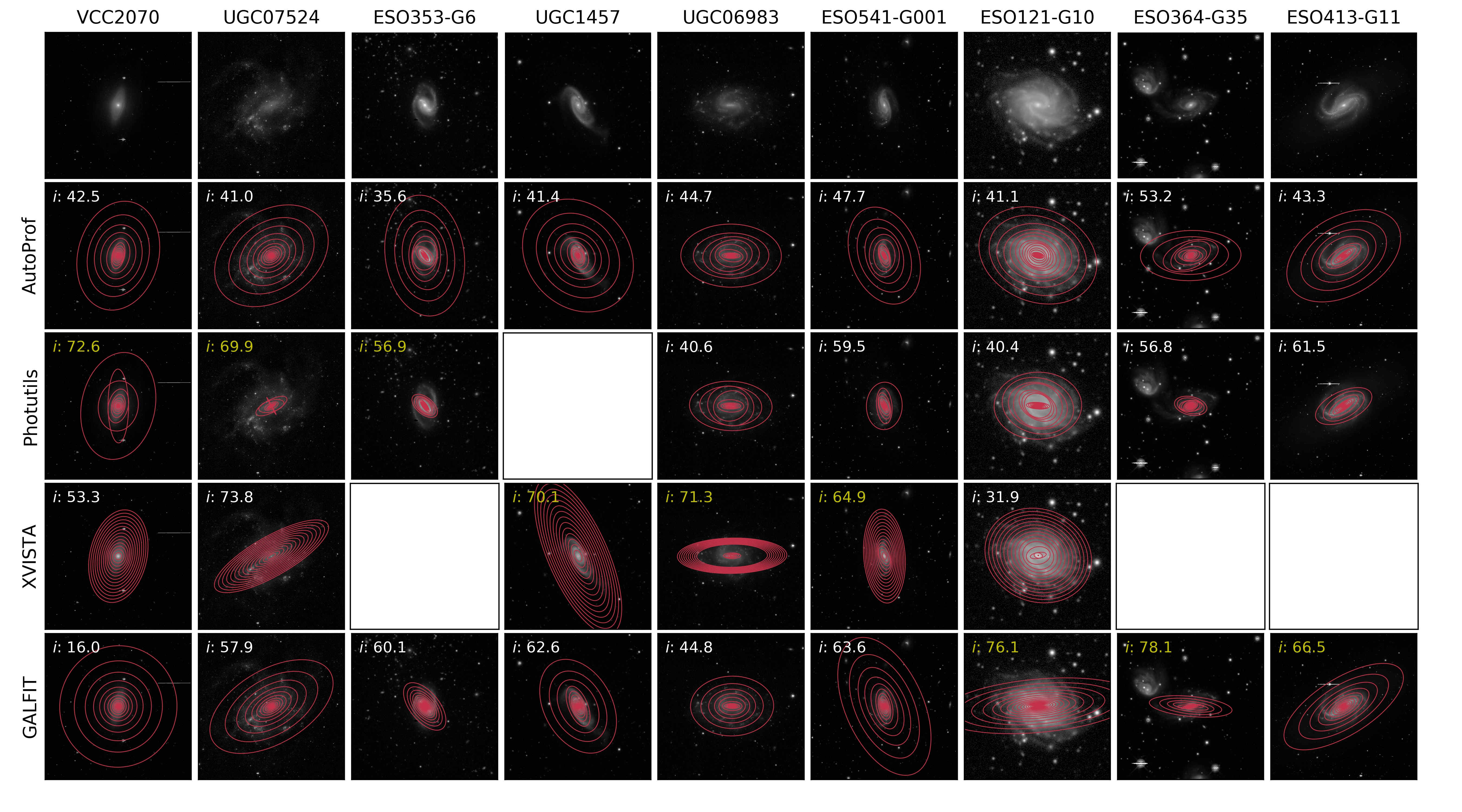}
    \caption{Comparison of most significant cases of discrepancy in inclination between {\scriptsize AUTOPROF} and the three other methods ({\scriptsize PHOTUTILS}, {\scriptsize XVISTA}, and {\scriptsize GALFIT}). 
    The top row shows an image of a challenging galaxy with its name overhead. The second row shows {\scriptsize AUTOPROF} with its fitted ellipses overlayed. 
    The third, fourth, and fifth rows show the fitted ellipse models for {\scriptsize PHOTUTILS}, {\scriptsize XVISTA}, and {\scriptsize GALFIT}, respectively. 
    The first three columns are galaxies selected for greatest discrepancy with {\scriptsize PHOTUTILS}, the middle three columns are discrepancies with {\scriptsize XVISTA}, and the last the columns are discrepancies with {\scriptsize GALFIT}. 
    The text inset gives the inclination (in degrees) from each method, with the cases of discrepancy highlighted in yellow. 
    Figures were left empty if the given isophotal fitter could not find a solution for the galaxy.} 
    \label{fig:allconflictI}
\end{figure*}

This appendix presents specific cases of profile disagreements.
\Fig{allconflictI} focuses on cases of most significant disagreement in galaxy inclination (evaluated at $R_{23.5}$) between {\scriptsize AUTOPROF} and {\scriptsize PHOTUTILS}, {\scriptsize XVISTA}, or {\scriptsize GALFIT}.
The first three columns show disagreement between {\scriptsize AUTOPROF} and {\scriptsize PHOTUTILS}, the middle set of three columns show disagreement between {\scriptsize AUTOPROF} and {\scriptsize XVISTA}, and the last three show disagreement between {\scriptsize AUTOPROF} and {\scriptsize GALFIT}.
The galaxies in this figure represent some of the most challenging in the PROBES data set, with nearly all of them including a bar, wide spiral arms, or both.
For each of the nine challenging galaxies we show the fitted isophotes from all four methods, in several cases there are multiple methods which fail to adequately represent the galaxy.

The first column shows \emph{VCC2070}, which has a large bar and several dust lanes through the disc.
In this case, the discrepancy comes from two un-masked stars on opposite sides of the galaxy which cause {\scriptsize PHOTUTILS} to seriously missalign intermediate isophotes (only one shown) that happen to be at $R_{23.5}$ where the inclination is measured.
This is a common issue for standard \citetalias{Jedrzejewski1987} fitting and can be circumvented with star masking or the regularization technique applied in {\scriptsize AUTOPROF}.
The second column shows \emph{UGC07524}, which has a bar-like feature and a faint, irregular, light distribution.
In this case {\scriptsize PHOTUTILS} is unable to model the light distribution beyond the bar feature and prematurely ends the light profile.
{\scriptsize AUTOPROF} fails to capture the bar feature as well, but adequately finds the global light distribution.
This galaxy also presented a challenge for {\scriptsize XVISTA}, which could only recover the bar feature, though this solution was extended to the optical edge of the galaxy.
The third column shows \emph{ESO353-G6}, a strongly barred system with a quick transition to two spiral arms and a wide opening angle.
Here, both {\scriptsize PHOTUTILS} and {\scriptsize GALFIT} could only recover the bar feature, while {\scriptsize AUTOPROF} fully represented the light distribution including the sharp transition from bar to spiral arms.

The fourth column shows \emph{UGC1457}, which has a bar-like feature at the centre and broad asymmetrical spiral arms.
{\scriptsize AUTOPROF} cannot adequately fit this galaxy, and this case was not identified by the fit checks.
{\scriptsize XVISTA}, however, succeeded in modelling the galaxy at all radii.
{\scriptsize PHOTUTILS} failed to find a solution for this galaxy and returned no result, while {\scriptsize GALFIT} recovered the global light distribution.
The fifth column shows \emph{UGC06983}, which has a central bar and faint spiral arms.
In this case, both {\scriptsize AUTOPROF} and {\scriptsize PHOTUTILS} obtained accurate models for the light distribution.
{\scriptsize XVISTA} fitted only the bar feature, though the solution is extended to the optical edge of the galaxy.
{\scriptsize GALFIT} (roughly) recovered the bar and accurately modelled the global light distribution using a two-component fit.
The sixth column shows \emph{ESO541-G001}, which has a bar, asymmetrical spiral arms and a faint extended disc.
Here {\scriptsize AUTOPROF} adequately recovered all features, including the faint extended dist.
{\scriptsize XVISTA} faithfully matched the main spiral structure of the galaxy, but failed to capture the faint extended disc.
{\scriptsize PHOTUTILS} accurately matched the features of this galaxy, but could not extend its solution as far as {\scriptsize AUTOPROF}; {\scriptsize GALFIT} fitted the bright central structures well, but missed the faint extended disc.

The seventh column shows \emph{ESO121-G10}, which has a bright central bar and prominent spiral arms.
All methods were able to recover the central bar feature, however the {\scriptsize GALFIT} solution fails to fit the disc, causing the large discrepancy in inclination with {\scriptsize AUTOPROF}.
The eighth column shows \emph{ESO364-G35}, which has a small central bar and faint, distinct spiral features at all scales of the disc.
{\scriptsize AUTOPROF} recovered the small central bar and the global structure of the galaxy while {\scriptsize GALFIT}'s solution was dominated by the bright central spiral arms causing it to miss the larger disc structure.
The ninth column shows \emph{ESO413-G11}, which has a strong bar and spiral arms with a wide opening angle.
Both {\scriptsize AUTOPROF} and {\scriptsize PHOTUTILS} accurately modelled the full galaxy light distribution, however {\scriptsize GALFIT} fitted only the large bar feature.

These cases of discrepancy between {\scriptsize AUTOPROF} and the other methods highlight a number of common problems encountered in elliptical isophote fitting.
With the exception of \emph{UGC1457}, {\scriptsize AUTOPROF} can handle these challenging cases in a fully automated fashion.
Overall, all four methods performed satisfactorily on most galaxies. They are differentiated by their treatment of challenging cases, their spatial extent (before reaching a nominal S/N level) in the galaxy's outskirts, their execution time, and other secondary considerations.

\section{Example Flagged {\scriptsize AUTOPROF} Fits}
\label{app:autoproffailedfits}

\begin{figure*}[b]
    \centering
    \includegraphics[width = 0.9\textwidth]{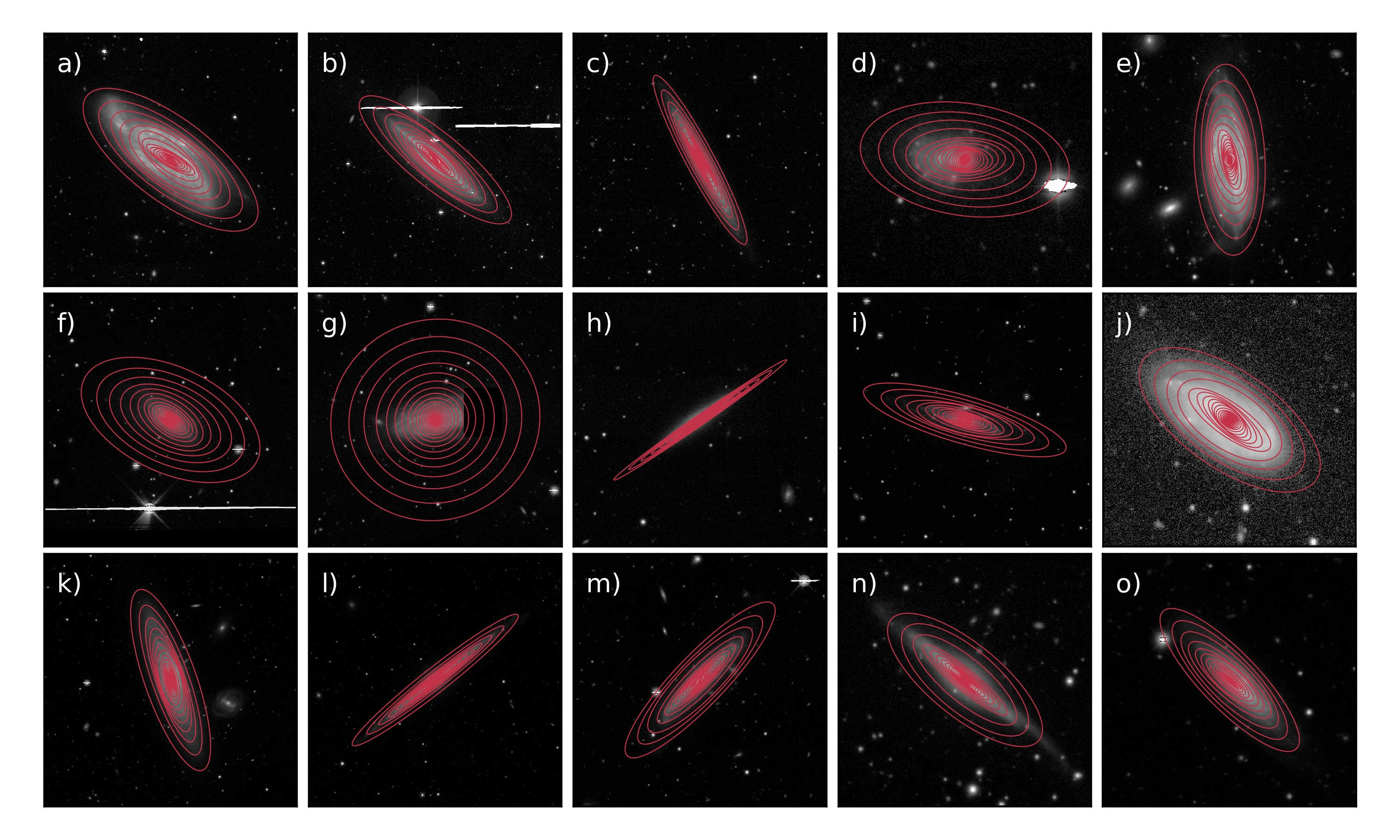}
    \caption{Cases of flagged {\scriptsize AUTOPROF} isophotal solutions as identified by the fit checks in \Sec{checkfit}. While the {\scriptsize AUTOPROF} algorithm may have converged to an optimal solution, some of these objects may simply not be adequately described by elliptical isophotes.}
    \label{fig:badfits}
\end{figure*}

We now present a random assortment of failed isophotal fits flagged by the \emph{checkfit} step of the {\scriptsize AUTOPROF} pipeline (see \Sec{checkfit}).
\Fig{badfits} presents a grid of failed fits demonstrating typical challenges that can be encountered.
The most common type of failed fits are shown in \subFig{badfits}{\it panels b, c, h, i, m, and n} where the isophotal analysis is (inappropriately) applied to edge-on galaxies.
Because edge-on galaxies do not present any aspect of their disc, surface photometry cannot and should not be attempted on those systems. 
{\scriptsize AUTOPROF} identifies failed fits as described in \Sec{checkfit}.
Since the fit checking step is not specifically designed for edge-on galaxies, the user should implement their own edge-on rejection method (e.g, via radial samples or an ellipticity cut).

\subFig{badfits}{\it panel d} is a case where the centre finding algorithm has failed.
No clear galaxy centre could be found; this is rather common for irregular galaxies and is reliably detected as a failed fit.
Cases in \subFig{badfits}{\it panels b, d, f, and o} all have a bright interloper along (or close to) the major axis.
In some cases, {\scriptsize AUTOPROF} can recover an adequate fit even though these are still flagged as potentially problematic. {\scriptsize AUTOPROF} cannot separate overlapping sources, and these cases require further attention and/or alternate analysis routines.
\subFig{badfits}{\it panels a, j, o, and k} are examples of asymmetric galaxies.
While the {\scriptsize AUTOPROF} isophotal solution follows the large scale features of these objects, elliptical isophotes should not be applied to strongly asymmetrical objects.
Finally, \subFig{badfits}{\it panel g} is a case of the {\scriptsize AUTOPROF} optimization algorithm being trapped in a local minimum and unable to find even an approximately correct solution.
The large block of missing data immediately adjacent to the galaxy is likely the culprit. 

\bsp	
\label{lastpage}
\end{document}